\documentclass[showpacs, oneside, twocolumn, prd, amsmath, amssymb, nofootinbib, superscriptaddress,preprintnumbers]{revtex4-1}

\usepackage[english]{babel}
\usepackage[letterpaper,top=2cm,bottom=2cm,left=3cm,right=3cm,marginparwidth=1.75cm]{geometry}
\usepackage{amsmath}
\usepackage{graphicx}
\usepackage[colorlinks=true, allcolors=blue]{hyperref}

\usepackage{cases}
\usepackage{amssymb}
\usepackage{amsfonts}
\usepackage{amssymb}
\usepackage{dcolumn}
\usepackage{bm}
\usepackage{bbm}
\usepackage{xcolor}
\usepackage{array}
\usepackage{subfigure}
\usepackage{wasysym}
\usepackage{amsmath}
\usepackage{float}
\usepackage[normalem]{ulem}

\def\l{\left}
\def\r{\right}
\def\ddd{\mathrm{d}}
\def\nn{\nonumber}

\begin{document}


\title{Flipped rotating axion non-minimally coupled to gravity: \\ Baryogenesis and Dark Matter}

\author{Chao Chen}
\email{cchao012@just.edu.cn}
\affiliation{School of Science, Jiangsu University of Science and Technology, Zhenjiang, 212100, China }

\author{Suruj Jyoti Das}
\email{surujjd@gmail.com}
\affiliation{Particle Theory  and Cosmology Group, Center for Theoretical Physics of the Universe,
Institute for Basic Science (IBS),
 Daejeon, 34126, Korea}

\author{Konstantinos Dimopoulos}
\email{konst.dimopoulos@lancaster.ac.uk}
\affiliation{Consortium for Fundamental Physics, Physics Department, Lancaster University, Lancaster LA1 4YB, United Kingdom}

\author{Anish Ghoshal}
\email{anish.ghoshal@fuw.edu.pl}
\affiliation{Institute of Theoretical Physics, Faculty of Physics, University of Warsaw, ul. Pasteura 5, 02-093 Warsaw, Poland}

\preprint{CTPU-PTC-25-04}

\begin{abstract}
We demonstrate that the co-genesis 
of baryon asymmetry and dark matter can be achieved through the rotation of an axion-like particle (but not the QCD axion), driven by a flip in the vacuum manifold's direction at the end of inflation.
This can occur if the axion has a periodic non-minimal coupling to gravity, while preserving the discrete shift symmetry. 
In non-oscillating inflation models, after inflation there is typically a period of kination (with $w = 1$). In this case, it is shown that the vacuum manifold of the axion is flipped and the axion begins rotating in field space, because it can slide across the decreasing potential barrier as in Ricci reheating. Such a rotating axion can generate the baryon asymmetry of the Universe through spontaneous baryogenesis, while at later epochs it can oscillate as dark matter. The period of kination makes the primordial gravitational waves (GW) generated during inflation sharply blue-tilted which constrains the parameter space due to GW overproduction, while being testable by next generation CMB experiments. 
As a concrete example, we show that such a cogenesis of baryon asymmetry and dark matter can be realized for the axion as the Majoron in the Type-I seesaw setup, predicting mass ranges for the Majoron  below sub eVs, with right-handed neutrino mass above $\mathcal{O}(10^{8})$ GeV. We also show that in order to avoid fragmentation of the axion condensate during the rotation, we require the non-minimal coupling \mbox{$\xi \sim (f/m_P)^2 $} or somewhat larger, where $f$ is the axion decay constant. 
\end{abstract}
\maketitle

\section{Introduction}





Cosmic inflation is 
a period of exponential increase in the size of the Universe which accounts for the otherwise fine-tuned initial conditions of the hot Big Bang history, and is responsible for generating the primordial density perturbations seeding structure formation~\cite{Starobinsky:1980te,Sato:1981qmu,Kazanas:1980tx,Guth:1980zm}. The measurements of the cosmic microwave background (CMB) radiation, help us unravel the microscopic model of inflation which come in many avatars. The recent CMB observations from the Planck satellite mission~\cite{Planck:2018vyg} have led to severe constraints on several inflationary models but neither it has been able to yet pin down upon a specific scenario nor it has given us definitive insights into embedding of the inflationary paradigm into fundamental particle physics theory like that of Grand Unification (GUT).
Upcoming CMB missions like the Simons Observatory~\cite{SimonsObservatory:2018koc}, the LiteBIRD satellite~\cite{Hazumi:2019lys, Sugai:2020pjw} or the ground-based CMB Stage 4 program~\cite{CMB-S4:2020lpa}, Ali CMB Polarization Telescope~\cite{Li:2017drr} and CMB-Bharat ~\cite{Adak:2021lbu} will be able to further investigate on this direction, particularly so concerning the measurement of the
tensor-to-scalar ratio $r$~\cite{Martin:2014rqa},  which basically depicts the energy scale of inflation which is determined once the BB-mode correlations are measured and dark radiation measurements of $\Delta N_{\rm eff}$. This leads to important science targets to go beyond this paradigm and also test the indirect evidence of such inflationary characteristics like for instance, observable signatures like primordial non-gaussianity which may point towards inflaton interactions or the microscopic details of the reheating epoch in the post-inflationary epoch~\cite{Bassett:2005xm,Martin:2010kz,Martin:2014nya,Meerburg:2019qqi,Allahverdi:2010xz,Amin:2014eta}.

Of particular interest are non-oscillatory models of inflation, which are driven by a runaway flat direction in field space \cite{Felder:1999pv}. 
These have been frequently envisaged when modeling quintessential inflation \cite{Peebles:1998qn} (for recent reviews see Refs.~\cite{Bettoni:2021qfs,deHaro:2021swo,Wetterich:2022brb,Jaman:2022bho,deHaro:2021swo}), where the inflaton and the quintessence \cite{Caldwell:1997ii} fields are unified and cosmic inflation is treated in the same theoretical framework as dark energy.

In quintessential inflation, because the inflaton has to survive until the present and explain dark energy, the Universe after inflation has to be reheated by means other than the decay of the inflaton field.
A prominent mechanism for this is Ricci reheating \cite{Dimopoulos:2018wfg,Opferkuch:2019zbd,Bettoni:2021zhq}. This amounts to a
minimal Hubble-induced reheating mechanism which is based on the time-dependence and change of sign of the Ricci scalar. 

In scenarios with the inflationary paradigm having non-oscillatory quintessential potential for the inflaton, 
rather generically 
in the post-inflationary era, 
the Universe becomes dominated by the kinetic energy density of the inflaton. This
period is called kination \cite{Joyce:1997fc,Gouttenoire:2021jhk}.
The essential idea of Ricci reheating in this framework is that, any spectator scalar field  
non-minimally coupled to gravity undergoes a second-order phase transition during the time when the Universe transitions from an inflating background to a kination dominated background when the Ricci scalar changes sign and turns negative. In Ricci reheating, after the phase transition, the non-minimally coupled scalar field undergoes coherent oscillations, which amount to particles that decay and reheat the Universe \cite{Dimopoulos:2018wfg,Opferkuch:2019zbd,Bettoni:2021zhq,Laverda:2023uqv}. In this paper, the above phase transition triggers a rotation in field space.


In the most generic context there exists several motivations of having spontaneously broken global symmetries involving solutions to a myriad of problems in the Standard Model (SM) of particle physics, to name a few,  the Peccei-Quinn (PQ) symmetry~\cite{Peccei:1977hh} as a solution to the strong CP problem, lepton number or baryon number symmetry~\cite{Chikashige:1980ui} in the context with the microscopic origin the SM neutrino masses, and flavor symmetries to explain the structure of CKM quark or PNMS neutrino mass  matrices as explanation of the observed pattern of the fermion masses and mixings~\cite{Froggatt:1978nt}.  Nonetheless, the breaking of any such symmetries inevitably leads to the existence of very light degrees of freedom, known as Nambu-Goldstone bosons (NGB) which have different names in different physics contexts, namely the QCD axion~\cite{Weinberg:1977ma,Wilczek:1977pj} for the PQ symmetry, a Majoron for lepton number~\cite{Chikashige:1980ui}, and a familon or flavon~\cite{Davidson:1981zd, Reiss:1982sq,Wilczek:1982rv,Davidson:1983fy,Davidson:1983fe} for the case involving flavor symmetry. For the purpose of this paper, we simply focus on any such pseudo 
NGB
or axion-like particle (ALP) generically but we dub it as ``axion" for the sake of simplicity (to avoid thermal corrections, we do not consider the QCD axion). The axions we study can even have a fundamental origin in the context of the string axiverse and can be present in more avatars \cite{Arvanitaki:2009fg}.

Such axion-like particles can couple to the Standard Model through various interactions following gauge-invariance \cite{Georgi:1986df, Jaeckel:2010ni, Ringwald:2014vqa, Bauer:2017ris, Brivio:2017ije}, in a manner similar to that of the QCD axion. However,  unlike the QCD axion, their mass is generally unrelated to the decay constant, leading to more possible parameter space in axion experiments. These interactions are suppressed by the symmetry breaking scale, and can occur with gluons, photons,  electroweak gauge bosons or fermions, and are being searched for in several experiments \cite{Choi:2020rgn, Giannotti:2022euq}.  In our study, we will consider such interactions with fermions carrying lepton number \cite{DallaValleGarcia:2023xhh, Domcke:2020kcp, Chun:2023eqc}, which, as we will see, has close connections with neutrino mass generation and leptogenesis \cite{Domcke:2020kcp, Chun:2023eqc, Fong:2023egk,Datta:2024xhg}.

Now, the common lore is that the axion field is initially static and misaligned at a non-zero field value, and later oscillates when its mass becomes comparable to the Hubble expansion rate of the Universe \cite{Preskill:1982cy,Abbott:1982af,Dine:1982ah}. However, the axion field may instead be initially rotating in the complex field space \cite{Gouttenoire:2021jhk,Li:2013nal,Co:2021lkc,Gouttenoire:2021wzu,Harigaya:2023mhl,Harigaya:2023pmw,Chung:2024ctx,Duval:2024jsg}. 
Such an initial kick in the angular direction is usually realized at high-radial field values, through operators explicitly breaking the global symmetry. These dynamics at high radial field values have been crucial in cosmology, for e.g., in realizing Affleck-Dine baryogenesis~\cite{Affleck:1984fy}.  Such setups have been explored further in the context of baryogenesis~\cite{Co:2019wyp}, for modifying the dark matter abundance through the 
kinetic misalignment~\cite{Co:2019jts,Co:2020dya}, for cogenesis of both dark matter and the baryon asymmetry~\cite{Co:2020xlh, Co:2020jtv} and also in several other related contexts~\cite{Jeong:2013axf, Jeong:2013xta, Higaki:2014ooa, Co:2021rhi,  Kawamura:2021xpu, Co:2021qgl, Co:2022aav, Barnes:2022ren, Co:2022kul, Badziak:2023fsc, Berbig:2023uzs, Chun:2023eqc, Chao:2023ojl, Barnes:2024jap, Datta:2024xhg}.

In this paper, we explore the possibility of generating such a rotation of the axion, without delving into the above route, such that the radial mode still remains at the minima of its potential. This can take place if the height of the axion potential barrier keeps on decreasing, such that the axion can slide over,in a similar manner to Ricci reheating 
\cite{Opferkuch:2019zbd,Bettoni:2021zhq} 
(see also Refs.~\cite{Figueroa:2016dsc,Nakama:2018gll}). Such an alternative was explored recently in~\cite{Chun:2024gvp}, with a time-dependent axion decay constant, realized through symmetry non-restoration.  Here, we propose a scenario where 
an axion couples non-minimally to gravity, in a periodic form, thereby respecting the shift-symmetry 
of any pNGB action~\cite{Ferreira:2018nav}. 
In contrast to Refs.~\cite{Gouttenoire:2021jhk,Li:2013nal,Co:2021lkc,Gouttenoire:2021wzu,Harigaya:2023mhl,Duval:2024jsg},
our rotating axion always remains a spectator field.
Because of its non-minimal coupling to gravity which depends on the cosmological background through the Ricci scalar $R$, the axion vacuum manifold evolves depending on the barotropic parameter $w$. While $w$ changes from $-1$ during inflation to 
$+1$ during kination
(right after inflation), the tilt of the axion vacuum manifold flips in the opposite direction \footnote{Refs. \cite{Takahashi:2019pqf, Huang:2020etx} considered a similar flip without utilising the non-minimal coupling to gravity.}, generating an initial kinetic energy density for the axion.  Interestingly, if the Universe happens to be kination-dominated after inflation, the potential barrier height being determined by the cosmological background, redshifts and decreases fast enough such that the axion can keep sliding across the barrier, and hence rotate in the complex field space. 

We utilise such a dynamics of the axion field to successfully generate both the baryon asymmetry and the dark matter abundance, dubbed as cogenesis. While the baryon asymmetry can be generated during the rotation through spontaneous baryogenesis~\cite{Cohen:1987vi, Cohen:1988kt}, the dark matter abundance can arise at a much later epoch, after the bare mass potential of the axion starts to dominate over its kinetic energy density. The observed value of baryon asymmetry constrains the non-minimal coupling $\xi$, while the dark matter abundance relates the axion decay constant to its mass. Importantly, because of the kination-dominated stiff era, the spectrum of the primordial gravitational waves (GWs) generated during inflation becomes blue-tilted for modes entering the horizon during kination. Such an increase in the GW amplitude tightly constrains our parameter space, because of an excess of GWs during Big Bang Nucleosynthesis (BBN).



The paper is organised as follows: We discuss our framework and idea in Section \ref{sec2}, along with the relevant constraints, especially on the non-minimal coupling $\xi$. Section \ref{sec:Baryo} discusses how the axion rotation can be used to generate the baryon asymmetry. In Section \ref{ALP}, we elaborate the axion dynamics at later epochs leading to the observed dark matter abundance. Section \ref{sec:GW} reviews the primordial gravitational wave spectra generated during inflation, with the important lower bound on the reheating temperature. In Section \ref{sec:Eg}, we demonstrate a viable particle physics setup to realize our scenario. We discuss briefly the issues related to axion fragmentation and Kibble problem, in Section \ref{sec:Fragmentation} and \ref{sec:Kibble} respectively, along with ways to overcome them. Finally, we present our conclusions in Section \ref{sec:conc}.

Throughout the paper we use natural units for which \mbox{$\hbar=k_B=c=1$} and \mbox{$8\pi G=m_P^{-2}$}, with \mbox{$m_P=2.43\times 10^{18}\,$GeV}
being the reduced Planck mass. The signature of the metric is positive.

\medskip

\section{The Framework}\label{sec2}
The Lagrangian density of the model is
\begin{equation}
    {\cal L}=\frac12 m_P^2\gamma^2(\phi)R-
    \frac12 
    (\partial\phi)^2-V(\phi)
    \label{L}
\end{equation}
introduced in Ref.~\cite{Ferreira:2018nav, Salvio:2021lka,Ghoshal:2023jvf}. In the above, \mbox{$(\partial\phi)^2\equiv\partial_\mu\phi\,\partial^\mu\phi$} and 
\begin{equation}
    \gamma^2(\phi)\equiv 1+\xi\left[1-\cos(\phi/f)\right]\,,
    \label{gamma}
\end{equation}
where the unity in the right-hand side denotes the usual Einstein-Hilbert term, and
\begin{equation}
    V(\phi)=M^4\left[1-\cos(\phi/f)\right]\,,
    \label{V}
\end{equation}
where $\xi$ the non-minimal coupling of the axion 
field $\phi$ to the Ricci scalar $R$, $f$ is the axion decay constant with \mbox{$0<f\ll m_P$} and $M$ is the 
symmetry breaking scale with \mbox{$0<M\ll f$}. The above Lagrangian density shows that the theory respects the discrete shift symmetry \mbox{$\phi\rightarrow\phi+2\pi f$}.  



\begin{figure*}[t]
    \centering
    \includegraphics[scale=0.45]{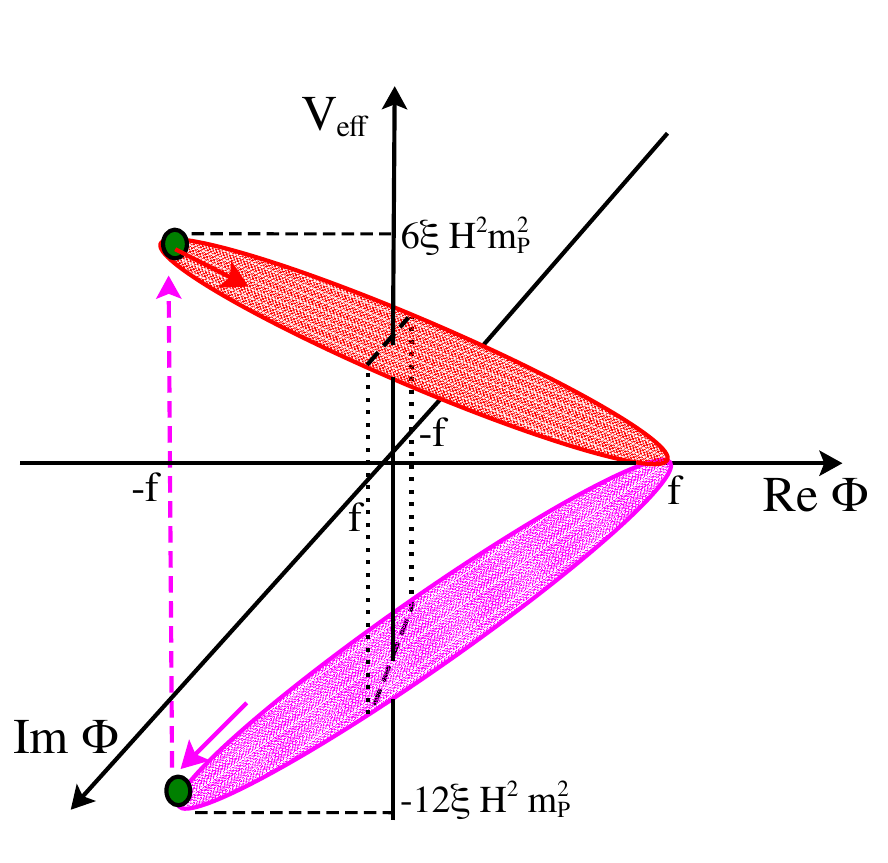}
    \includegraphics[scale=0.45]{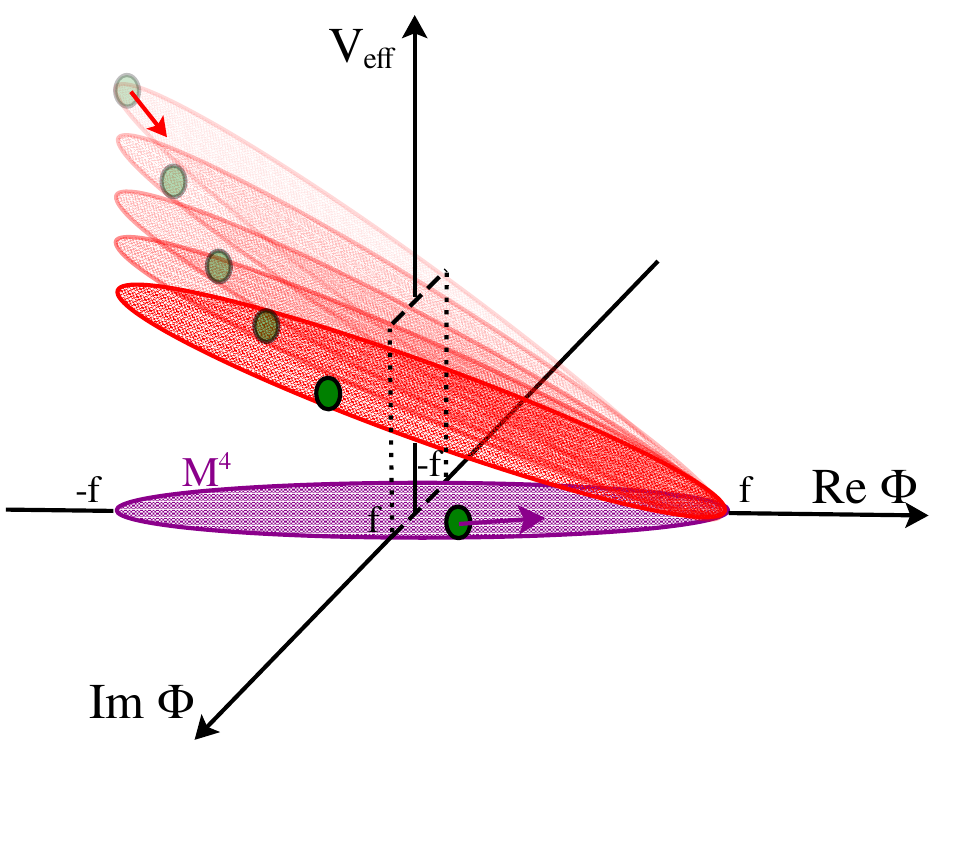}
    \caption{Schematic diagram to visualise the evolution of the axion vacuum manifold through the cosmic history.
    {\it Left panel}: Tilt of the vacuum manifold at the end of inflation. During inflation, the vacuum manifold is depicted by the purple ellipse. The axion is driven to its minimum at \mbox{$\phi=\pi\,f$}. After inflation, kination begins. The vacuum manifold is tilted in the opposite direction, and it is now depicted by the red ellipse. Right after the tilt changes, the axion finds itself at the maximum and starts rolling towards the new minimum, which is now at \mbox{$\phi=0$}. {\it Right panel}: Evolution after inflation. While the axion rolls towards the minimum at \mbox{$\phi=0$}, the tilt of the vacuum manifold diminishes with time (red ellipses), in tandem with the axion kinetic energy density. This means that, the rotating axion can overcome the potential hill when climbing back the vacuum manifold, because the hill has diminished accordingly. The figure also depicts the disappearing of the tilt at reheating, when $R=0$. The axion continues to rotate around an almost horizontal vacuum manifold (purple ellipse), given that its potential is given by Eq.~\eqref{V}, with $M$ very small.}
    \label{fig:tilt}
\end{figure*}

When \mbox{$|\phi|\ll f$}, 
\mbox{$[1-\cos(\phi/f)]\simeq\frac12(\phi/f)^2$}. Then, the theory in Eq.~\eqref{L} becomes
    \begin{equation}
{\cal L}\simeq\frac12 m_P^2R+\frac{1}{4}\!\left(\frac{m_P}{f}\right)^2 \!\! \xi R\phi^2-\frac12(\partial\phi)^2-\frac12\frac{M^4}{f^2}\phi^2\,,
\label{Lsmallphi}
\end{equation}
which is reminiscent of the usual Lagrangian density of a scalar field with a non-minimal coupling to gravity\footnote{Refs. \cite{Takahashi:2015waa,Berbig:2024ufe} considered such a non-minimal coupling of the axion to gravity, where the latter anticipated a similar dynamics as in our scenario.}. However, for a rotating axion the approximation \mbox{$|\phi|\ll f$} is not valid (\mbox{$0\leq \phi< 2\pi f$}), and we have to consider the full Lagrangian density in Eq.~\eqref{L} without this approximation.

Equation~\eqref{L} can be written as
\begin{eqnarray}
{\cal L} & = & \frac12m_P^2 R-\frac12(\partial\phi)^2\nonumber\\
    & & -\left(M^4-\frac12\xi m_P^2 R\right)[1-\cos(\phi/f)] ~.
    \label{Lphi}
\end{eqnarray}
The non-minimal coupling term is much smaller than the Einstein-Hilbert term (which is dominant) for \mbox{$\xi\ll 1$}, but it can be compared with $M^4$ that can be very small (e.g., \mbox{$M^4\ll\xi m_P^2R$} during inflation). Therefore, for small $\xi$, we are effectively in Einstein gravity with the non-minimal coupling practically only contributing to the effective potential, as in Ricci reheating \cite{Dimopoulos:2018wfg,Opferkuch:2019zbd,Bettoni:2021zhq}.

In FRW spacetime \mbox{$R=3(1-3w)H^2$}, where $w$ and $H$ are the barotropic and the Hubble parameter, respectively. We consider that our axion is a spectator field in a non-oscillatory inflationary scenario, where inflation (with \mbox{$w=-1$}) is followed by a period of kination (with \mbox{$w=1$}). We see, therefore, that during inflation, the axion effective potential is 
\begin{equation}
    V_{\rm eff}(\phi)\simeq -6\xi m_P^2 H^2[1-\cos(\phi/f)] ~,
    \label{Veffinf}
\end{equation}
while during kination we have
\begin{equation}
    V_{\rm eff}(\phi)\simeq 3\xi m_P^2 H^2[1-\cos(\phi/f)] ~,
    \label{Veffkin}
\end{equation}
where we have ignored $M$ as it can be very small as mentioned above. We see that the prefactor in $V_{\rm eff}$ changes sign. This means that the sinusoidal axion effective potential changes phase, such that the minimum, which is at \mbox{$\phi=\pi\,f$} during inflation becomes a maximum in kination, when the minimum is at zero.  

One can picture this effect as follows, see Fig.~\ref{fig:tilt}. During inflation, the sinusoidal potential corresponds to a tilted circle in field space. The axion field gradually rolls down to its minimum at $\pi\,f$ (The rolling will be justified later, see Eq.~\eqref{xirange}). At the end of inflation, as kination begins, the circular vacuum manifold is tilted in the opposite direction, such that the axion finds itself at the peak of its potential, and starts rolling towards the new minimum which is at zero. 

We consider that the rolling axion remains subdominant. After inflation, Eq.~\eqref{Veffkin} suggests that the maximum potential density is \mbox{$V_{\rm eff}^{\rm max}=6\xi m_P^2H^2$}. Thus, we require
\begin{equation}
    1\gg\frac{V_{\rm eff}^{\rm max}}{\rho}=2\xi ~,
    \label{ximax}
\end{equation}
where we considered \mbox{$\rho=3m_P^2H^2$}. Therefore, \mbox{$\xi\ll 1$}, consistent with our previous assumptions.

Following the same logic as Ref.~\cite{Opferkuch:2019zbd} (see also
Refs.~\cite{Bettoni:2021zhq,Bettoni:2018utf,Bettoni:2024ixe}), we expect that the rolling axion is not halted by the effective potential hill at \mbox{$\phi=\pi\,f$} after each cycle, because the size of this hill is decreasing with time in the same way that the energy density of the rolling axion is decreasing with time. Indeed, for the rolling axion, if its roll is not impeded and does not lead to oscillations, we have \mbox{$\rho_\phi\propto a^{-6}\propto t^{-2}$}, where we considered that \mbox{$a\propto t^{1/3}$} during kination. In the same way, the size of the effective potential hill is \mbox{$V_{\rm eff}^{\rm max}=6\xi m_P^2H^2\propto t^{-2}$} because \mbox{$H=1/3t$} during kination. 

For the rolling of the axion to occur we need that 
\mbox{$||V_{\rm eff}''||>(\frac{3}{2}H)^2$},
where \mbox{$||V_{\rm eff}''|| = 3\xi m_P^2 H^2/f^2$}. This condition suggests 
\mbox{$\xi>\frac{3}{4}(f/m_P)^2$}, such that the range of $\xi$ is
\begin{equation}
    \frac{3}{4}\left(\frac{f}{m_P}\right)^2<\xi\ll \frac{1}{2} ~.
    \label{xirange}
\end{equation}
This condition also makes sure that during inflation the axion is driven towards the minimum expectation value \mbox{$\phi=\pi\,f$}. 
Because the axion is heavy during inflation, it does not undergo particle production. As a result, it does not introduce an isocurvature perturbation. 

At some point reheating takes place, when a subdominant radiation energy density dominates the Universe and the usual radiation era of the hot Big Bang begins. During the radiation era \mbox{$R=0$}, which means that the axion becomes exactly canonical, with the potential given in Eq.~\eqref{V}.

As we have assumed that $M$ is very small, we expect that the effective axion mass is \mbox{$m_\phi^2=||V''(\phi)||\simeq M^4/f^2\ll H^2$} after reheating. 
This means that the rotating axion eventually freezes with \mbox{$\phi={\cal O}(f)$}, until the decreasing \mbox{$H(t)=1/2t$} catches up with $m_\phi$. 
At that point, the axion unfreezes and begins coherent oscillations in its potential, which soon becomes quadratic with mass \mbox{$m_\phi=M^2/f$}. The energy density of the oscillating axion condensate decreases as pressureless matter such that \mbox{$\rho_\phi\propto a^{-3}$} from now on. As a result, the axion condensate eventually dominates the radiation background. 

After that, a matter era begins, when \mbox{$R=3H^2(t)$}. However, this occurs very late, such that \mbox{$m_\phi^2\gg R$} and the axion remains approximately canonical. In principle, modifications of gravity are switched on, but we can safely ignore them because, when \mbox{$\phi\ll f$}, Eq.~\eqref{gamma} suggests that \mbox{$\gamma^2\simeq 1+\frac12\xi(\phi/f)^2\approx 1$}. If the oscillating axion dominates the Universe at the time of matter-radiation equality, then it can be the dark matter.

A lower bound on the axion decay constant $f$ can be estimated as follows. The axion is the angular direction of the complex field 
\mbox{$\Phi=|\Phi|e^{i\theta}$}, with 
\mbox{$\theta=\phi/f$}. Perturbatively, we expect the potential of $|\Phi|$ to be
\begin{equation}
    V(|\Phi|)\sim\lambda(|\Phi|^2-f^2)^2 ~,
\end{equation}
where $0<\lambda\lesssim1$ is its self-coupling constant. To the above we add a temperature correction due to interaction with the thermal bath
\begin{equation}
    \Delta V(|\Phi|)\sim g^2T^2|\Phi|^2\,,\label{eq: intrn}
\end{equation}
where $0<g\lesssim1$ is the interaction coupling constant. Therefore, the effective mass of $|\Phi|$ is 
\begin{equation}
    (m_{|\Phi|}^{\rm eff})^2\sim-\lambda f^2+g^2T^2 ~.
\end{equation}
During inflation, the Universe is supercooled and \mbox{$T\rightarrow 0$}.
Consequently, \mbox{$|\Phi|=f$} and the axion exists. We would like our axion to keep existing so that, high temperature after inflation is not enough to render $(m_{|\Phi|}^{\rm eff})^2$ positive and send \mbox{$|\Phi|\rightarrow 0$}. This means that there must be an upper bound on $T$ such that
\begin{equation}
    T<T_{\rm max}\sim (\sqrt\lambda/g)f ~.
    \label{Tmax}
\end{equation}
Thus, assuming that \mbox{$\lambda\sim g\sim 1$}  and because \mbox{$T\geq T_{\rm reh}$}, a crude estimate of a lower bound on $f$ is \mbox{$f>T_{\rm reh}$}.
However, we can do better than that.

Assuming that the thermal bath is generated at the end of inflation (there are many mechanisms that do so naturally, e.g., by instant preheating \cite{Felder:1998vq,Dimopoulos:2017tud} or Ricci reheating \cite{Dimopoulos:2018wfg,Opferkuch:2019zbd,Bettoni:2021zhq}), we obtain an estimate of the maximum temperature $T_{\rm max}$ in the following way. After the end of inflation there is a period of kination, when the density of the Universe scales as \mbox{$\rho\propto a^{-6}$}, while radiation, which appears at the end of inflation, scales as \mbox{$\rho_r\propto a^{-4}$}. This means that
\begin{equation}
    \left.\frac{\rho}{\rho_r}\right|_{\rm end}\!\!\!\!=\left.\frac{\rho}{\rho_r}\right|_{\rm reh}\!\!\left(\frac{a_{\rm reh}}{a_{\rm end}}\right)^2\!\!\Rightarrow
    \left(\frac{a_{\rm reh}}{a_{\rm end}}\right)^2\sim\frac{H_{\rm end}^2m_P^2}{T_{\rm max}^4}\,.
    \label{rhobound1}
\end{equation}
Now, for radiation we find 
\begin{equation}
    \left(\frac{a_{\rm reh}}{a_{\rm end}}\right)^2\simeq\sqrt{\frac{\rho_r^{\rm end}}{\rho_r^{\rm reh}}}\sim\frac{H_{\rm end}m_P}{T_{\rm reh}^2}\,.
    \label{rhobound2}
\end{equation}
Combining Eqs.~\eqref{rhobound1} and \eqref{rhobound2}, we find the bound
\begin{equation}
    f>T_{\rm max}\sim\left(H_{\rm end}m_P T_{\rm reh}^2\right)^{1/4}\,.
    \label{Tmaxbound}
\end{equation}
This bound ensures that \mbox{$|\Phi|\neq 0$} and the axion always exists. Considering GUT-scale inflation \mbox{$H_{\rm end}\sim 10^{-5}m_P$}. 
Then, because \mbox{$T_{\rm reh}\gtrsim 10^7\,$GeV} to avoid overproduction of gravitational radiation at BBN (cf. Eq.~\eqref{BBNbound}), we obtain 
\begin{equation}
    f\gtrsim 10^{11}\,{\rm GeV}\,.
\end{equation}
This constraint is relaxed when we consider a reheating mechanism with $T_{\rm max}$ closer to $T_{\rm reh}$ (e.g., curvaton reheating~\cite{Feng:2002nb,BuenoSanchez:2007jxm}
or reheating because of primordial black hole evaporation~\cite{Dalianis:2021dbs}) or if the interaction of $\Phi$ with the thermal bath is suppressed, $g\ll$1.



Before closing this section, we verify our claim of axion rotation above by numerically solving the equation of motion for the non-minimally coupled axion in a kination background.  The equation of motion for $\theta=\phi /f$ is given by 
\begin{align}
   \ddot{\theta}+3H\dot{\theta}+ V'(\phi)/f=0 \,, \label{eq:EOMthet}
\end{align}
where $V'(\phi)/f=\frac{3\xi m_P^2 H^2}{f^2} \sin\theta$. Now, after the end of inflation and flip of the axion vacuum manifold, $\theta$ is located at $\theta_i=-\pi$ as explained earlier. The axion receives an initial kick due to the tachyonic mass at the top of the potential. The magnitude of this quantum kick can be estimated as $\delta \phi = m/2\pi$ per Hubble time $\delta t= H^{-1}$ \cite{Felder:2001kt}, where $m^2=-V''(\theta_i)$ indicates the tachyonic mass-squared at the top of the potential. Thus, the initial kinetic energy at $\theta_i= -\pi$ can be estimated as
\begin{align}
\frac12 f^2\dot\theta_i^2
  = \frac{1}{2} \left(\frac{\delta\phi}{\delta t}\right)^2\!\!\simeq \frac{3 \xi}{8 \pi^2}\left(\frac{H_{\rm end}}{f}\right)^2\!\! H_{\rm end}^2 m_P^2\;,\label{eq:thetdotin} 
\end{align}
where we considered Eq.~\eqref{Veffkin}. In the limit of large $\xi$, i.e. $\xi \gg \left(f/m_P\right)^2$, the maximum velocity attained can be estimated as
\begin{align}
  \frac{1}{2} f^2 \dot{\theta}_{\rm m}^2 &\simeq 6 \xi m_P^2 H_{\rm end}^2 \\
\implies \dot{\theta}_{\rm m}&\simeq \frac{ \sqrt{12\xi} m_P H_{\rm end}}{f} ~,\label{eq:dotthet}   
\end{align}
where $H_{\rm end}$ denotes the Hubble scale at the end of inflation (\mbox{$H_{\rm end}\simeq 10^{13}\,$GeV} for GUT-scale inflation).\footnote{To avoid excessive isocurvature perturbations, since the tachyonic kick is stochastic, we require \mbox{$\dot\theta_i<\dot\theta_{\rm m}$}. This translates into the bound \mbox{$2f\gtrsim H/2\pi$}, which is comparable to the bound discussed after Eq.~\eqref{xirange+}.}
When $\xi$ is close to the lower limit $\left(f/m_P\right)^2$ (cf. Eq. \eqref{xirange}), the Hubble friction becomes comparable to the effective mass, and the axion is slowed down when it moves to the minimum. By this time, the potential energy also decreases, and the above estimate of the velocity $\dot{\theta}_m$ no longer holds.

\begin{figure*}
\includegraphics[width=0.4
\textwidth]{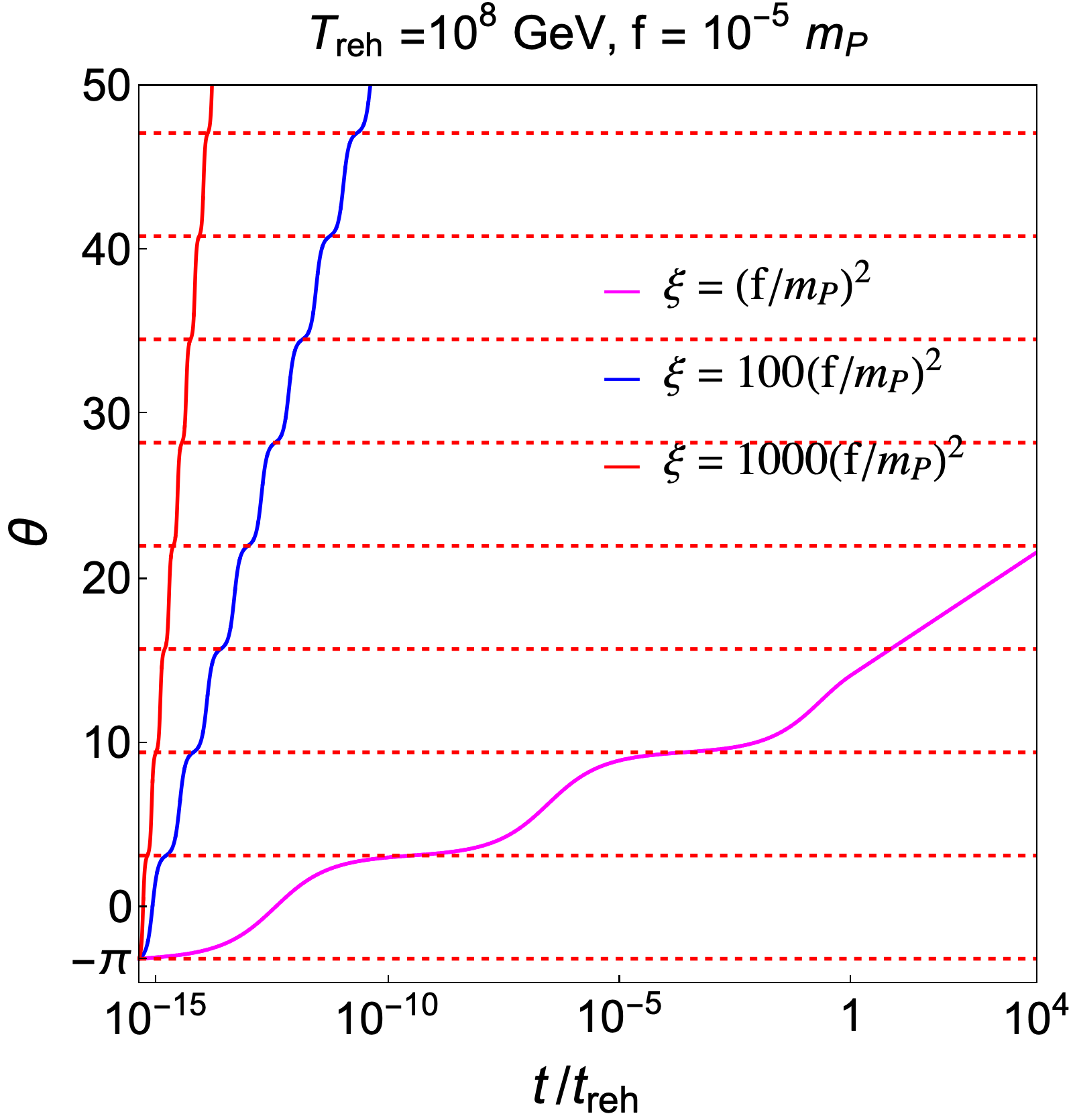}~~
\includegraphics[width=0.4
\textwidth]{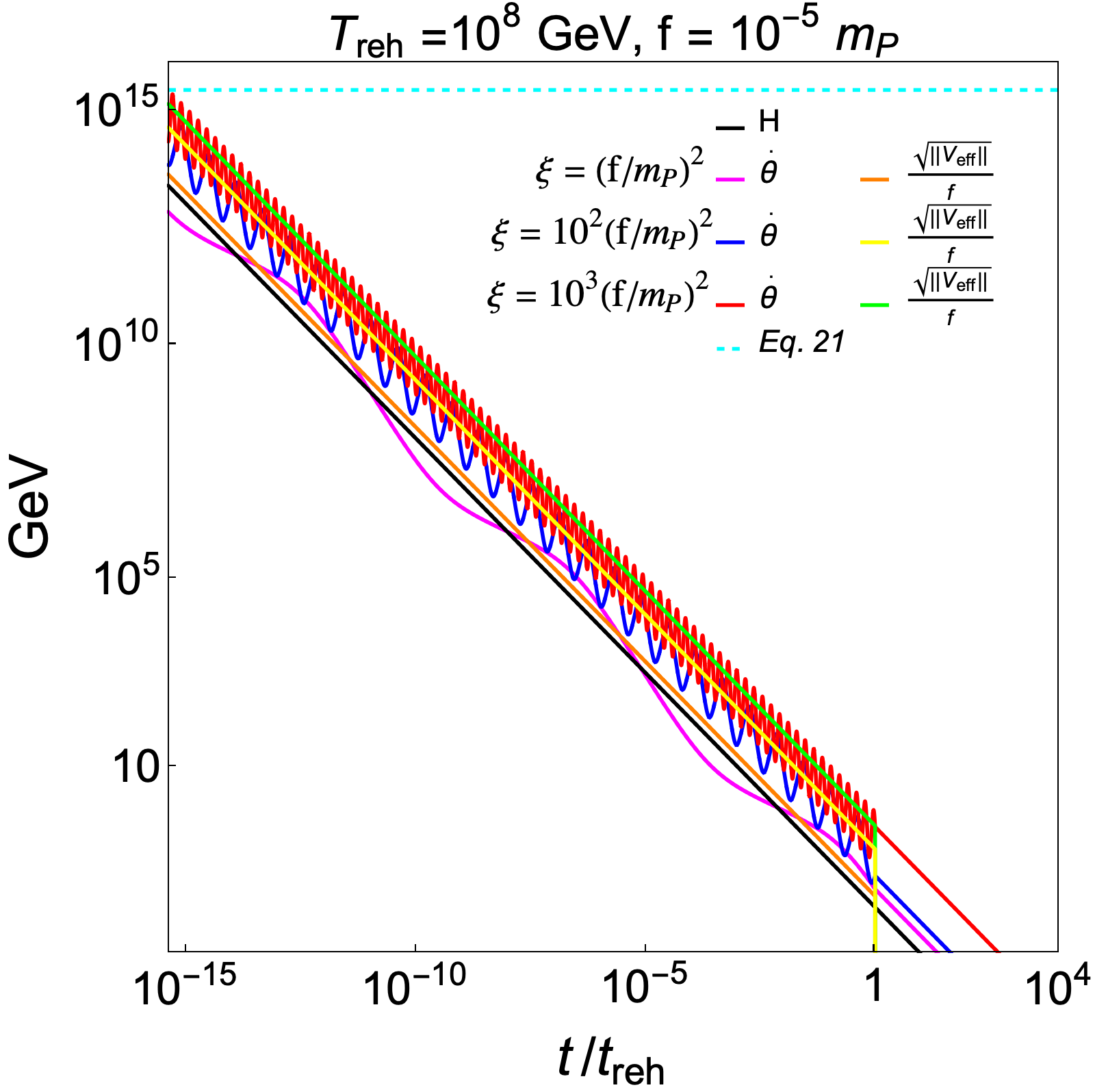}
    \caption{\textit{Left panel:} The rotation of the axion ($\theta$), with the red dashed lines indicating the potential barriers at $-\pi,\pi, 3 \pi, 5 \pi, ...~$. The different colors denote different values of $\xi$, with rotation taking place even for $\xi=\left(f/m_P\right)^2$. \textit{Right panel:} Evolution of $\dot{\theta}, H, \sqrt{||V_{\rm eff}||}/f$ 
     from the end of inflation until reheating. The black line denotes the Hubble parameter $H$, while the colored lines indicate $\dot{\theta}$ and $\sqrt{||V_{\rm eff}||}/f$ for several values of $\xi$. The velocity $\dot{\theta}$ oscillates as the axion travels along the potential barriers. For large $\xi$, for instance $\xi= 1000\left(f/m_P\right)^2$, the maximum velocity after half rotation $\theta_m$ is approximately  estimated in Eq. \eqref{eq:dotthet}, which is shown by the dashed horizontal line. This is because the rotation period is much smaller than a Hubble time so that the variation of the height of the potential hill is negligible. Note that the kinetic energy determined by $\dot{\theta}$ becomes larger than the barrier height $\sqrt{||V_{\rm eff}||}/f$, which leads to successful rotation. At $t=t_{\rm reh}$, the effective potential diminishes to the bare potential given by Eq. \eqref{V}.}
    \label{fig:numthet}
\end{figure*}

However, we find that even for values of $\xi$ close to this lower range, the axion still rotates, albeit very slowly. Recall that after reheating, the axion continues to rotate in an almost vanishing potential determined by $M$ (cf. Eq. \eqref{V}). We solve Eq.~\eqref{eq:EOMthet} for $\theta(t)$ using the above value of $\dot\theta_i$ (from Eq. \eqref{eq:thetdotin}) as the initial condition at \mbox{$\theta_i = -\pi$}. The numerical results are demonstrated in Fig.~\ref{fig:numthet}, considering some benchmark values of $\xi$. The plot in the left panel shows the evolution of $\theta$, starting from $-\pi$ and subsequently crossing the barriers at $\pi, 3 \pi, 5 \pi, ...$, which are shown by the red dashed lines. The plot in the right panel shows the evolution of $\dot{\theta}, H, \sqrt{||V_{\rm eff}||}/f$. Note that all these quantities scale similarly ($\propto 1/t$), as explained earlier. Thus, as long as the kinetic energy determined by $\dot{\theta}$ is larger than the effective potential barrier $\sqrt{||V_{\rm eff}||}/f$, with the Hubble rate subdominant, the rotation of the axion can persist. Note that, after reheating, the axion continues to rotate until $\dot{\theta}$ reaches $M^2/f$, as discussed above.

\section{Baryogenesis}\label{sec:Baryo}

The rotating axion can play a role in generating the baryon asymmetry of the Universe~\cite{Cohen:1987vi, Cohen:1988kt, Domcke:2020kcp}. 
The important feature here is that the required rotation can be generated from the effective potential flipping direction after inflation, rather than explicit $U(1)$ breaking operators in the scalar potential with high radial field values as required in the usual Affleck-Dine scenarios~\cite{Affleck:1984fy}. Because of this tilt,  the axion can attain a non-zero $\dot{\theta}$ and also climb up the potential barrier, as we saw earlier (cf. Fig. \ref{fig:numthet}). The amplitude of $\dot{\theta}$ scales as $a^{-3}$, i.e. $||\dot{\theta}(t)||= \dot{\theta}_m \left(a_m/a(t) \right)^3$. Recall that in the limit $\xi\gg \left(f/m_P\right)^2$, $\dot{\theta}_m$ is given by Eq. \eqref{eq:dotthet}. The presence of this non-zero velocity spontaneously breaks CPT in the expanding Universe, which in the presence of baryon number violating interactions in equilibrium can lead to the generation of baryon asymmetry, through spontaneous baryogenesis\footnote{Here, we don't consider transfer of asymmetry from decay of the rotating condensate \cite{Harigaya:2019emn, Barrie:2021mwi,Barrie:2022cub,Barrie:2024yhj}, since we require it to behave as dark matter at late times.}~\cite{Cohen:1987vi, Cohen:1988kt, Domcke:2020kcp}.

The basic idea of spontaneous baryogenesis is that a dynamical pseudo-scalar field (the axion in our case) can generate an external chemical potential for quarks and/or leptons. This can happen through derivative couplings of the axion with fermion currents of the form $x_\psi \partial_\mu \phi j^{\mu}_{\psi}/ f$, where $\psi$ indicates a fermion with $B-L$ charge $x_\psi$ and $j^{\mu}_{\psi}=\bar{\psi}\gamma_\mu \psi$.\footnote{The backreaction from such a term to the axion equation of motion is suppressed for large $f$ \cite{Domcke:2020kcp}, which is of our interest (cf. Eq. \eqref{Tmaxbound}).} 
The presence of this coupling with a non-zero $\dot{\theta}$, accompanied by a B (or L) violating interaction in thermal equilibrium, causes an energy shift in particles (anti-particles) proportional to $\dot{\theta}(-\dot{\theta})$, giving rise to equilibrium values of the baryon or lepton asymmetry as $n_B\sim n_L\sim\dot{\theta}T^2$. The essential part of spontaneous baryogenesis is to generate a non-zero $\dot{\theta}$ sufficiently large for generating the observed asymmetry.

The production of asymmetry continues as long as the baryon or lepton number violating interaction remains in thermal equilibrium.  When such an interaction decouples, say at a temperature $T_{\rm B-L}$, the asymmetry gets frozen to a constant value, with the final yield of baryon asymmetry given by
\begin{align}
      Y_B
\simeq \frac{n_B}{s}&\simeq \frac{45 c_B  T_{\rm B-L}^2 \dot{\theta}(T_{\rm B-L})}{2 \pi^2 g_{*s} T_{\rm B-L}^3} \nonumber\\ 
&\simeq \frac{45 c_B   \dot{\theta}_{\rm m
}}{2 \pi^2 g_{*s} T_{\rm B-L}} \left(\frac{a_{\rm m}}{a_{\rm B-L}}\right)^3 \,, \label{eq:YB0}
\end{align}  
where $s$ is the entropy energy density and $g_{*s}$ denotes the entropy degrees of freedom. $c_B$ is an $\mathcal{O}(1)$ factor to be calculated from the transport equations \cite{Domcke:2020kcp}. Now, the decoupling can take place before or after reheating. Note that the thermal bath can exist even before  reheating, where the reheating temperature $T_{\rm reh}$, is defined as the temperature when the radiation energy density starts to dominate the Universe.  

First, we consider 
\mbox{$T_{\rm max}>T_{\rm B-L}> T_{\rm reh}$}, where we have
\begin{align}
    Y_B &\simeq \frac{45 c_B   \dot{\theta}_{\rm m
}}{2 \pi^2 g_{*s} T_{\rm B-L}} \frac{H(t_{\rm B-L})}{H(t_{\rm m})},\label{eqn:YB1}
\end{align}
where we used that during kination \mbox{$a\propto H^{-1/3}$}.
Now, the modified Hubble parameter during the kination era can be written as
\begin{align}
    H_{\rm kin}(T)=\frac{\pi\sqrt{g_*}}{\sqrt{90}} \,\frac{T^2}{m_P} \left[1+ \mathcal{G}\left(\frac{T} {T_{\rm reh}}\right)^2\right]^{1/2}\,,\label{eq:Hubble}
\end{align}
where \mbox{$\mathcal{G}= \left(\frac{g_*(T_{\rm reh})}{g_*(T)}\right)\left(\frac{g_{*s}(T_{\rm reh)}} {g_{*s}(T)}\right)^2$} and $\frac{\pi\sqrt{g_*}}{\sqrt{90}} \,T^2/m_P$ is the Hubble parameter in a radiation-dominated Universe.
Here, $g_{*}$ is the effective relativistic degrees of freedom and considering ${\cal G}\simeq1$, which holds at high temperatures, we have
\begin{align}
    H(T_{\rm B-L}) \simeq {\pi\sqrt{ g_{*} } \over \sqrt{90}}  \frac{T_{\rm B-L}^3}{m_P T_{\rm reh}} ~.
\end{align}
Using the above in Eq. \eqref{eqn:YB1}, gives us
\begin{align}
    Y_B \simeq  \frac{3 \sqrt{30\xi} c_B }{2 \pi \sqrt{g_*} } \frac{ T_{\rm B-L}^2}{ f ~T_{\rm reh}} \,,\label{eq:YB2}
\end{align}
 where we used the value of $\dot{\theta}_{\rm m}$ given by Eq.~\eqref{eq:dotthet} and took \mbox{$H_{\rm end}\simeq H(t_{\rm m})$}, because the time $t_{\rm m}$ of maximum kinetic energy density of our rotating axion occurs half a rotation after the end of inflation. Considering the observed value of baryon asymmetry, \mbox{$Y_B^{(0)} \simeq 8.7 \times 10^{-11}$} and the range of $\xi$ given in Eq.~\eqref{xirange}, we find that the decoupling temperature obeys
 \begin{align}
    {Y_B^{(0)} 2 \pi \sqrt{g_*} \over 3\sqrt{15}}  f  \ll  c_B \frac{T_{\rm B-L}^2}{T_{\rm reh}} \lesssim  {Y_B^{(0)} 4 \pi \sqrt{g_*} \over 9\sqrt{10}} m_P~,
\end{align}
or
\begin{align}
    4.8 \times 10^{-10}  \left(\frac{f}{\text{GeV}}\right) 
    \ll  \frac{c_B}{\rm GeV} \frac{T_{\rm B-L}^2}{T_{\rm reh}} \lesssim 9.5 \times 10^8\,,
    \label{eq:xirange2}
\end{align}
where we considered $g_*\simeq 100$ in the second line. On the other hand, in the case \mbox{$T_{\rm B-L} <T_{\rm reh}$}, we have 
\begin{align}
      Y_B &\simeq \frac{45 c_B   \dot{\theta}_{\rm m
}}{2 \pi^2 g_* T_{\rm B-L}} \frac{H(t_{\rm reh})}{H(t_{\rm m})} \left(\frac{H(t_{\rm B-L})}{H(t_{\rm reh})}\right)^{3/2} \nonumber \\
&\simeq   \frac{3\sqrt{30\xi} c_B }{2 \pi \sqrt{g_*} } \frac{ T_{\rm B-L}^2}{ f~ T_{\rm reh} } \,.
\end{align}
where we now used that during radiation domination \mbox{$a\propto H^{-1/2}$}. Note that the above result coincides with Eq.~\eqref{eq:YB2}. This is because $\dot{\theta}_m\propto H_m$
and there is no extra source of entropy injection into the bath.



\section{axion dark matter}\label{ALP}

Let us look into the requirements on $M$ such that our axion can be the dark matter. We assume that the tilt of the axion vacuum manifold already exists when $V(\phi)$ in Eq.~\eqref{V} takes over from the effective potential in Eq.~\eqref{Veffkin}. Otherwise, after reheating our axion would continue to rotate in a flat vacuum manifold with \mbox{$V(\phi)=0$} until a phase transition produces the tilt, as is the case with the QCD transition regarding the QCD axion. In effect, we assume that, if a phase transition is responsible for the tilt, then this occurs before the time when \mbox{$\rho_\phi\sim M^4$}.

\begin{figure*}
    \centering   \includegraphics[scale=0.5]{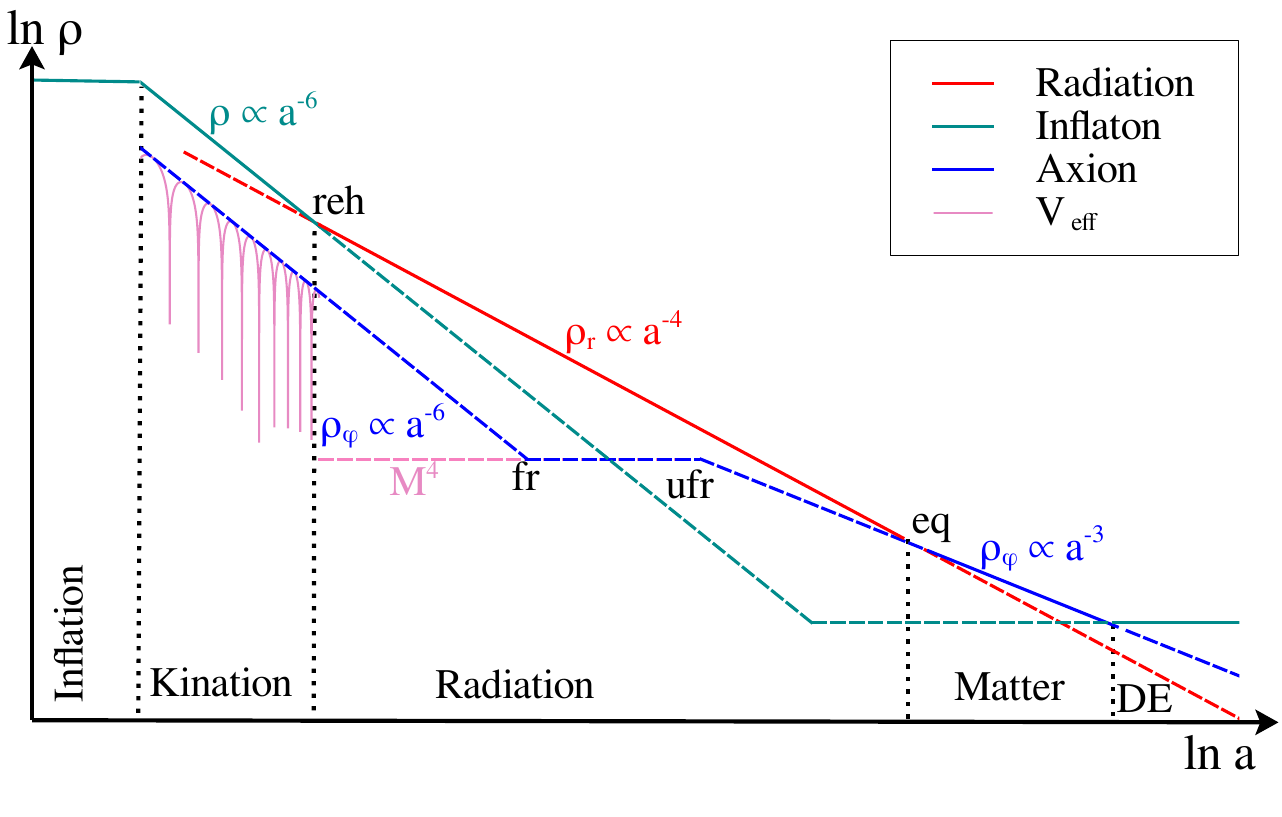}
    \caption{
    Schematic showing the evolution of the energy densities of the different components of the Universe in our scenario. The solid lines indicate the dominant component. We have assumed a model of quintessential
    inflation, such that the inflaton condensate gives rise eventually to the dark energy (DE) at present, but this is merely a possibility; our mechanism would operate with any non-oscillatory model of inflation followed by a period of kination.}
    \label{fig:sketch}
\end{figure*}

As mentioned already, during kination the energy density of the rotating axion decreases as \mbox{$\rho_\phi\propto t^{-2}$}, which is also true of the dominant background density \mbox{$\rho\propto t^{-2}$}. Thus, we expect \mbox{$\rho_\phi/\rho=\,$constant} during kination. 
As in Ref.~\cite{Bettoni:2021zhq}, we expect that these rotations are not halted by the potential hill of $V_{\rm eff}$ in Eq.~\eqref{Veffkin}, because the latter is diminishing in height as \mbox{$||V_{\rm eff}''||\propto t^{-2}$}.

After our axion starts rotating, we have
\begin{equation}
    \frac{\rho_\phi}{\rho}=\frac{6\xi m_P^2H^2}{3m_P^2H^2}=2\xi\ll 1\,,
    \label{Omegakin}
\end{equation}
where we considered Eq.~\eqref{ximax}.
This remains (approximately) constant until the moment of reheating, after which the energy density of the dominant radiation background decreases as \mbox{$\rho\propto a^{-4}$}. The axion energy density continues to be dominated by the kinetic energy density,
even though the effective potential is no more (because \mbox{$R=0$} during the radiation era\footnote{To be precise, quantum effects arising because of the trace anomaly of non-Abelian gauge theories \cite{Kajantie:2002wa} induce a non-zero $R$, which for the SM can be approximated as \cite{Davoudiasl:2004gf, Takahashi:2015waa, Saikawa:2018rcs, Berbig:2024ufe} $ R\simeq \frac{486}{19 \pi^2}\alpha_s(T)^2 H^2\,\simeq 0.037\,H^2$, where $\alpha_s(T\,\sim{\rm GeV}){\,\simeq 0.12}$ is the strong coupling constant 
\cite{Huston:2023ofk, Deur:2022msf}. 
The non-minimal coupling introduces a mass-squared of order
\mbox{$\sim\xi(m_P/f)^2R$} (cf. Eq.~\eqref{Lsmallphi}). For \mbox{$\xi\sim(f/m_P)^2$}, as suggested by Eq.~\eqref{xi}, we find that the trace anomaly gives rise to an effective mass \mbox{${\cal O}(10^{-1})H$}. Being smaller than $H$, such mass does not affect the dynamics of the axion, namely the freezing at $H_{\rm fr}$ (cf. Eq. \eqref{Hfr}) and thawing when \mbox{$H_{\rm ufr}\simeq M^2/f$ (cf. Eq. \eqref{Hufr})}.
}), such that
\mbox{$\rho_\phi\propto a^{-6}$}. Thus, during this epoch we have
\begin{equation}
    \frac{\rho_\phi}{\rho}=\left(\frac{a_{\rm reh}}{a}\right)^2\left.\frac{\rho_\phi}{\rho}\right|_{\rm reh}=\frac{H(t)}{H_{\rm reh}}\,2\xi\,,
\end{equation}
where we used that, in the radiation era \mbox{$a\propto t^{1/2}\propto H^{-1/2}$}, where \mbox{$H(t)=1/2t$}.

The above remains valid until the moment when the axion potential in Eq.~\eqref{V} overwhelms the decreasing kinetic energy density of the rotating axion. The moment, denoted by `fr', when this occurs is obtained as
\begin{eqnarray}
    M^4 & \simeq & \rho_\phi^{\rm fr}=
    \left(\frac{a_{\rm reh}}{a_{\rm fr}}\right)^6\rho_\phi^{\rm reh}
    \nonumber\\
    & = & \left(\frac{H_{\rm fr}}{H_{\rm reh}}\right)^3 2\xi\rho_{\rm reh}=
    6\xi\,\frac{H_{\rm fr}^3m_P^2}{H_{\rm reh}}\nonumber\\
\Rightarrow\; H_{\rm fr} & = &
    \left(\frac{M^4H_{\rm reh}}{6\xi m_P^2}\right)^{1/3}.
    \label{Hfr}
\end{eqnarray}

Provided $M$ is small enough, once it ceases to be kinetic dominated, the rotating axion freezes with \mbox{$\phi_{\rm fr}\sim f$} so that its residual potential energy density is 
\mbox{$\rho_\phi^{\rm fr}\simeq V(\phi_{\rm fr})\simeq M^4$}. The axion remains frozen until $H(t)$ decreases down to a value comparable to the axion mass
\begin{equation}
   H_{\rm ufr}\simeq m_\phi\simeq \frac{M^2}{f}\,,
   \label{Hufr}
\end{equation}
when the axion thaws and begins oscillating around its VEV at zero.\footnote{We do not consider any other interactions of our axion to the matter sector.}
Soon after the axion stars oscillating in the potential in Eq.~\eqref{V} its density decreases as \mbox{$\rho_\phi\propto a^{-3}$}. Therefore, for the remaining radiation era we have \mbox{$\rho_\phi/\rho\propto a$}. If the axion is to become the dark matter, it has to dominate the Universe 
at the time of equality of matter and radiation energy densities, denoted by `eq'. Then, we have
\begin{eqnarray}
    1 & \simeq & \left.\frac{\rho_\phi}{\rho}\right|_{\rm eq}=\left(\frac{a_{\rm eq}}{a_{\rm ufr}}\right)\left.\frac{\rho_\phi}{\rho}\right|_{\rm ufr}\nonumber\\
    & = & \frac{H_{\rm ufr}^{1/2}}{H_{\rm eq}^{1/2}}
    \frac{M^4}{3H_{\rm ufr}^2m_P^2}
    \nonumber\\
    & \sim & 
    \sqrt{t_{\rm eq}}\frac{M}{\sqrt f}
    \left(\frac{f}{m_P}\right)^2
    \nonumber\\
    \Rightarrow\;M & \sim & 
    \sqrt{\frac{m_P}{t_{\rm eq}}}\left(\frac{m_P}{f}\right)^{3/2}.
\end{eqnarray}
Using that 
\mbox{$t_{\rm eq}\simeq 3.14\times 10^{35}\,$GeV$^{-1}$}, we obtain\footnote{Note that a large value of $f$, needed for the backreaction to be negligible, also leads to a small value of $M$.}
\begin{equation}
    M\sim 10^{-9}\,{\rm GeV}\left(\frac{m_P}{f}\right)^{3/2}.
    \label{M}
\end{equation}
A schematic showing the evolution of all the energy density components of the Universe is shown in Fig.~\ref{fig:sketch}. From the above discussion
it is clear that the moment of freezing is unrelated to the moment of unfreezing. A high reheating temperature only ensures that the axion remains frozen for a longer time since 
a large $H_{\rm reh}$ corresponds to a large $H_{\rm fr}$, as determined by Eq.~\eqref{Hfr}. 

For this freezing period to exist at all, we need \mbox{$H_{\rm fr}>H_{\rm ufr}$}. In view of Eqs.~\eqref{Hfr} and~\eqref{Hufr}, this results in the bound
\begin{equation}
    T_{\rm reh}>T_{\rm reh}^{\rm min}\equiv
    \sqrt\xi\,M\left(\frac{m_P}{f}\right)^{3/2}\,,
    \label{Trehbound}
\end{equation}
where we used \mbox{$\sqrt{H_{\rm reh}}\sim T_{\rm reh}/\sqrt{m_P}$}.

Let us consider 
a specific example.
Assume that the axion decay constant is at the scale of grand unification (GUT-scale), such that \mbox{$f/m_P\sim 10^{-2}$}. Assume also that inflation takes place at GUT-scale, 
such that \mbox{$H\sim 10^{-5}\,m_P$}. Then, Eq.~\eqref{xirange} suggests that
\mbox{$10^{-4}\ll\xi<1$}. Eqs.~\eqref{Hufr}  and \eqref{M} give \mbox{$M\sim 10^{-6}\,$GeV} and \mbox{$m_{\phi}\sim 10^{-20}\,$eV}. 
Taking \mbox{$\xi\sim 10^{-4}$}, 
the bound in Eq.~\eqref{Trehbound} suggests that \mbox{$T_{\rm reh}>10^{-5}\,$GeV}. 
The latter is well satisfied, because the reheating temperature must be higher than the temperature at the time of BBN, 
\mbox{$T_{\rm reh}> 10^{-3}\,$GeV}.
In terms of baryogenesis, when choosing the extreme value of $\xi$ as in the above 
example, producing the observed baryon asymmetry requires (see Eq. \eqref{eq:xirange2}) $T_{\rm B-L}\simeq 8 \times 10^{7}$ GeV, for $T_{\rm reh}\sim 10^{7}$ GeV. Later in Sec.~\ref{sec:Eg}, we show an explicit model with such order of $T_{\rm B-L}$.  

\begin{figure*}[ht]
    \centering
    \includegraphics[width=0.8\textwidth]{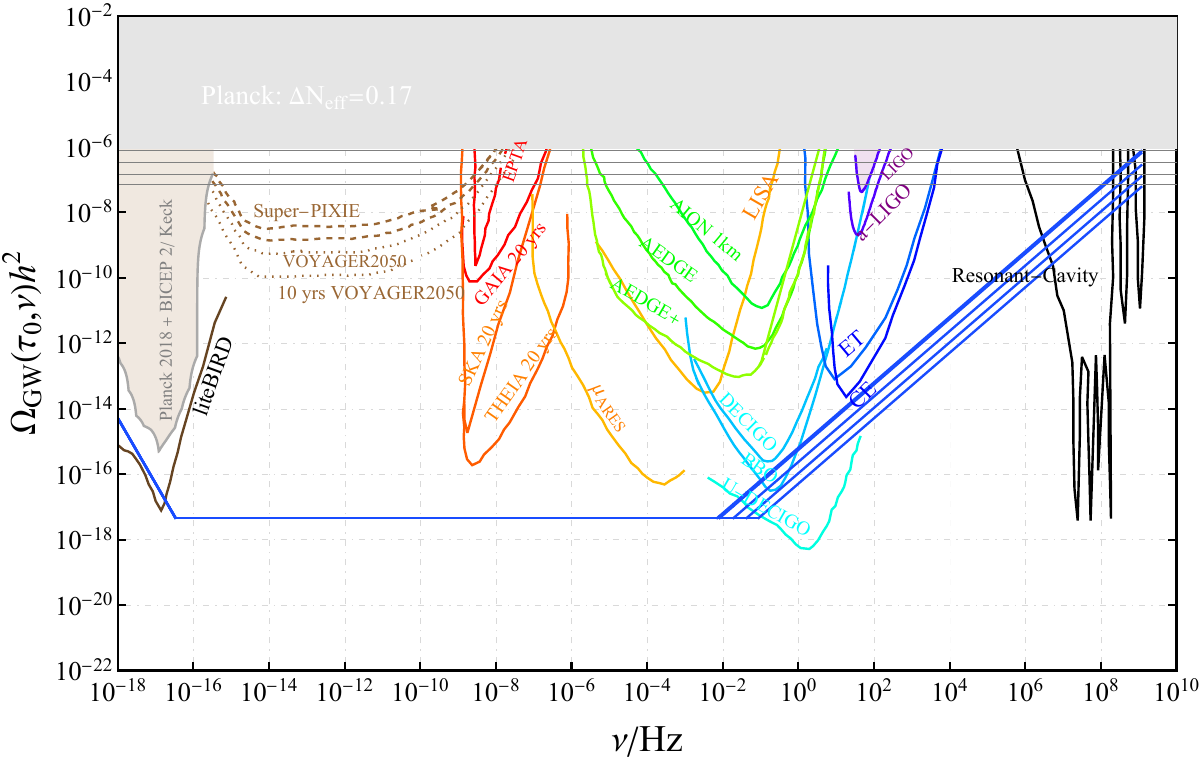}
    \caption{
    The current GW spectrum (blue), choosing values of reheating temperature $T_{\rm reh}$ given in Table \ref{tab:DNeff}. The different gray solid lines indicate the future sensitivity reaches of several experiments for $\Delta N_{\rm eff}$, i.e.  BBN+CMB, CMB-S4/PICO, CMB-HD, COrE/EUCLID (see Table \ref{tab:DNeff}), and hence to the peak of the GW energy spectra given by Eq.~\eqref{eq:bbn_constraint}. In the figure, the lowest possible values of $T_{\rm reh}$ are shown. If $T_{\rm reh}$ were even lower (and kination lasted longer) then the peak in the spectrum of GWs would violate the BBN bound (Eq. \eqref{eq:peak_bbn}), depicted by the horizontal gray band on top of the figure.}
    \label{fig:gw}
\end{figure*}

If the bound in Eq.~\eqref{Trehbound} is violated though, it does not mean that our mechanism for axion dark matter does not work. It simply means that the field never freezes. This can happen if $m_{\phi}$ catches up with Hubble (\mbox{$H<m_\phi\sim M^2/f$})  before the kinetic energy reaches the potential barrier $M^4$, thus switching to the vacuum potential in Eq.~\eqref{V}. In this case, while it is still rotating, the axion becomes subject to the potential in Eq.~\eqref{V}, stops rotating and begins coherent oscillations around its minimum at $\phi=0$. We expect this to occur if the reheating temperature is lower than $T_{\rm reh}^{\rm min}$ in Eq.~\eqref{Trehbound}.


\medskip

\section{Gravitational Waves}\label{sec:GW}

It is well known that the observed energy spectrum of primordial gravitational waves (GW) generated during inflation is affected by the thermal history after inflation~\cite{Giovannini:1998bp,Gouttenoire:2021jhk,Haque:2021dha,Figueroa:2019paj,Chen:2024roo}. 
The GW amplitudes are damped on subhorizon scales, i.e., \mbox{$h_k(\tau) \propto k^{-3/2} a_{k}/a(\tau)$}, where $k^{-3/2}$ is due to the fact that the GW power spectrum is nearly scale-invariant in the vanilla slow-roll inflation (as in our model). 
Here the subscript `$k$' denotes the moment that the length scale of wavenumber $k$ reeneters the horizon during the hot Big Bang.
Using the relationship\footnote{Obtained using $a\propto t^{2/3(1+w)}$ and $H\propto 1/t$.} 
$k = a_k H_k \propto a^{-(1+3w)/2}$, namely, replacing $a_k$ by $k^{-2/(1+3w)}$, we derive the GW energy spectrum as\footnote{For the detailed calculations, readers may refer to, e.g., Ref.~\cite{Chen:2024roo}.}
\begin{equation}
\Omega_{\rm GW}(\nu) \propto \nu^\beta ~,
\quad {\rm where} \quad
\beta = 2\, {w - 1/3 \over w + 1/3} ~,
\label{fbeta}
\end{equation}
where $\nu$ is the GW frequency.
Hence, for modes that reenter the Hubble horizon during radiation domination (RD), when $w = 1/3$, the observed GW energy spectrum is flat, while for modes that reenter the horizon during kination, with $w=1$, correspond to a blue-tilted spectrum. For the extremely low-frequency GWs whose modes reenter the Hubble horizon during matter domination ($w=0$), its energy spectrum is red-tilted. 
We can simply parametrize the observed GW energy spectrum as 
\begin{widetext}
\begin{equation} \label{eq:omegaGW_1}
\Omega_{\rm GW}(\tau_0, \nu) \simeq \Omega_{\rm GW}^{\rm RD}
\l\{
\begin{matrix}
    &\nu/\nu_{\rm reh}
    ~, \quad &&\nu_{\rm reh} < \nu < \nu_{\rm end}
    \\ \\
    &1 ~,  &&\nu_{\rm eq} < \nu < \nu_{\rm reh}
    \\ \\
    &(\nu_{\rm eq}/\nu)^2 ~,  &&\nu_{0} < \nu < \nu_{\rm eq}
\end{matrix}
\r. ~,
\end{equation}    
\end{widetext}
where $\Omega_{\rm GW}^{\rm RD}$ is a constant representing the GW density parameter of modes that reenter the horizon during RD. For GUT-scale inflation ($H_{\rm end} \sim 10^{-5} m_{P}$), we have \mbox{$\Omega_{\rm GW}^{\rm RD} \sim 10^{-17}$}. 
The observed frequencies $\nu_{\rm end}$, $\nu_{\rm reh}$, $\nu_{\rm eq}$, and $\nu_{0}$, correspond to the GW modes that reenter the horizon at the end of inflation, the onset of RD, the radiation-matter equality, and the present horizon, respectively.

The lowest frequency of the stochastic GW background is estimated as
\begin{equation}
\nu_0 = {H_{\rm 0} \over 2\pi}
\sim 10^{-19} ~{\rm Hz} ~,
\end{equation}
where $H_0 \sim 10^{-33} ~{\rm eV}$ is the Hubble constant. Similarly,
\begin{equation}
\nu_{\rm eq} 
= {H_{\rm eq} \over 2\pi} {a_{\rm eq} \over a_0}
\simeq {\sqrt{\rho_{\rm eq}} \over 2\pi \sqrt{3}\, m_P} {a_{\rm eq} \over a_0}
\sim 10^{-18} ~{\rm Hz} ~,
\end{equation}
where we ignored the dark energy domination and considered \mbox{$T_{\rm eq} \sim 1 ~{\rm eV}$}. 
Given that \mbox{$H_{\rm end} \simeq 10^{-5} m_{P}$} (GUT-scale inflation) and \mbox{$H_{\rm reh} = {\pi\over \sqrt{90}m_P} \sqrt{ g_{*} }\, T_{\rm reh}^2$}, as well as \mbox{$a_{\rm end}/a_0 \simeq$} \mbox{$ T_0/T_{\rm end} \simeq T_{\rm CMB}/(\rho_{\rm end})^{1/4} \sim 10^{-28}$}, where \mbox{$T_{\rm CMB} \sim 10^{-13}~{\rm GeV}$}, we derive
\begin{align}
    \nu_{\rm end} = {H_{\rm end} \over 2\pi} {a_{\rm end} \over a_0}
    \sim 10^9 ~{\rm Hz} ~.
\end{align}
The frequency $\nu_{\rm reh}$ is related to the reheating temperature $T_{\rm reh}$ as
\begin{align}
    \nu_{\rm reh} &= \nu_{\rm end} {H_{\rm reh} a_{\rm reh} \over H_{\rm end} a_{\rm end}}
    = \nu_{\rm end} \l( {H_{\rm reh} \over H_{\rm end}} \r)^{2/3}
    \nn\\&\sim 10^{-12} ~{\rm Hz} \l( { T_{\rm reh} \over {\rm GeV} } \r)^{4/3} ~.\label{eq:freh}
\end{align}

According to Eq.~\eqref{eq:omegaGW_1}, the peak amplitude is
\begin{align} \label{eq:omega_gw_peak}
    \Omega_{\rm peak} &\equiv \Omega_{\rm GW}(\tau_0, \nu_{\rm end})
    = \Omega_{\rm GW}^{\rm RD} \nu_{\rm end}/\nu_{\rm reh}
    \nn\\&\sim 10^{4} \l( { {\rm GeV} \over T_{\rm reh} } \r)^{4/3}.
\end{align}
Now, since the GW background acts as an extra radiation component, they can contribute to extra relativistic degrees of freedom during BBN, parametrised by the quantity $\Delta N_{\rm eff}$. Since the value of $\Delta N_{\rm eff}$ is being measured by several experiments, it can be used to put a constraint on the GW amplitude and subsequently on the value of the reheating temperature $T_{\rm reh}$. We require~\cite{Maggiore:1999vm,Boyle:2007zx,Caprini:2018mtu}
\begin{align}  \label{eq:bbn_int}
    \int_{\nu_{\rm BBN}}^{\nu_{\rm end}}\frac{\ddd \nu}{\nu}\Omega_{_{\rm GW}}(\nu) \,h^2 \leq \frac{7}{8}\left(\frac{4}{11}\right)^{4/3}\Omega_{\rm \gamma}h^2\,\Delta N_{\rm eff} ~,
\end{align}
where $\Omega_{\rm \gamma}h^2\simeq 2.47\times10^{-5}$ corresponds to 
the relic density of the radiation measured today, and $\nu_{\rm BBN} \sim 10^{-11} {\rm Hz}$.
Applying Eqs.~\eqref{eq:omegaGW_1} and~\eqref{eq:omega_gw_peak}, we 
obtain
\begin{align} \label{eq:bbn_constraint}
    \Omega_{\rm peak}\,h^2 \lesssim \frac{7}{8}\left(\frac{4}{11}\right)^{4/3}\Omega_{\rm \gamma}h^2\,\Delta N_{\rm eff} ~,
\end{align}
or
\begin{align}
    \Omega_{\rm peak}\,h^2 \lesssim 5.6 \times 10^{-6} \Delta N_{\rm eff} ~.
\end{align}
Thus, Eq.~\eqref{eq:bbn_constraint} gives
\begin{align}
    T_{\rm reh} \gtrsim 1.7 \times 10^{16} {\rm GeV} \l( {\Omega_{\rm GW}^{\rm RD} \over \Omega_{\gamma} \Delta N_{\rm eff}} \r)^{3/4} ~,
\end{align}
or
\begin{align} \label{eq:treh_bbn}
    T_{\rm reh} \gtrsim {5.8 \times 10^6~\text{GeV} \over \Delta N_{\rm eff}^{3/4}} ~,
\end{align}
where we used Eq.~\eqref{eq:freh}. Using the current value from the Planck observations ($\Delta N_{\rm eff}=0.17$), we get the lower bound on the reheating temperature and the peak of GW spectrum~\eqref{eq:omega_gw_peak} as
\begin{align}
     T_{\rm reh} &\gtrsim 2.2 \times 10^{7} ~\text{GeV} ~,
     \label{BBNbound}
     \\ \label{eq:peak_bbn}
     \Omega_{\rm peak} h^2 &\lesssim 7.3 \times 10^{-7} ~.
\end{align}
  
While this constraints our parameter space, several other future experiments can be sensitive to lower values of $\Delta N_{\rm eff}$ and thus can test our predicted value of $T_{\rm reh}$. In Table~\ref{tab:DNeff}, we present the current bounds and future sensitivities of several experiments which can probe $\Delta N_{\rm eff}$, along with the corresponding $T_{\rm reh}$ (from Eq.~\eqref{eq:treh_bbn}) that can be tested.
\begin{table*}[t!]
    \begin{center}
        \begin{tabular}{|c||c|c|}
            \hline
            $\Delta N_{\rm eff}$ & Experiments & $T_{\rm reh}$ \\ 
            \hline\hline
            $0.17$ & Planck legacy data (combining BAO)~\cite{Planck:2018vyg} & $2.2 \times 10^{7}\,\mathrm{GeV}$\\
            \hline
            $0.14$ & BBN+CMB combined~\cite{Yeh:2022heq} & $2.5 \times 10^{7}\,\mathrm{GeV}$\\
            \hline
            $0.06$ & CMB-S4~\cite{Abazajian:2019eic},PICO~\cite{NASAPICO:2019thw}, CMB-Bharat ~\cite{Adak:2021lbu} & $4.8 \times 10^{7}\,\mathrm{GeV}$ \\
            \hline
            $0.027$ & CMB-HD~\cite{CMB-HD:2022bsz} & $8.7 \times 10^{7}\,\mathrm{GeV}$ \\
            \hline
            $0.013$ & COrE~\cite{COrE:2011bfs}, EUCLID~\cite{EUCLID:2011zbd} & $1.5 \times 10^{8}\,\mathrm{GeV}$ \\
            \hline
        \end{tabular}
    \end{center}
    \caption{Current bounds and future sensitivity reaches for $\Delta N_{\rm eff}$ along with the corresponding reheating temperatures $T_{\rm reh}$ in Eq.~\eqref{eq:treh_bbn}.}
    \label{tab:DNeff}
\end{table*}

\section{A Concrete Example}\label{sec:Eg}

As an explicit example to realise B$-$L violating interactions, we consider the \textit{Type I seesaw} framework, where the Majorana mass term of right handed neutrinos (RHNs) provides the source for the lepton number violation. Considering the RHN mass to be generated from a spontaneous breaking of global $U_{\rm B-L}$ symmetry, we have the Majoron as the pseudo Nambu Goldstone Boson~\cite{Chikashige:1980ui,Gelmini:1980re}. Such a Type I seesaw framework has been utilised in~\cite{Chun:2023eqc}  for spontaneous baryogenesis using conventional and kinetic misalignment, and also recently in~\cite{Chun:2024gvp}. Note that we consider the contribution to the baryon asymmetry from the CP-violating decay of RHNs, i.e., the vanilla thermal leptogenesis to be absent or suppressed.

\begin{figure*}
\includegraphics[width=0.4
\textwidth]{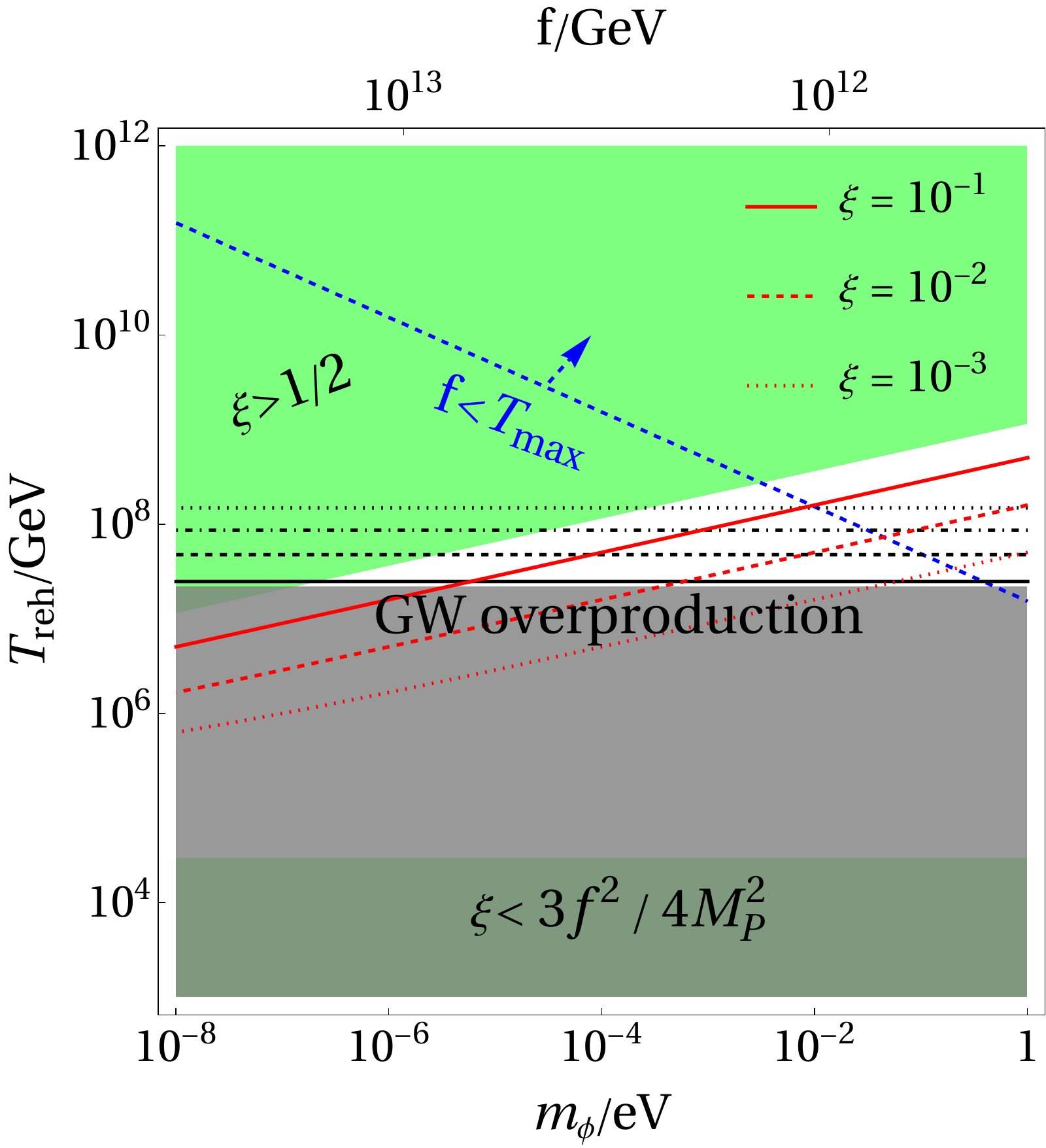}~~
\includegraphics[width=0.4
\textwidth]{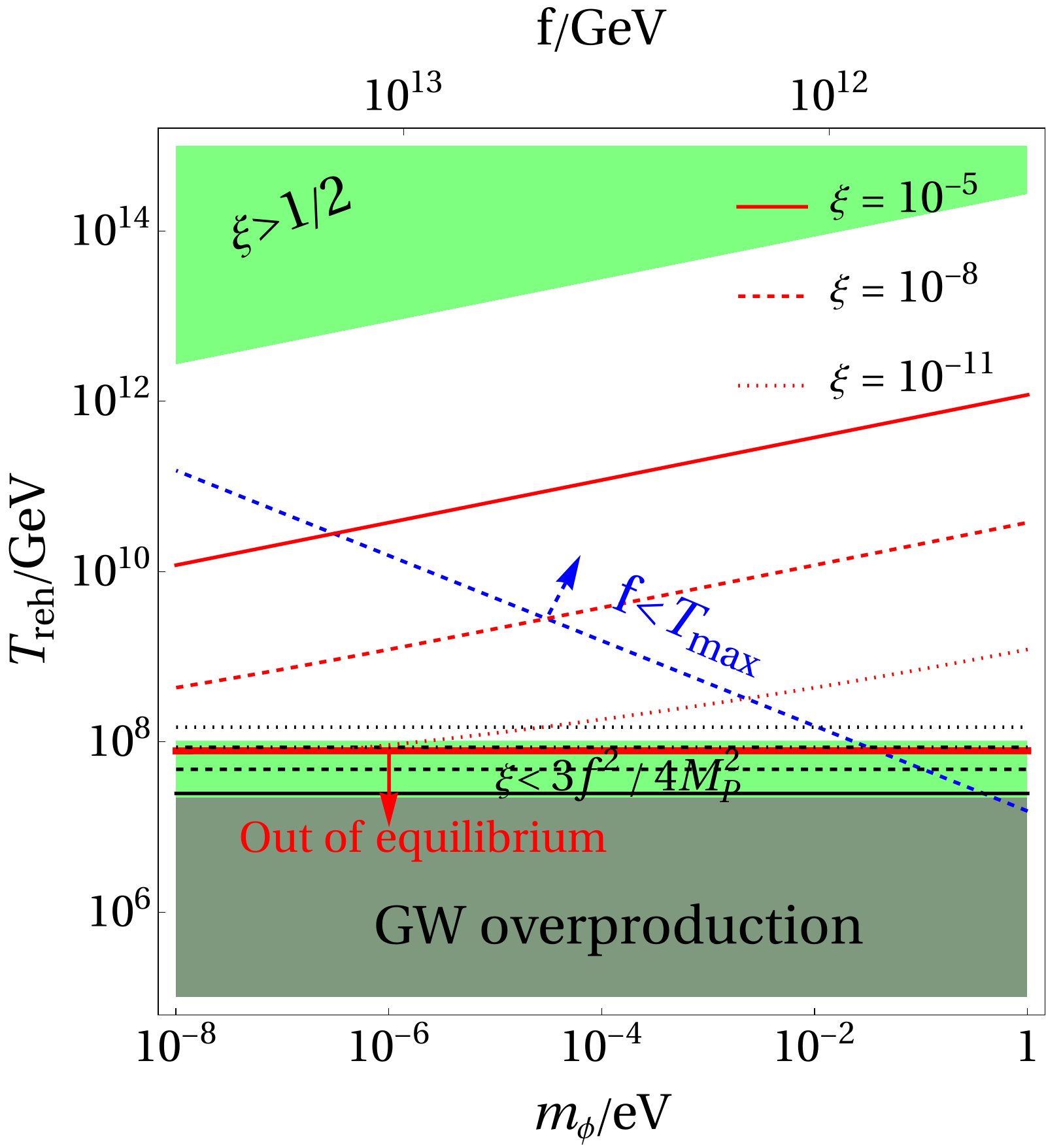}
    \caption{
    Predictions for cogenesis of dark matter and baryon asymmetry in the Type I seesaw setup with Majoron, considering $M_N = 10^7, 5 \times 10^{9}$ GeV in the left and right panel, respectively. The gray-shaded regions are ruled out from the BBN bound on GW (see Eq. \eqref{eq:peak_bbn}), whereas in the green-shaded regions $\xi$ lies outside the allowed range given by Eq. \eqref{xirange}. Above the blue-dashed line, the symmetry of the scalar potential might get restored (see discussion at the end of Sec. \ref{sec2}).  Below the red solid contour in the right panel plot, the B-L violating interactions are out of equilibrium because of faster expansion of the Universe (see text). The different black lines indicate the future sensitivity reaches of several experiments for $\Delta N_{\rm eff}$ with solid, dashed, dot-dashed and dotted corresponding to BBN+CMB, CMB-S4/PICO/CMB-Bharat, CMB-HD, COrE/EUCLID respectively (see Table \ref{tab:DNeff}).}
    \label{fig:parspace}
\end{figure*}

The relevant seesaw Lagrangian is given by
\begin{align}
    - {\cal L}_{\rm seesaw} = \sum_{i,j} Y_{N_i} \Phi \bar{N}^c_i N^c_j +  \sum_{i,\alpha} Y_{D_{\alpha i}} \bar{l}_\alpha \Tilde{\mathcal{H}} N^c_i +h.c.\,, 
\end{align}
where $N$ indicates the right-handed neutrinos and $\mathcal{H},l$ are the SM Higgs and lepton doublets respectively. $i,j,\alpha$ run from 1 to 3. Note that the angular mode of the complex scalar $\Phi$ plays the role of Majoron, and undergoes the dynamics discussed in the preceding sections. The Majorana masses of $N$ are generated from $\Phi$ as $M_N= Y_N f /\sqrt{2}$, giving light neutrino mass $m_\nu$ through the usual seesaw mechanism as $m_\nu\sim Y_D^2 ~v^2/M_N$,  where $v$ is the vacuum expectation value (VEV)  of the standard model Higgs. 

In such a framework, any lepton number violating interactions such as the inverse decays ($l\mathcal{H}\rightarrow N$) tend to erase or ``wash-out" the created asymmetry in the conventional thermal leptogenesis. However, in the case of spontaneous baryogenesis, the inverse decays of RHNs in equilibrium can provide the required B$-$L violating interaction required for the generation of asymmetry, acting rather as a ``wash-in" term\footnote{For details, readers may refer to Ref. \cite{Chun:2023eqc}.}. In the strong-washout regime, the asymmetry is effectively frozen when the inverse decays go out of equilibrium at a temperature around \mbox{$T_{\rm dec}\sim M_N / z_{\rm fo}$}, where $z_{\rm fo}$ is around $\mathcal{O}(10)$, depending on the model parameters and the background cosmology. The created lepton asymmetry finally  gets converted to the baryon asymmetry when the electroweak sphaleron decouples at \mbox{$T_{\rm EW}\simeq 130\,$GeV}. 

In Fig.~\ref{fig:parspace}, we show the viable parameter space for generating the observed DM relic along with the baryon asymmetry in the $m_\phi-T_{\rm reh}$ plane, considering two benchmark values of the RHN mass: $M_N=10^{7}$ GeV, and $5 \times 10^9$ GeV in the left and right panels, respectively. For these values of mass, considering  $m_\nu\sim 0.05$ eV, the Dirac Yukawa couplings are fixed through seesaw as $Y_D\sim 7 \times 10^{-5}$ and $10^{-3}$, respectively. For a given value of the Majoron mass, the decay constant is fixed through Eq.~\eqref{M}, which is denoted in the upper X-axes. The gray shaded regions are ruled out from overproduction of GWs, as given by Eq.~\eqref{eq:peak_bbn}. The red contours indicate different values of $\xi$ along which we can have cogenesis of DM and the baryon asymmetry, following Eqs.~\eqref{eq:YB2} and~\eqref{M}. 

The inverse decay rate can be written as~\cite{Buchmuller:2004nz}
\begin{align}
    \Gamma_{\rm ID}\simeq \Gamma_N \frac{K_1 (z)}{K_2 (z)}\frac{n_N^{\rm eql}(z)}{n_l^{\rm eql}(z)} ~,
\end{align}
where $z=M_N/T$. $n_N^{\rm eql}, n_l^{\rm eql}$ denotes the equilibrium number densities of RHN  and SM leptons respectively and $\Gamma_N\simeq |Y_{D_{\alpha i}}|^2 M_N / 16 \pi$  is the RHN decay rate. $K_1(z)$ and $K_2 (z)$ are the Bessel functions of the first and second kind respectively. Thus, the inverse decays are in equilibrium when the above rate is larger than the Hubble expansion rate, until $z_{\rm fo}$ as mentioned above. For interactions in a kination-dominated Universe, the decoupling occurs earlier, since the Universe expands faster, with a larger Hubble rate (cf. Eq. \eqref{eq:Hubble}). Thus, comparing the inverse decay rate with the Hubble \cite{Chen:2019etb, Chun:2023eqc}, we find $z_{\rm fo}$ decreases slightly for $T_{\rm B-L} > T_{\rm reh}$. 

For sufficiently low reheating temperatures, the Hubble rate turns large such that the inverse decays cannot be in equilibrium. The region below the red thick contour in the right panel plot for $M_N = 5 \times 10^{9}$ GeV indicates this, while in the left panel the bound from GW overproduction is much stronger. In the green-shaded regions $\xi$ lies outside the allowed range given by Eqs.~\eqref{xirange} and \eqref{eq:xirange2}. The lower bound on $\xi$ becomes important for a higher value of decoupling temperature (cf. Eq. \eqref{eq:xirange2}), as seen in the right panel plot for the heavier RHN mass. Above the blue-dashed contour, the value of $f$ becomes lower than the maximum temperature of the bath $T_{\rm max}$, which can potentially disrupt the VEV of the scalar potential as discussed earlier. As we will see in the next two sections, it is favorable for $\xi$ to lie close to the lower bound of the range in Eq. \eqref{xirange}. Thus, the benchmark chosen in the left-panel ($M_N=10^{7}$ GeV) becomes unfavorable due to constraints from GW overproduction.

Additionally, we need to have the decoupling of B$-$L violating interaction during the axion rotation, i.e. $T_{\rm B-L}> T_{\rm fr}$, which we find to be trivially satisfied for this model. 
Finally, note that in this parameter space, the freezing period may or may not exist depending on whether $H_{\rm fr}> H_{\rm ufr}\simeq m_{\phi}$ is satisfied or not. However, this does not change our results. 

The constraints discussed above are model-independent, once the decoupling temperature $T_{\rm B-L}$ is given, which in our case is primarily determined by the RHN mass scale $M_{N}$. Apart from these constraints, in order to have the Majoron as the dark matter candidate in our model, it needs to be cosmologically stable. The Majoron can decay into light neutrinos with a decay rate given by $\Gamma_{\phi\rightarrow \nu \nu}\simeq \frac {m_{\phi}}{16 \pi f^2}\sum_{j} m_{\nu_{j}}^2$. Thus, the stability criteria is given by  $\Gamma_{\phi\rightarrow \nu \nu}^{-1}\gtrsim \tau_{U}$, where $\tau_{U}\simeq 250$~Gyr \cite{Audren:2014bca,Enqvist:2019tsa,Nygaard:2020sow,Alvi:2022aam,Simon:2022ftd}. Considering the current upper bound on the sum of neutrino mass $\sum_{j} m_{\nu_{j}}^2\simeq (0.05 \text{eV})^2$, we get the upper bound on the Majoron mass as $m_{\phi}\lesssim 30$ MeV. Since, the interactions of Majoron with the standard model originate from the $Y_D$ term, the interactions are suppressed by the light neutrino mass~$m_\nu$. 

Hence, testing our above predictions of light Majoron mass of sub-eVs becomes highly challenging. However, as discussed in Section \ref{sec:GW}, our scenario can be probed through the blue-tilted gravitational wave spectra appearing because of the kination domination era. The requirement of cogenesis  of baryon asymmetry and dark matter abundance predicts specific values of the reheating temperature (see Fig. \ref{fig:parspace}), which determines the frequency $\nu_{\rm reh}$ given by Eq.~\eqref{eq:freh}, which in turn indicates the start of the blue-tilted spectra. Although tightly constrained by BBN bounds as discussed earlier, such a positive slope can be tested \cite{Chen:2024roo} by future experiments like BBO \cite{Crowder:2005nr,Corbin:2005ny}, UDECIGO \cite{Sato:2017dkf, Ishikawa:2020hlo}, and also by proposed high-frequency  resonant-cavity experiments \cite{Ringwald:2020ist,Ringwald:2022xif} in the MHz-GHz regime, see Ref. \cite{Aggarwal:2020olq} for a review. Moreover, the $\Delta N_{\rm eff}$ values from the GW background, can be within reach of several near-future CMB experiments (cf. Table \ref{tab:DNeff}), which are shown by different black lines in Fig.~\ref{fig:parspace}.

Finally, our example also has some predictions for light neutrino mass as follows. Using the seesaw formula ($m_\nu\sim Y_D^2 ~v^2/M_N$) and the condition for the dark matter abundance (cf. Eq.~\eqref{M}), the light-neutrino mass can be related to the Majoron mass as
\begin{align}
    m_{\nu}\sim 3 \times 10^{4}\,v\,\frac{Y_D^2 }{Y_N}\left(\frac{m_{\phi}}{\text{GeV}}\right)^{1/4} \left(\frac{\text{GeV}}{m_P}\right)^{3/4} 
    \frac{v}{\text{GeV}}
\end{align}
From the experimental constraints of neutrinoless double $\beta$-decay from KATRIN, the direct neutrino mass measurement gives \cite{KATRIN:2021uub}, \mbox{$m_{\nu} \leq 0.8$ eV} and future sensitivity can reach upto \mbox{$m_{\nu} \leq 0.2$ eV}. Thus, Majoron masses below $\mathcal{O}$ (GeV), can be indirectly tested at these neutrino experiments. The scenario becomes more predictive considering that it gives rise to the baryon asymmetry during the rotation phase, since $Y_N$ is determined (\mbox{$T_{\rm B-L}\sim \frac{M_N}{z_{\rm fo}}\sim \frac{Y_N f}{z_{\rm fo}}$}) imposing Eq.~\eqref{eq:YB2}. Considering 
\mbox{$\xi\sim(f/m_P)^2 $} 
(cf. Eq.~\eqref{xi}), the light neutrino mass is predicted to be 
\begin{align}
    \hspace{-.4cm}
    m_\nu \sim c_B^{1/2} \left(\frac{v}{246~\text{GeV}}\right)\left(\frac{10}{z_{\rm fo}}\right)\left(\frac{10^{7}~\text{GeV}}{T_{\rm reh}}\right)^{1/2}\nonumber\\\times\left(\frac{Y_D}{6.3 \times 10^{-2}}\right)^2\text{eV}.\label{eq:neumass}
\end{align}
Hence, the setup has complementary tests in GW, CMB and neutrino experiments.

\section{Axion Fragmentation}\label{sec:Fragmentation}

Our setup can lead to the enhancement of axion fluctuations, which can lead to a significant loss of kinetic energy of the rotating axion, eventually stopping the variation of the axion. This can happen through parametric resonance of certain values of wave numbers. This phenomenon known as axion fragmentation was first studied in detail in Ref.~\cite{Fonseca:2019ypl} and further explored in Refs.~\cite{Eroncel:2022vjg, Eroncel:2022efc}. In Ref.~\cite{Eroncel:2022vjg}, it was shown that the axion can be fragmented even before getting trapped, if the mass is much larger than the Hubble rate during trapping, i.e., $m_{\phi}/H_{\rm ufr}\gg \mathcal{O}(1)$. Basically, this means that when $H$ reaches $m_{\phi}$, the axion still has larger kinetic energy than the barrier, thus delaying the conventional oscillation time.  Recall that in our case the trapping potential given by Eq.~\eqref{V} is different from the initial potential (Eq.~\eqref{Veffkin}) over which the axion starts to roll, and we have $m_{\phi}\simeq H_{\rm ufr}$ during trapping, at least for the case when $H_{\rm fr} > H_{\rm ufr}$ (subject to Eq.~\eqref{Trehbound}). However, during the initial rotation in the kination domination, parametric resonance may be important, 
as it can stop the rotation of our axion. We
investigate  this danger below.

The axion field can be decomposed into the homogeneous mode and fluctuations as
\begin{align}
    \Theta(t,\textbf{x})= \theta(t) + \delta \theta (t,\textbf{x})\,.
\end{align}
The equation of motion for the Fourier modes $\delta\theta_k(t)$ is given by~\cite{Fonseca:2019ypl, Eroncel:2022vjg}
\begin{align}
\delta\ddot{\theta}_k+3H\delta \dot{\theta}_k+\left[\frac{k^2}{a^2}+ V''(\phi)\right]\delta\theta_k=0\,,\label{eq:thetkeom}
\end{align}
where $k$ here denotes the comoving momentum and $V''(\phi)=\frac{3\xi m_P^2 H^2}{f^2} \cos(\frac{\phi}{f})$ with $H=1/3t$ during kination. We closely follow Refs.~\cite{Fonseca:2019ypl, Eroncel:2022vjg}, and derive some approximate analytical conditions for fragmentation, and compare our analytical estimates with the numerical solution of the above equation. 

In the limit of $H=0$, the above equation has instability bands around $\dot{\theta}/2$, with a width depending on $V''(\phi)$, obeying
\begin{align}
    \frac{\dot{\theta}^2}{4} - \frac{3\xi m_P^2 H^2}{2f^2}\lesssim \frac{k^2}{a^2}\lesssim \frac{\dot{\theta}^2}{4} + \frac{3\xi m_P^2 H^2}{2f^2}\,.\label{eq:insblbnd}
\end{align}
Expanding the square root, the above band can be approximated as $|\frac{k}{a}-\frac{k_{\rm cr}}{a}|\lesssim\frac{\delta k_{\rm cr}}{a}$, where
\begin{align}
\frac{k_{\rm cr}}{a}= \frac{\dot{\theta}}{2} \,,
\quad
{\rm and}\quad\frac{\delta k_{\rm cr}}{a}= \frac{3\xi m_P^2 H^2}{2f^2\dot{\theta}}\,. \label{eq:insblbnd2} 
\end{align}
Note that the band moves because of redshift as well as change in $\dot{\theta}$. Additionaly, the physical momentum also redshifts as $k/a$. 

The exponential growth of $\delta\theta_k$ is given by \mbox{$\delta\theta_k\propto \exp{\left[\sqrt{(\frac{\delta k_{\rm cr}}{a})^2-(\frac{k}{a}-\frac{k_{\rm cr}}{a})^2}\;t\right]}$}. Thus, the maximum growth is around $\exp{(\frac{3\xi m_P^2 H^2}{2f^2\dot{\theta}}t)}$. Now,  the inclusion of the Hubble term in the equation of motion \eqref{eq:thetkeom} provides a source of friction for suppression of the growth. Thus, in order for the modes to grow, we should have
\begin{align}
    H< \frac{3\xi m_P^2 H^2}{2f^2\dot{\theta}}~ 
{\;\Rightarrow\;}
    \frac{3\xi m_P^2 H}{2f^2\dot{\theta}}>1.
\end{align}
The above condition gives us a conservative upper bound on $\xi$ as 
\begin{align}
    \xi \lesssim 5.3\left(\frac{f}{m_P}\right)^2\,\label{eq:xifragbnd}
\end{align}
where we used Eq.~\eqref{eq:dotthet} for the initial velocity~$\dot{\theta}_m$ and 12(2/3)$^2\simeq\,$5.3. Note that violating the above condition does not necessarily imply that the axion would be fragmented. To see if the modes would grow, one has to integrate the growth factor within the exponent over the time that the modes stay inside the instability band \cite{Fonseca:2019ypl, Eroncel:2022vjg}, which can be very small. In this work, we don't derive a full analytical expression for this growth but rather solve Eq.~\eqref{eq:thetkeom} numerically along with the background homogeneous equation (cf. Eq. \eqref{eq:EOMthet})for $\theta$, to see if our condition 
in
Eq.~\eqref{eq:xifragbnd} holds\footnote{We consider adiabatic initial conditions for Eq.~\eqref{eq:thetkeom}, sourced by the curvature perturbations \cite{Mukhanov:2005sc, Eroncel:2022vjg, Eroncel:2022efc}.}.

\begin{figure*}
\includegraphics[width=0.4
\textwidth]{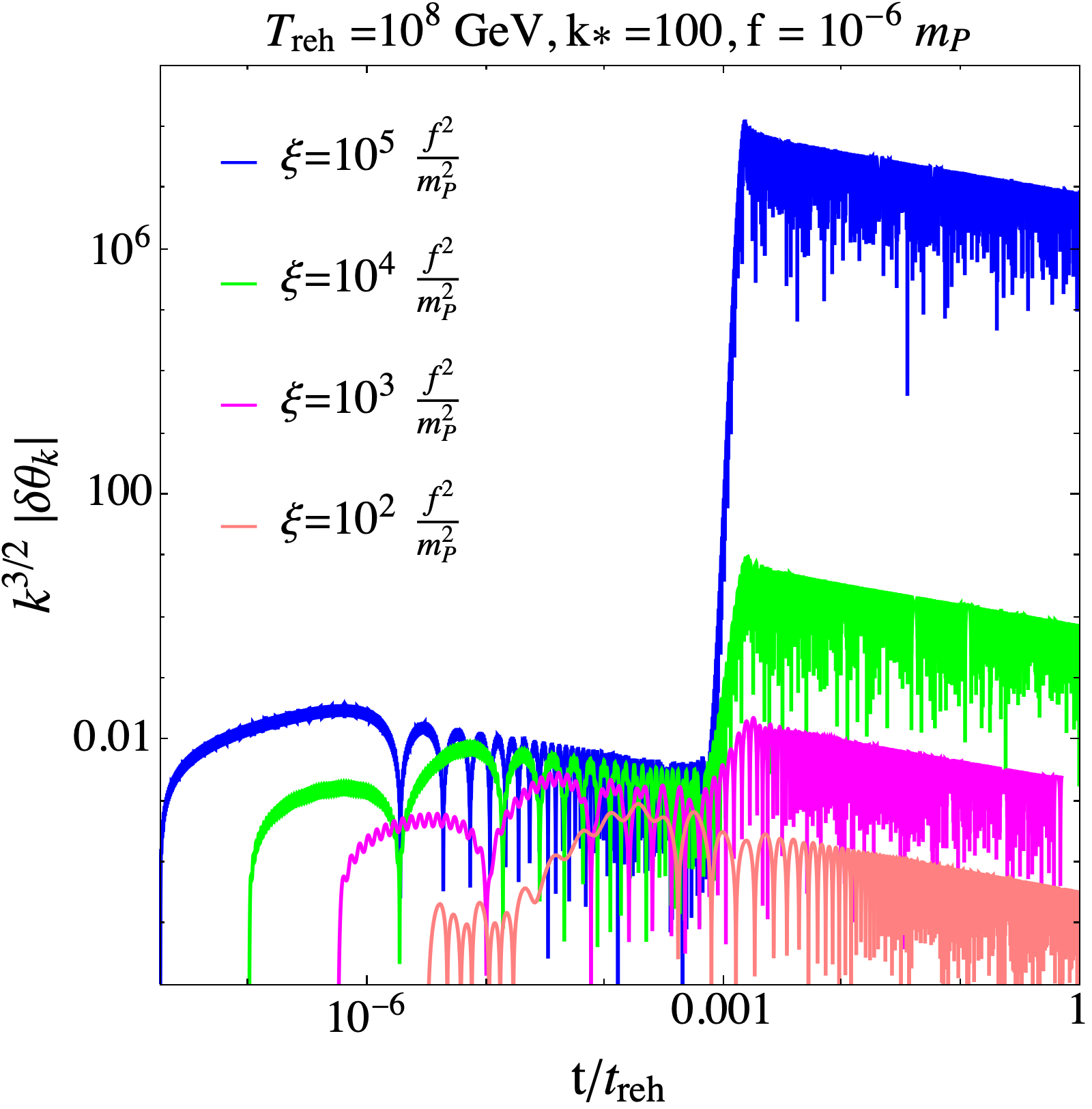}~~
\includegraphics[width=0.4
\textwidth]{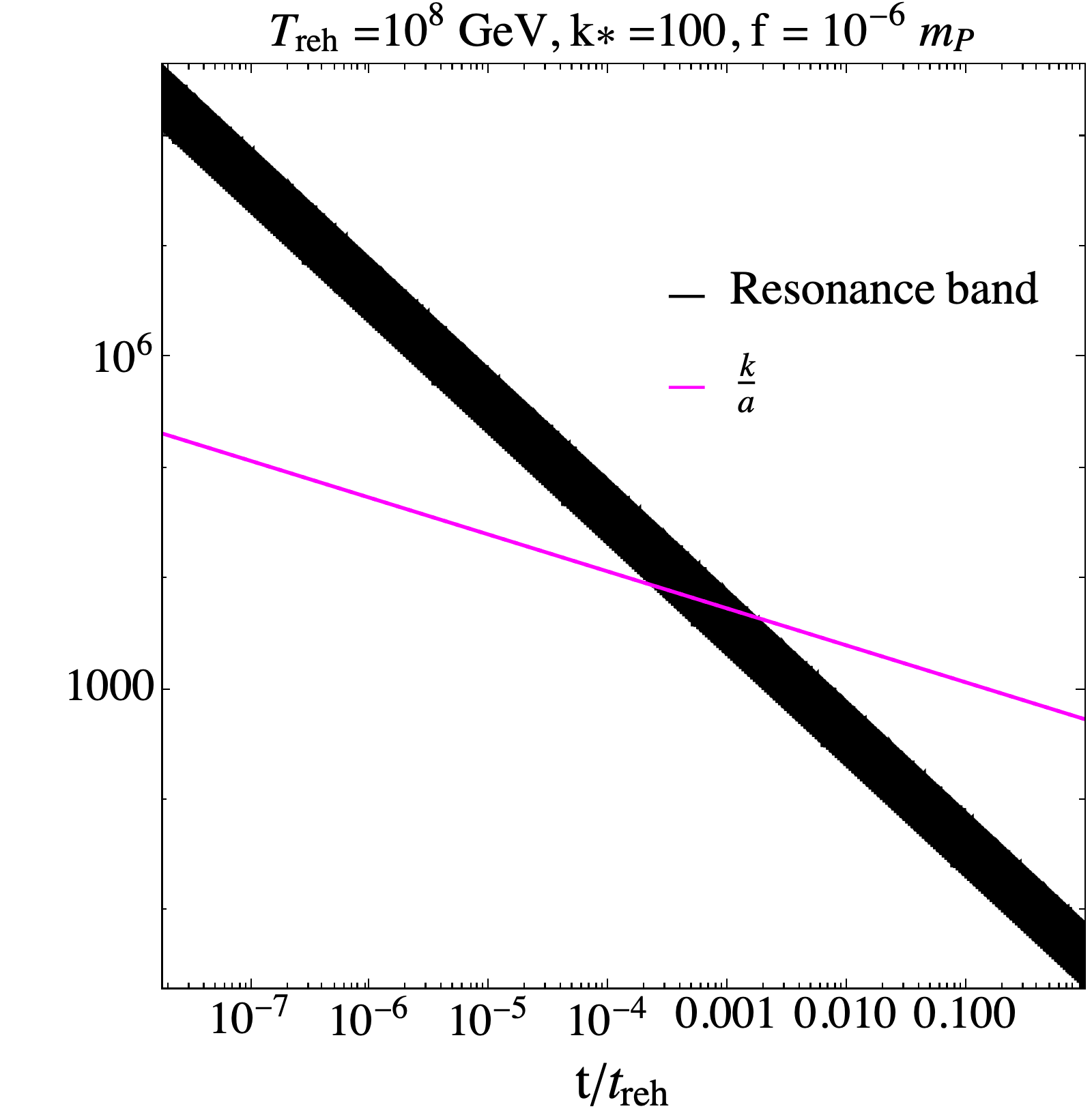}
    \caption{\textit{Left panel:} Evolution of the axion fluctuations for different values of $\xi$, exceeding \mbox{$f^2/m_P^2$}. Substantial growth can be seen with the increase in the value of $\xi$. \textit{Right panel:} Evolution of the instability band along with the physical momentum (see Eq. \eqref{eq:insblbnd}). Note that the fluctuations in the left panel grow around the time when the mode enters the band.}
    \label{fig:parres}
\end{figure*}

Recall that, after reheating, the effect of the non-minimal $\xi$-term vanishes and mass term from the vacuum potential in Eq.~\eqref{V} is not large enough to lead to growth as discussed above. We define \mbox{$k_* = \frac{k}{m_* a_*}$}, where `$*$' indicates quantities evaluated at the reheating temperature. In Fig.~\ref{fig:parres}, we show our numerical results for a benchmark set of values $f= 10^{-6} m_P,~ T_{\rm reh}= 10^{8}~\text{GeV},~ k_* =100$.  The left panel shows the evolution of $\delta \theta_k$ (multiplied by $k^{3/2}$, making it dimensionless) for several values of $\xi$. Interestingly, as anticipated, we see that the growth happens only if $\xi$ is greater than \mbox{$\left[
\mathcal{O}(10^2)-\mathcal{O}(10^3)
\right] 
\times(f/m_P)^2
$}, which is significantly relaxed compared to blue the bound in Eq.~\eqref{eq:xifragbnd}. We find similar behavior for other benchmark parameters.  In the right panel, we show the evolution of the instability band, given by Eq. \eqref{eq:insblbnd}, along with the physical momentum 
\mbox{$k/a$}, for \mbox{$\xi = 10^5 
(f/m_P)^2$}. We see that the band is very narrow as expected, and so the time spent by the k-mode inside the band is 
very little.

An upper bound on $\xi$ close to our numerical results can be derived as follows \cite{Eroncel:2022vjg}. During the amplification, the velocity and the scale factor can be written as
\begin{align}
     \dot{\theta}(t)=\dot{\theta}(t_k) +\ddot{\theta}(t_k)(t-t_k) \,\,,\,\,\nonumber\\a(t) = a(t_k) [1+ H(t_k) (t-t_k)]\,,
\end{align}
where the subscript `$k$' indicates the time when the mode is at the center of the instability band given by Eq. \eqref{eq:insblbnd2}. Here, we consider $\ddot{\theta}$ and $H$ to be constant during the amplification. Using the above in Eq. \eqref{eq:insblbnd}, the total amplification time, i.e. the time spent inside the instability band, is found to be
\begin{align}
    t_{\rm amp}=2|\Delta t|\simeq\frac{2 \times 3\xi m_P^2 H^2 }{f^2\dot{\theta}(t_k)|H(t_k)\dot{\theta}(t_k)+\ddot{\theta}(t_k)|}\,.
\end{align}
where $\Delta t (-\Delta t)$ corresponds to the time taken to reach the band's upper (lower) edge. Thus, the maximum growth during the amplification is approximately given by
\begin{align}
     N_k \simeq \exp{\left( \int_{-t_{\rm amp}/2}^{t_{\rm amp}/2}\frac{3\xi m_P^2 H^2}{2f^2\dot{\theta}(t_k)} d t\right)} \\  \simeq \exp{\left(\frac{(3\xi m_P^2 H^2)^2}{f^4\dot{\theta}(t_k)^2|H(t_k)\dot{\theta}(t_k)+\ddot{\theta}(t_k)|}\right)}.
\end{align}
Hence, for the fluctuations to grow significantly, we need to have the term inside the exponent to be greater than unity. Using $\dot{\theta}(t)= \dot{\theta}_m \frac{H(t)}{H_m}$,  $\ddot{\theta} (t)=- 3 H(t)\dot{\theta}(t)$ and our relation for $\dot{\theta}_m$ (see Eq. \eqref{eq:dotthet}), we arrive at the following bound on $\xi$
\begin{align}
    \xi \lesssim \frac{4}{9}(12)^3 \left(\frac{f}{m_P}\right)^2 \simeq 768\left(\frac{f}{m_P}\right)^2\,.
\end{align}
Comparing with Fig. \ref{fig:parres}, we can see that it agrees quite well with our numerical results\footnote{This bound is even more relaxed because the growth of the fluctuations does not necessarily imply fragmentation of the axion, where the latter takes place only when the energy density of the fluctuations become comparable to that of the background homogeneous mode.}.
Therefore, the lower part of the range in Eq.~\eqref{xirange} is 
quite
safe against fragmentation of the rotating axion condensate,
for which we can have successfull baryogenesis as well, as discussed in Section \ref{sec:Baryo}. In fact, 
in the 
next section we argue that it may be
preferable for $\xi$ to be 
not much larger than 
\mbox{$(f/m_P)^2$}. 

We leave a detailed study of axion fragmentation including improved analytical results and numerical solutions with backreaction effects in such a scenario for future studies.


\section{The Kibble issue}\label{sec:Kibble}

Our setup considers an ever-existing axion, during and after inflation, with the radial mode \mbox{$|\Phi|=f\neq 0$} always. This means that there is no phase transition which can give rise to topological defects through the Kibble mechanism.

However, 
our rotating axion suffers from 
the Kibble mechanism problem in a different way.
This problem has to do with the fact that, after the end of inflation, the axion rotation  can be clockwise or anticlockwise, at random. Consequently, originally causally disconnected areas at the end of inflation may correspond to opposite orientations for the axion rotation. What would happen when the growth of these originally causally disconnected areas brings them in contact? Most probably, the rotation will be halted.

However, we can use to our advantage the fact that our axion is also present during inflation.\footnote{In contrast to Ref.~\cite{Bettoni:2018utf}, which also considers the Kibble problem for a rotating axion.}
Indeed, we can avoid the Kibble problem by considering a slight misalignment of the expectation value of the axion at the end of inflation from the value \mbox{$\phi=\pi\,f$}.

To achieve this, we have to assume that the effective mass-squared of the axion during inflation is not larger than $H^2$ after all. As a result, the axion undergoes particle production and its condensate diffuses around the minimum $\pi\,f$, such that the mean-squared field is~\cite{Bunch:1978yq,Mijic:1994vv}
\begin{equation}
    \langle\bar\phi^2\rangle=\frac{3H^4}{8\pi^2 m_\phi^2}\simeq\frac{1}{16\pi^2\xi}\left(\frac{f}{m_P}\right)^2H^2 ~,
    \label{mean}
\end{equation}
where \mbox{$\bar\phi\equiv\phi-\pi\,f$} and we used that near the effective minimum \mbox{$\bar\phi\approx 0$} we have that \mbox{$V_{\rm eff}\simeq\frac12(6\xi m_P^2/f^2)H^2\bar\phi^2$} as suggested by Eq.~\eqref{Veffinf}. Requiring the value of $\langle\bar\phi^2\rangle$ to be larger than 
$H^2/4\pi^2$ at the end of inflation, so that quantum fluctuations cannot lift the field up the peak of the effective potential, results in the bound \mbox{$\xi<\frac14(f/m_P)^2$}. Thus, the lower bound in Eq.~\eqref{xirange} is mildly violated. 

There is a lower bound on $\xi$, obtained by the requirement that \mbox{$\langle\bar\phi^2\rangle\ll (\pi\,f)^2$} for otherwise the axion condensate would spread all around its vacuum manifold.This is not necessarily bad, but we would like to avoid it, to be close to the minimum during inflation, and obtain the maximum kinetic energy density for the axion, after inflation ends and the vacuum manifold is flipped. In view of Eq.~\eqref{mean}, this requirement is equivalent to \mbox{$\xi\gg\frac{1}{16\pi^4}(H/m_P)^2$}. Thus, the range of $\xi$ is now
\begin{equation}
    \frac{1}{16\pi^4}\left(\frac{H}{m_P}\right)^2\ll\xi<\frac14\left(\frac{f}{m_P}\right)^2.
    \label{xirange+}
\end{equation}
For the above range to exist we simply require \mbox{$\pi\,f\gg H/2\pi$}, which means that the axion cannot be pushed over the effective potential hill in a few (about 60) e-folds, where the typical quantum jump is \mbox{$\delta\phi=H/2\pi$} per e-fold. 

Thus, at the end of inflation, after the flip of the axion vacuum manifold, which switches the minimum from $\pi\,f$ to zero, our axion does not find itself quite at the top of the effective potential hill, but displaced by the amount \mbox{$\phi=\pi\,f-\sqrt{\langle\bar\phi^2\rangle}$}, where $\langle\bar\phi^2\rangle$ is given in Eq.~\eqref{mean} and, without loss of generality, we assumed that the quantum diffusion during inflation has pushed the condensate to values smaller (not larger) than the $\pi\,f$. Importantly, even though quantum diffusion is random and the condensate spreads equiprobably in the domain \mbox{$(\pi\,f-\sqrt{\langle\bar\phi^2\rangle}, \pi\,f+\sqrt{\langle\bar\phi^2\rangle})$} this is from a global perspective. In our observable Universe, inflation homogenises the value of the condensate, as is the case in the curvaton paradigm \cite{Lyth:2002my,Lyth:2003ip}.

Therefore, we expect the axion to roll only in one direction (e.g. clockwise) in the observable Universe, which evades the concerns of the Kibble problem. It also avoids another possible criticism of our setup. This has to do with the fact that the second-order phase transition which follows the change of tilt of the axion vacuum manifold, would be a waterfall transition, as in Ricci reheating, and is expected to proceed in a stochastic manner \cite{Bettoni:2021zhq,Figueroa:2016dsc,Nakama:2018gll}, that might not end up with the dynamics we discussed in Sec,~\ref{sec2}. Having our axion displaced from the maximum of the effective potential, as suggested by Eq.~\eqref{mean}, implies that the transition is not waterfall, and our treatment is better justified.
The price to pay
is that we have to make sure that excessive isocurvature perturbations are avoided.\footnote{This is quite feasible for a rotating axion, e.g. see Ref.~\cite{Chung:2024ctx}.}

Another important point is that $\xi$ has to gradually increase during kination if the axion is to roll at all, because the latter requires \mbox{$||V_{\rm eff}''||>H^2$}. For this to happen $\xi$ needs to grow by a couple of orders of magnitude, if we are close to the upper bound in Eq.~\eqref{xirange+}. One possibility is to consider a running $\xi(\sigma)$ as \cite{Bezrukov:2014bra,Hamada:2014wna,Ezquiaga:2017fvi,Drees:2019xpp,Cheong:2021vdb,Ghoshal:2024hfk}
\begin{equation}
\xi(\sigma)=\xi_0\left[1+\beta\ln\left(\frac{\sigma^2}{\mu^2}+1\right)\right]\,,\label{xiofsigma}
\end{equation}
where $\sigma$ is the inflaton field, $\mu$ is a constant energy scale and $\xi_0,\beta$ are dimensionless constants which depend upon the microscopic details of the model.

During slow-roll inflation, since the inflaton is light, $\sigma$ is almost invariant so that $\xi$ is approximately constant, as assumed. During kination, however, $\sigma$
fast-rolls such that \mbox{$\sigma\propto\ln(t/t_{\rm end})$}, so that \mbox{$\xi\propto\ln[\ln(a)]$}, were \mbox{$a\propto t^{1/3}$}. This means that $\xi$ is not changing fast, and considering it approximately constant is a good approximation. The total roll of the inflaton depends on how early reheating occurs, and it can be several orders of magnitude. It then depends on the values of $\mu$ and more importantly of $\beta$ in Eq.~\eqref{xiofsigma} to see how much $\xi$ changes. However, changing $\xi$ by an order of magnitude seems quite realistic. 

Therefore, we should tune the required value of $\xi$ to be close to the boundary value 
\begin{equation}
    \xi\sim\left(\frac{f}{m_P}\right)^2\,,
    \label{xi}
\end{equation}
i.e. near the edge of both ranges in Eqs.~\eqref{xirange}
and \eqref{xirange+}.
It is such a value that was assumed in the example at the end of Sec.~\ref{ALP}, while, as shown in Sec.~\ref{sec:Fragmentation}, it is also safe from fragmentation of the axion condensate. 

Recall that in this low range of $\xi$, our estimate of $\dot{\theta}_m$ given by Eq.~\eqref{eq:dotthet} is unreliable. However, the relevant physical quantity to track is the average value of $\dot{\theta}$, which remains approximately equal to the potential barrier height (cf. Fig.~\ref{fig:numthet}). This is what we 
used in Eq.~\eqref{eq:dotthet}.  Hence, our estimate of the baryon asymmetry  is expected to remain valid as an order of magnitude, 
even with $\xi$ around the value given by Eq.~\eqref{xi}. We leave a detailed numerical simulation for future studies.

Now, using the above value of $\xi$, we can find a lower bound on the RHN mass scale as follows. Combining Eqs.~\eqref{eq:YB2} and \eqref{xi}, it is straightforward to find
\begin{equation}
    \frac{T_{\rm B-L}^2}{T_{\rm reh}}\sim Y_B\, m_P\sim 10^8\,{\rm GeV}\,,\label{eq:barpredtn}
\end{equation}
where we used \mbox{$Y_B\sim 10^{-10}$}. Thus, if for example \mbox{$T_{\rm reh}\sim 10^{10}\,$GeV} we find \mbox{$T_{\rm B-L}\sim 10^9\,$GeV}.
Considering the BBN bound $T_{\rm reh}\gtrsim10^{7}$ GeV, we get $T_{\rm B-L}\gtrsim 3 \times 10^{7}$ GeV. In the case of our concrete example, discussed in Sec.~\ref{sec:Eg}, for $z_{\rm fo}$ around $\mathcal{O}(10)$, this gives a lower bound on the RHN mass as \mbox{$M_N \gtrsim 3\times10^{8}\,$GeV}.


\section{Conclusion}\label{sec:conc}



In this work, we investigated an attractive way to generate rotation for an axion-like  particle (which is not the QCD axion), if it is non-minimally coupled to gravity, respecting the shift-symmetry. Such a rotation is generated when the effective potential of the axion flips at the end of inflation, providing the axion enough kinetic energy to surpass the potential barrier. This becomes possible if inflation is followed by a kination-dominated epoch ($w=1$), when the axion  potential barrier redshifts in the same way as its kinetic energy, i.e. as $a^{-6}$. We find that for the axion to roll and remain sub-dominant, the value of $\xi$ is constrained to be in the range given by Eq.~\eqref{xirange}. We utilize such an axion rotation to generate the baryon asymmetry through spontaneous baryogenesis. The final baryon asymmetry, given by Eq.~\eqref{eq:YB2}, is found to depend on the initial kinetic energy of the axion determined crucially by the non-minimal coupling $\xi$ and the decay constant $f$, and its subsequent redshift which depends on the reheating temperature $T_{\rm reh}$. Additionally, the asymmetry is determined by the decoupling temperature of the baryon or lepton number violating interaction $T_{\rm B-L}$, which depends on the details of the particle physics setup. The axion continues to rotate even after reheating until the kinetic energy becomes comparable to the height of the bare mass potential (given in Eq.~\eqref{V}), determined by $M$, after which the coherent oscillation of the axion can behave as dark matter, relating the axion decay constant to its mass through Eq.~\eqref{M}.

The presence of the kination-dominated epoch leads to a substantial enhancement of the stochastic gravitational wave background from inflation. The blue-tilt of such a GW spectrum is determined by the reheating temperature. In order not to overproduce GWs during BBN, the reheating temperature is constrained to be greater than $\mathcal{O}(10^7)\,$GeV (Eq. \eqref{BBNbound}), which substantially constrains the viable parameter space. 

We also investigated the danger of the fragmentation of the axion condensate as well as effects of the Kibble issue, which could destroy our mechanism. In both cases we found that
we need the non-minimal coupling to be \mbox{$\xi\sim(f/m_P)^2$} to avoid difficulties. For successful baryogenesis in this case, we find the condition
\mbox{$T_{\rm B-L}^2\sim Y_B^{(0)}m_PT_{\rm reh}$}, where
\mbox{$Y_B^{(0)}=8.7\times 10^{-11}$} is the final yield of baryon asymmetry.

We applied our mechanism to a concrete example of Majoron in the well-motivated Type I seesaw model, where our framework can be utilized for the cogenesis of baryon asymmetry and dark matter, with the viable parameter space shown in Fig.~\ref{fig:parspace}. It is important to note that in our mechanism of flipped axion rotation, although the same axion field is responsible for cogenesis, the potential (shown in Eq.~\eqref{Veffkin}) playing role in baryogenesis during rotation differs from that (shown in Eq.~\eqref{V}) generating the dark matter abundance during oscillation. Thus, the axion mass $m_{\phi}$ is unconstrained from the observed baryon asymmetry (cf. Eq. \eqref{eq:YB2}), which opens up a larger parameter space unlike in other axion rotation scenarios. In our example, the Majoron mass is generally predicted to take any value  in the range below sub-eVs. On the other hand, the decoupling temperature $T_{\rm B-L}$ is tightly constrained by bounds from GW overproduction and issues related to fragmentation and the Kibble problem discussed earlier. As a result, in our example, the RHN mass which determines $T_{\rm B-L}$ is constrained to be larger than around $10^{8}$ GeV (see the discussion after Eq.~\eqref{eq:barpredtn}).  While considering $m_{\nu}\sim 0.05$ eV and the Yukawa couplings to obey the perturbativity bound, $M_{N}$ and $T_{\rm max}$ can be upto the GUT-scale.

Our scenario is very predictive and favors specific values of the non-minimal coupling $\xi$, which for instance, considering GUT-scale decay constant $f\sim10^{-2}~m_P$ is around $10^{-4}$. These predictions can be indirectly tested by GW experiments like BBO, UDECIGO and resonant cavities, which can be sensitive to the blue tilt (see Fig.~\ref{fig:gw}). Morever, the predicted values of $\Delta N_{\rm eff}$  coming from the GW background can be within the sensitivities of several experiments such as CMB-S4, CMB-HD, EUCLID (cf. Table \ref{tab:DNeff}, Figs.~\ref{fig:gw} and \ref{fig:parspace}). In the example we considered, one can also have complementary tests in neutrino experiments (cf. Eq.~ \eqref{eq:neumass}). Our idea can also have interesting phenomenology for other particle physics models including the QCD axion, which we leave for future exploration.




\section*{Acknowledgements}
CC is supported by National Natural Science Foundation of China (No. 12433002) and Start-up Funds for Doctoral Talents of Jiangsu University of Science and Technology. SJD is supported by IBS under the project code IBS-R018-D1. KD is supported (in part) by the STFC consolidated Grant:
ST/X000621/1. Authors thank Pawel Kozow for discussion. CC thanks the supports from The Asia Pacific Center for Theoretical Physics, The Center for Theoretical Physics of the Universe at Institute for Basic Science during his visits. The authors also thank Maximilian Berbig for some useful comments.

\bibliographystyle{apsrev4-1}
\bibliography{sample}

\begin{thebibliography}{155}%
\makeatletter
\providecommand \@ifxundefined [1]{%
 \@ifx{#1\undefined}
}%
\providecommand \@ifnum [1]{%
 \ifnum #1\expandafter \@firstoftwo
 \else \expandafter \@secondoftwo
 \fi
}%
\providecommand \@ifx [1]{%
 \ifx #1\expandafter \@firstoftwo
 \else \expandafter \@secondoftwo
 \fi
}%
\providecommand \natexlab [1]{#1}%
\providecommand \enquote  [1]{``#1''}%
\providecommand \bibnamefont  [1]{#1}%
\providecommand \bibfnamefont [1]{#1}%
\providecommand \citenamefont [1]{#1}%
\providecommand \href@noop [0]{\@secondoftwo}%
\providecommand \href [0]{\begingroup \@sanitize@url \@href}%
\providecommand \@href[1]{\@@startlink{#1}\@@href}%
\providecommand \@@href[1]{\endgroup#1\@@endlink}%
\providecommand \@sanitize@url [0]{\catcode `\\12\catcode `\$12\catcode `\&12\catcode `\#12\catcode `\^12\catcode `\_12\catcode `\%12\relax}%
\providecommand \@@startlink[1]{}%
\providecommand \@@endlink[0]{}%
\providecommand \url  [0]{\begingroup\@sanitize@url \@url }%
\providecommand \@url [1]{\endgroup\@href {#1}{\urlprefix }}%
\providecommand \urlprefix  [0]{URL }%
\providecommand \Eprint [0]{\href }%
\providecommand \doibase [0]{http://dx.doi.org/}%
\providecommand \selectlanguage [0]{\@gobble}%
\providecommand \bibinfo  [0]{\@secondoftwo}%
\providecommand \bibfield  [0]{\@secondoftwo}%
\providecommand \translation [1]{[#1]}%
\providecommand \BibitemOpen [0]{}%
\providecommand \bibitemStop [0]{}%
\providecommand \bibitemNoStop [0]{.\EOS\space}%
\providecommand \EOS [0]{\spacefactor3000\relax}%
\providecommand \BibitemShut  [1]{\csname bibitem#1\endcsname}%
\let\auto@bib@innerbib\@empty
\bibitem [{\citenamefont {Starobinsky}(1980)}]{Starobinsky:1980te}%
  \BibitemOpen
  \bibfield  {author} {\bibinfo {author} {\bibfnamefont {A.~A.}\ \bibnamefont {Starobinsky}},\ }\href {\doibase 10.1016/0370-2693(80)90670-X} {\bibfield  {journal} {\bibinfo  {journal} {Phys. Lett. B}\ }\textbf {\bibinfo {volume} {91}},\ \bibinfo {pages} {99} (\bibinfo {year} {1980})}\BibitemShut {NoStop}%
\bibitem [{\citenamefont {Sato}(1981)}]{Sato:1981qmu}%
  \BibitemOpen
  \bibfield  {author} {\bibinfo {author} {\bibfnamefont {K.}~\bibnamefont {Sato}},\ }\href {\doibase 10.1093/mnras/195.3.467} {\bibfield  {journal} {\bibinfo  {journal} {Mon. Not. Roy. Astron. Soc.}\ }\textbf {\bibinfo {volume} {195}},\ \bibinfo {pages} {467} (\bibinfo {year} {1981})}\BibitemShut {NoStop}%
\bibitem [{\citenamefont {Kazanas}(1980)}]{Kazanas:1980tx}%
  \BibitemOpen
  \bibfield  {author} {\bibinfo {author} {\bibfnamefont {D.}~\bibnamefont {Kazanas}},\ }\href {\doibase 10.1086/183361} {\bibfield  {journal} {\bibinfo  {journal} {Astrophys. J. Lett.}\ }\textbf {\bibinfo {volume} {241}},\ \bibinfo {pages} {L59} (\bibinfo {year} {1980})}\BibitemShut {NoStop}%
\bibitem [{\citenamefont {Guth}(1981)}]{Guth:1980zm}%
  \BibitemOpen
  \bibfield  {author} {\bibinfo {author} {\bibfnamefont {A.~H.}\ \bibnamefont {Guth}},\ }\href {\doibase 10.1103/PhysRevD.23.347} {\bibfield  {journal} {\bibinfo  {journal} {Phys. Rev. D}\ }\textbf {\bibinfo {volume} {23}},\ \bibinfo {pages} {347} (\bibinfo {year} {1981})}\BibitemShut {NoStop}%
\bibitem [{\citenamefont {Aghanim}\ \emph {et~al.}(2020)\citenamefont {Aghanim} \emph {et~al.}}]{Planck:2018vyg}%
  \BibitemOpen
  \bibfield  {author} {\bibinfo {author} {\bibfnamefont {N.}~\bibnamefont {Aghanim}} \emph {et~al.} (\bibinfo {collaboration} {Planck}),\ }\href {\doibase 10.1051/0004-6361/201833910} {\bibfield  {journal} {\bibinfo  {journal} {Astron. Astrophys.}\ }\textbf {\bibinfo {volume} {641}},\ \bibinfo {pages} {A6} (\bibinfo {year} {2020})},\ \bibinfo {note} {[Erratum: Astron.Astrophys. 652, C4 (2021)]},\ \Eprint {http://arxiv.org/abs/1807.06209} {arXiv:1807.06209 [astro-ph.CO]} \BibitemShut {NoStop}%
\bibitem [{\citenamefont {Ade}\ \emph {et~al.}(2019)\citenamefont {Ade} \emph {et~al.}}]{SimonsObservatory:2018koc}%
  \BibitemOpen
  \bibfield  {author} {\bibinfo {author} {\bibfnamefont {P.}~\bibnamefont {Ade}} \emph {et~al.} (\bibinfo {collaboration} {Simons Observatory}),\ }\href {\doibase 10.1088/1475-7516/2019/02/056} {\bibfield  {journal} {\bibinfo  {journal} {JCAP}\ }\textbf {\bibinfo {volume} {02}},\ \bibinfo {pages} {056} (\bibinfo {year} {2019})},\ \Eprint {http://arxiv.org/abs/1808.07445} {arXiv:1808.07445 [astro-ph.CO]} \BibitemShut {NoStop}%
\bibitem [{\citenamefont {Hazumi}\ \emph {et~al.}(2019)\citenamefont {Hazumi} \emph {et~al.}}]{Hazumi:2019lys}%
  \BibitemOpen
  \bibfield  {author} {\bibinfo {author} {\bibfnamefont {M.}~\bibnamefont {Hazumi}} \emph {et~al.},\ }\href {\doibase 10.1007/s10909-019-02150-5} {\bibfield  {journal} {\bibinfo  {journal} {J. Low Temp. Phys.}\ }\textbf {\bibinfo {volume} {194}},\ \bibinfo {pages} {443} (\bibinfo {year} {2019})}\BibitemShut {NoStop}%
\bibitem [{\citenamefont {Sugai}\ \emph {et~al.}(2020)\citenamefont {Sugai} \emph {et~al.}}]{Sugai:2020pjw}%
  \BibitemOpen
  \bibfield  {author} {\bibinfo {author} {\bibfnamefont {H.}~\bibnamefont {Sugai}} \emph {et~al.},\ }\href {\doibase 10.1007/s10909-019-02329-w} {\bibfield  {journal} {\bibinfo  {journal} {J. Low. Temp. Phys.}\ }\textbf {\bibinfo {volume} {199}},\ \bibinfo {pages} {1107} (\bibinfo {year} {2020})},\ \Eprint {http://arxiv.org/abs/2001.01724} {arXiv:2001.01724 [astro-ph.IM]} \BibitemShut {NoStop}%
\bibitem [{\citenamefont {Abazajian}\ \emph {et~al.}(2022)\citenamefont {Abazajian} \emph {et~al.}}]{CMB-S4:2020lpa}%
  \BibitemOpen
  \bibfield  {author} {\bibinfo {author} {\bibfnamefont {K.}~\bibnamefont {Abazajian}} \emph {et~al.} (\bibinfo {collaboration} {CMB-S4}),\ }\href {\doibase 10.3847/1538-4357/ac1596} {\bibfield  {journal} {\bibinfo  {journal} {Astrophys. J.}\ }\textbf {\bibinfo {volume} {926}},\ \bibinfo {pages} {54} (\bibinfo {year} {2022})},\ \Eprint {http://arxiv.org/abs/2008.12619} {arXiv:2008.12619 [astro-ph.CO]} \BibitemShut {NoStop}%
\bibitem [{\citenamefont {Li}\ \emph {et~al.}(2019)\citenamefont {Li} \emph {et~al.}}]{Li:2017drr}%
  \BibitemOpen
  \bibfield  {author} {\bibinfo {author} {\bibfnamefont {H.}~\bibnamefont {Li}} \emph {et~al.},\ }\href {\doibase 10.1093/nsr/nwy019} {\bibfield  {journal} {\bibinfo  {journal} {Natl. Sci. Rev.}\ }\textbf {\bibinfo {volume} {6}},\ \bibinfo {pages} {145} (\bibinfo {year} {2019})},\ \Eprint {http://arxiv.org/abs/1710.03047} {arXiv:1710.03047 [astro-ph.CO]} \BibitemShut {NoStop}%
\bibitem [{\citenamefont {Adak}\ \emph {et~al.}(2022)\citenamefont {Adak}, \citenamefont {Sen}, \citenamefont {Basak}, \citenamefont {Delabrouille}, \citenamefont {Ghosh}, \citenamefont {Rotti}, \citenamefont {Mart\'\i{}nez-Solaeche},\ and\ \citenamefont {Souradeep}}]{Adak:2021lbu}%
  \BibitemOpen
  \bibfield  {author} {\bibinfo {author} {\bibfnamefont {D.}~\bibnamefont {Adak}}, \bibinfo {author} {\bibfnamefont {A.}~\bibnamefont {Sen}}, \bibinfo {author} {\bibfnamefont {S.}~\bibnamefont {Basak}}, \bibinfo {author} {\bibfnamefont {J.}~\bibnamefont {Delabrouille}}, \bibinfo {author} {\bibfnamefont {T.}~\bibnamefont {Ghosh}}, \bibinfo {author} {\bibfnamefont {A.}~\bibnamefont {Rotti}}, \bibinfo {author} {\bibfnamefont {G.}~\bibnamefont {Mart\'\i{}nez-Solaeche}}, \ and\ \bibinfo {author} {\bibfnamefont {T.}~\bibnamefont {Souradeep}},\ }\href {\doibase 10.1093/mnras/stac1474} {\bibfield  {journal} {\bibinfo  {journal} {Mon. Not. Roy. Astron. Soc.}\ }\textbf {\bibinfo {volume} {514}},\ \bibinfo {pages} {3002} (\bibinfo {year} {2022})},\ \Eprint {http://arxiv.org/abs/2110.12362} {arXiv:2110.12362 [astro-ph.CO]} \BibitemShut {NoStop}%
\bibitem [{\citenamefont {Martin}\ \emph {et~al.}(2014)\citenamefont {Martin}, \citenamefont {Ringeval},\ and\ \citenamefont {Vennin}}]{Martin:2014rqa}%
  \BibitemOpen
  \bibfield  {author} {\bibinfo {author} {\bibfnamefont {J.}~\bibnamefont {Martin}}, \bibinfo {author} {\bibfnamefont {C.}~\bibnamefont {Ringeval}}, \ and\ \bibinfo {author} {\bibfnamefont {V.}~\bibnamefont {Vennin}},\ }\href {\doibase 10.1088/1475-7516/2014/10/038} {\bibfield  {journal} {\bibinfo  {journal} {JCAP}\ }\textbf {\bibinfo {volume} {10}},\ \bibinfo {pages} {038} (\bibinfo {year} {2014})},\ \Eprint {http://arxiv.org/abs/1407.4034} {arXiv:1407.4034 [astro-ph.CO]} \BibitemShut {NoStop}%
\bibitem [{\citenamefont {Bassett}\ \emph {et~al.}(2006)\citenamefont {Bassett}, \citenamefont {Tsujikawa},\ and\ \citenamefont {Wands}}]{Bassett:2005xm}%
  \BibitemOpen
  \bibfield  {author} {\bibinfo {author} {\bibfnamefont {B.~A.}\ \bibnamefont {Bassett}}, \bibinfo {author} {\bibfnamefont {S.}~\bibnamefont {Tsujikawa}}, \ and\ \bibinfo {author} {\bibfnamefont {D.}~\bibnamefont {Wands}},\ }\href {\doibase 10.1103/RevModPhys.78.537} {\bibfield  {journal} {\bibinfo  {journal} {Rev. Mod. Phys.}\ }\textbf {\bibinfo {volume} {78}},\ \bibinfo {pages} {537} (\bibinfo {year} {2006})},\ \Eprint {http://arxiv.org/abs/astro-ph/0507632} {arXiv:astro-ph/0507632} \BibitemShut {NoStop}%
\bibitem [{\citenamefont {Martin}\ and\ \citenamefont {Ringeval}(2010)}]{Martin:2010kz}%
  \BibitemOpen
  \bibfield  {author} {\bibinfo {author} {\bibfnamefont {J.}~\bibnamefont {Martin}}\ and\ \bibinfo {author} {\bibfnamefont {C.}~\bibnamefont {Ringeval}},\ }\href {\doibase 10.1103/PhysRevD.82.023511} {\bibfield  {journal} {\bibinfo  {journal} {Phys. Rev. D}\ }\textbf {\bibinfo {volume} {82}},\ \bibinfo {pages} {023511} (\bibinfo {year} {2010})},\ \Eprint {http://arxiv.org/abs/1004.5525} {arXiv:1004.5525 [astro-ph.CO]} \BibitemShut {NoStop}%
\bibitem [{\citenamefont {Martin}\ \emph {et~al.}(2015)\citenamefont {Martin}, \citenamefont {Ringeval},\ and\ \citenamefont {Vennin}}]{Martin:2014nya}%
  \BibitemOpen
  \bibfield  {author} {\bibinfo {author} {\bibfnamefont {J.}~\bibnamefont {Martin}}, \bibinfo {author} {\bibfnamefont {C.}~\bibnamefont {Ringeval}}, \ and\ \bibinfo {author} {\bibfnamefont {V.}~\bibnamefont {Vennin}},\ }\href {\doibase 10.1103/PhysRevLett.114.081303} {\bibfield  {journal} {\bibinfo  {journal} {Phys. Rev. Lett.}\ }\textbf {\bibinfo {volume} {114}},\ \bibinfo {pages} {081303} (\bibinfo {year} {2015})},\ \Eprint {http://arxiv.org/abs/1410.7958} {arXiv:1410.7958 [astro-ph.CO]} \BibitemShut {NoStop}%
\bibitem [{\citenamefont {Meerburg}\ \emph {et~al.}(2019)\citenamefont {Meerburg} \emph {et~al.}}]{Meerburg:2019qqi}%
  \BibitemOpen
  \bibfield  {author} {\bibinfo {author} {\bibfnamefont {P.~D.}\ \bibnamefont {Meerburg}} \emph {et~al.},\ }\href@noop {} {\bibfield  {journal} {\bibinfo  {journal} {Bull. Am. Astron. Soc.}\ }\textbf {\bibinfo {volume} {51}},\ \bibinfo {pages} {107} (\bibinfo {year} {2019})},\ \Eprint {http://arxiv.org/abs/1903.04409} {arXiv:1903.04409 [astro-ph.CO]} \BibitemShut {NoStop}%
\bibitem [{\citenamefont {Allahverdi}\ \emph {et~al.}(2010)\citenamefont {Allahverdi}, \citenamefont {Brandenberger}, \citenamefont {Cyr-Racine},\ and\ \citenamefont {Mazumdar}}]{Allahverdi:2010xz}%
  \BibitemOpen
  \bibfield  {author} {\bibinfo {author} {\bibfnamefont {R.}~\bibnamefont {Allahverdi}}, \bibinfo {author} {\bibfnamefont {R.}~\bibnamefont {Brandenberger}}, \bibinfo {author} {\bibfnamefont {F.-Y.}\ \bibnamefont {Cyr-Racine}}, \ and\ \bibinfo {author} {\bibfnamefont {A.}~\bibnamefont {Mazumdar}},\ }\href {\doibase 10.1146/annurev.nucl.012809.104511} {\bibfield  {journal} {\bibinfo  {journal} {Ann. Rev. Nucl. Part. Sci.}\ }\textbf {\bibinfo {volume} {60}},\ \bibinfo {pages} {27} (\bibinfo {year} {2010})},\ \Eprint {http://arxiv.org/abs/1001.2600} {arXiv:1001.2600 [hep-th]} \BibitemShut {NoStop}%
\bibitem [{\citenamefont {Amin}\ \emph {et~al.}(2014)\citenamefont {Amin}, \citenamefont {Hertzberg}, \citenamefont {Kaiser},\ and\ \citenamefont {Karouby}}]{Amin:2014eta}%
  \BibitemOpen
  \bibfield  {author} {\bibinfo {author} {\bibfnamefont {M.~A.}\ \bibnamefont {Amin}}, \bibinfo {author} {\bibfnamefont {M.~P.}\ \bibnamefont {Hertzberg}}, \bibinfo {author} {\bibfnamefont {D.~I.}\ \bibnamefont {Kaiser}}, \ and\ \bibinfo {author} {\bibfnamefont {J.}~\bibnamefont {Karouby}},\ }\href {\doibase 10.1142/S0218271815300037} {\bibfield  {journal} {\bibinfo  {journal} {Int. J. Mod. Phys. D}\ }\textbf {\bibinfo {volume} {24}},\ \bibinfo {pages} {1530003} (\bibinfo {year} {2014})},\ \Eprint {http://arxiv.org/abs/1410.3808} {arXiv:1410.3808 [hep-ph]} \BibitemShut {NoStop}%
\bibitem [{\citenamefont {Felder}\ \emph {et~al.}(1999{\natexlab{a}})\citenamefont {Felder}, \citenamefont {Kofman},\ and\ \citenamefont {Linde}}]{Felder:1999pv}%
  \BibitemOpen
  \bibfield  {author} {\bibinfo {author} {\bibfnamefont {G.~N.}\ \bibnamefont {Felder}}, \bibinfo {author} {\bibfnamefont {L.}~\bibnamefont {Kofman}}, \ and\ \bibinfo {author} {\bibfnamefont {A.~D.}\ \bibnamefont {Linde}},\ }\href {\doibase 10.1103/PhysRevD.60.103505} {\bibfield  {journal} {\bibinfo  {journal} {Phys. Rev. D}\ }\textbf {\bibinfo {volume} {60}},\ \bibinfo {pages} {103505} (\bibinfo {year} {1999}{\natexlab{a}})},\ \Eprint {http://arxiv.org/abs/hep-ph/9903350} {arXiv:hep-ph/9903350} \BibitemShut {NoStop}%
\bibitem [{\citenamefont {Peebles}\ and\ \citenamefont {Vilenkin}(1999)}]{Peebles:1998qn}%
  \BibitemOpen
  \bibfield  {author} {\bibinfo {author} {\bibfnamefont {P.~J.~E.}\ \bibnamefont {Peebles}}\ and\ \bibinfo {author} {\bibfnamefont {A.}~\bibnamefont {Vilenkin}},\ }\href {\doibase 10.1103/PhysRevD.59.063505} {\bibfield  {journal} {\bibinfo  {journal} {Phys. Rev. D}\ }\textbf {\bibinfo {volume} {59}},\ \bibinfo {pages} {063505} (\bibinfo {year} {1999})},\ \Eprint {http://arxiv.org/abs/astro-ph/9810509} {arXiv:astro-ph/9810509} \BibitemShut {NoStop}%
\bibitem [{\citenamefont {Bettoni}\ and\ \citenamefont {Rubio}(2022)}]{Bettoni:2021qfs}%
  \BibitemOpen
  \bibfield  {author} {\bibinfo {author} {\bibfnamefont {D.}~\bibnamefont {Bettoni}}\ and\ \bibinfo {author} {\bibfnamefont {J.}~\bibnamefont {Rubio}},\ }\href {\doibase 10.3390/galaxies10010022} {\bibfield  {journal} {\bibinfo  {journal} {Galaxies}\ }\textbf {\bibinfo {volume} {10}},\ \bibinfo {pages} {22} (\bibinfo {year} {2022})},\ \Eprint {http://arxiv.org/abs/2112.11948} {arXiv:2112.11948 [astro-ph.CO]} \BibitemShut {NoStop}%
\bibitem [{\citenamefont {de~Haro}\ and\ \citenamefont {Sal\'o}(2021)}]{deHaro:2021swo}%
  \BibitemOpen
  \bibfield  {author} {\bibinfo {author} {\bibfnamefont {J.}~\bibnamefont {de~Haro}}\ and\ \bibinfo {author} {\bibfnamefont {L.~A.}\ \bibnamefont {Sal\'o}},\ }\href {\doibase 10.3390/galaxies9040073} {\bibfield  {journal} {\bibinfo  {journal} {Galaxies}\ }\textbf {\bibinfo {volume} {9}},\ \bibinfo {pages} {73} (\bibinfo {year} {2021})},\ \Eprint {http://arxiv.org/abs/2108.11144} {arXiv:2108.11144 [gr-qc]} \BibitemShut {NoStop}%
\bibitem [{\citenamefont {Wetterich}(2022)}]{Wetterich:2022brb}%
  \BibitemOpen
  \bibfield  {author} {\bibinfo {author} {\bibfnamefont {C.}~\bibnamefont {Wetterich}},\ }\href {\doibase 10.3390/galaxies10020050} {\bibfield  {journal} {\bibinfo  {journal} {Galaxies}\ }\textbf {\bibinfo {volume} {10}},\ \bibinfo {pages} {50} (\bibinfo {year} {2022})},\ \Eprint {http://arxiv.org/abs/2201.12213} {arXiv:2201.12213 [astro-ph.CO]} \BibitemShut {NoStop}%
\bibitem [{\citenamefont {Jaman}\ and\ \citenamefont {Sami}(2022)}]{Jaman:2022bho}%
  \BibitemOpen
  \bibfield  {author} {\bibinfo {author} {\bibfnamefont {N.}~\bibnamefont {Jaman}}\ and\ \bibinfo {author} {\bibfnamefont {M.}~\bibnamefont {Sami}},\ }\href {\doibase 10.3390/galaxies10020051} {\bibfield  {journal} {\bibinfo  {journal} {Galaxies}\ }\textbf {\bibinfo {volume} {10}},\ \bibinfo {pages} {51} (\bibinfo {year} {2022})},\ \Eprint {http://arxiv.org/abs/2202.06194} {arXiv:2202.06194 [gr-qc]} \BibitemShut {NoStop}%
\bibitem [{\citenamefont {Caldwell}\ \emph {et~al.}(1998)\citenamefont {Caldwell}, \citenamefont {Dave},\ and\ \citenamefont {Steinhardt}}]{Caldwell:1997ii}%
  \BibitemOpen
  \bibfield  {author} {\bibinfo {author} {\bibfnamefont {R.~R.}\ \bibnamefont {Caldwell}}, \bibinfo {author} {\bibfnamefont {R.}~\bibnamefont {Dave}}, \ and\ \bibinfo {author} {\bibfnamefont {P.~J.}\ \bibnamefont {Steinhardt}},\ }\href {\doibase 10.1103/PhysRevLett.80.1582} {\bibfield  {journal} {\bibinfo  {journal} {Phys. Rev. Lett.}\ }\textbf {\bibinfo {volume} {80}},\ \bibinfo {pages} {1582} (\bibinfo {year} {1998})},\ \Eprint {http://arxiv.org/abs/astro-ph/9708069} {arXiv:astro-ph/9708069} \BibitemShut {NoStop}%
\bibitem [{\citenamefont {Dimopoulos}\ and\ \citenamefont {Markkanen}(2018)}]{Dimopoulos:2018wfg}%
  \BibitemOpen
  \bibfield  {author} {\bibinfo {author} {\bibfnamefont {K.}~\bibnamefont {Dimopoulos}}\ and\ \bibinfo {author} {\bibfnamefont {T.}~\bibnamefont {Markkanen}},\ }\href {\doibase 10.1088/1475-7516/2018/06/021} {\bibfield  {journal} {\bibinfo  {journal} {JCAP}\ }\textbf {\bibinfo {volume} {06}},\ \bibinfo {pages} {021} (\bibinfo {year} {2018})},\ \Eprint {http://arxiv.org/abs/1803.07399} {arXiv:1803.07399 [gr-qc]} \BibitemShut {NoStop}%
\bibitem [{\citenamefont {Opferkuch}\ \emph {et~al.}(2019)\citenamefont {Opferkuch}, \citenamefont {Schwaller},\ and\ \citenamefont {Stefanek}}]{Opferkuch:2019zbd}%
  \BibitemOpen
  \bibfield  {author} {\bibinfo {author} {\bibfnamefont {T.}~\bibnamefont {Opferkuch}}, \bibinfo {author} {\bibfnamefont {P.}~\bibnamefont {Schwaller}}, \ and\ \bibinfo {author} {\bibfnamefont {B.~A.}\ \bibnamefont {Stefanek}},\ }\href {\doibase 10.1088/1475-7516/2019/07/016} {\bibfield  {journal} {\bibinfo  {journal} {JCAP}\ }\textbf {\bibinfo {volume} {07}},\ \bibinfo {pages} {016} (\bibinfo {year} {2019})},\ \Eprint {http://arxiv.org/abs/1905.06823} {arXiv:1905.06823 [gr-qc]} \BibitemShut {NoStop}%
\bibitem [{\citenamefont {Bettoni}\ \emph {et~al.}(2022)\citenamefont {Bettoni}, \citenamefont {Lopez-Eiguren},\ and\ \citenamefont {Rubio}}]{Bettoni:2021zhq}%
  \BibitemOpen
  \bibfield  {author} {\bibinfo {author} {\bibfnamefont {D.}~\bibnamefont {Bettoni}}, \bibinfo {author} {\bibfnamefont {A.}~\bibnamefont {Lopez-Eiguren}}, \ and\ \bibinfo {author} {\bibfnamefont {J.}~\bibnamefont {Rubio}},\ }\href {\doibase 10.1088/1475-7516/2022/01/002} {\bibfield  {journal} {\bibinfo  {journal} {JCAP}\ }\textbf {\bibinfo {volume} {01}},\ \bibinfo {pages} {002} (\bibinfo {year} {2022})},\ \Eprint {http://arxiv.org/abs/2107.09671} {arXiv:2107.09671 [hep-ph]} \BibitemShut {NoStop}%
\bibitem [{\citenamefont {Joyce}\ and\ \citenamefont {Prokopec}(1998)}]{Joyce:1997fc}%
  \BibitemOpen
  \bibfield  {author} {\bibinfo {author} {\bibfnamefont {M.}~\bibnamefont {Joyce}}\ and\ \bibinfo {author} {\bibfnamefont {T.}~\bibnamefont {Prokopec}},\ }\href {\doibase 10.1103/PhysRevD.57.6022} {\bibfield  {journal} {\bibinfo  {journal} {Phys. Rev. D}\ }\textbf {\bibinfo {volume} {57}},\ \bibinfo {pages} {6022} (\bibinfo {year} {1998})},\ \Eprint {http://arxiv.org/abs/hep-ph/9709320} {arXiv:hep-ph/9709320} \BibitemShut {NoStop}%
\bibitem [{\citenamefont {Gouttenoire}\ \emph {et~al.}(2021{\natexlab{a}})\citenamefont {Gouttenoire}, \citenamefont {Servant},\ and\ \citenamefont {Simakachorn}}]{Gouttenoire:2021jhk}%
  \BibitemOpen
  \bibfield  {author} {\bibinfo {author} {\bibfnamefont {Y.}~\bibnamefont {Gouttenoire}}, \bibinfo {author} {\bibfnamefont {G.}~\bibnamefont {Servant}}, \ and\ \bibinfo {author} {\bibfnamefont {P.}~\bibnamefont {Simakachorn}},\ }\href@noop {} {\  (\bibinfo {year} {2021}{\natexlab{a}})},\ \Eprint {http://arxiv.org/abs/2111.01150} {arXiv:2111.01150 [hep-ph]} \BibitemShut {NoStop}%
\bibitem [{\citenamefont {Laverda}\ and\ \citenamefont {Rubio}(2024)}]{Laverda:2023uqv}%
  \BibitemOpen
  \bibfield  {author} {\bibinfo {author} {\bibfnamefont {G.}~\bibnamefont {Laverda}}\ and\ \bibinfo {author} {\bibfnamefont {J.}~\bibnamefont {Rubio}},\ }\href {\doibase 10.1088/1475-7516/2024/03/033} {\bibfield  {journal} {\bibinfo  {journal} {JCAP}\ }\textbf {\bibinfo {volume} {03}},\ \bibinfo {pages} {033} (\bibinfo {year} {2024})},\ \bibinfo {note} {[Erratum: JCAP 06, E01 (2024)]},\ \Eprint {http://arxiv.org/abs/2307.03774} {arXiv:2307.03774 [astro-ph.CO]} \BibitemShut {NoStop}%
\bibitem [{\citenamefont {Peccei}\ and\ \citenamefont {Quinn}(1977)}]{Peccei:1977hh}%
  \BibitemOpen
  \bibfield  {author} {\bibinfo {author} {\bibfnamefont {R.~D.}\ \bibnamefont {Peccei}}\ and\ \bibinfo {author} {\bibfnamefont {H.~R.}\ \bibnamefont {Quinn}},\ }\href {\doibase 10.1103/PhysRevLett.38.1440} {\bibfield  {journal} {\bibinfo  {journal} {Phys. Rev. Lett.}\ }\textbf {\bibinfo {volume} {38}},\ \bibinfo {pages} {1440} (\bibinfo {year} {1977})}\BibitemShut {NoStop}%
\bibitem [{\citenamefont {Chikashige}\ \emph {et~al.}(1981)\citenamefont {Chikashige}, \citenamefont {Mohapatra},\ and\ \citenamefont {Peccei}}]{Chikashige:1980ui}%
  \BibitemOpen
  \bibfield  {author} {\bibinfo {author} {\bibfnamefont {Y.}~\bibnamefont {Chikashige}}, \bibinfo {author} {\bibfnamefont {R.~N.}\ \bibnamefont {Mohapatra}}, \ and\ \bibinfo {author} {\bibfnamefont {R.~D.}\ \bibnamefont {Peccei}},\ }\href {\doibase 10.1016/0370-2693(81)90011-3} {\bibfield  {journal} {\bibinfo  {journal} {Phys. Lett. B}\ }\textbf {\bibinfo {volume} {98}},\ \bibinfo {pages} {265} (\bibinfo {year} {1981})}\BibitemShut {NoStop}%
\bibitem [{\citenamefont {Froggatt}\ and\ \citenamefont {Nielsen}(1979)}]{Froggatt:1978nt}%
  \BibitemOpen
  \bibfield  {author} {\bibinfo {author} {\bibfnamefont {C.~D.}\ \bibnamefont {Froggatt}}\ and\ \bibinfo {author} {\bibfnamefont {H.~B.}\ \bibnamefont {Nielsen}},\ }\href {\doibase 10.1016/0550-3213(79)90316-X} {\bibfield  {journal} {\bibinfo  {journal} {Nucl. Phys. B}\ }\textbf {\bibinfo {volume} {147}},\ \bibinfo {pages} {277} (\bibinfo {year} {1979})}\BibitemShut {NoStop}%
\bibitem [{\citenamefont {Weinberg}(1978)}]{Weinberg:1977ma}%
  \BibitemOpen
  \bibfield  {author} {\bibinfo {author} {\bibfnamefont {S.}~\bibnamefont {Weinberg}},\ }\href {\doibase 10.1103/PhysRevLett.40.223} {\bibfield  {journal} {\bibinfo  {journal} {Phys. Rev. Lett.}\ }\textbf {\bibinfo {volume} {40}},\ \bibinfo {pages} {223} (\bibinfo {year} {1978})}\BibitemShut {NoStop}%
\bibitem [{\citenamefont {Wilczek}(1978)}]{Wilczek:1977pj}%
  \BibitemOpen
  \bibfield  {author} {\bibinfo {author} {\bibfnamefont {F.}~\bibnamefont {Wilczek}},\ }\href {\doibase 10.1103/PhysRevLett.40.279} {\bibfield  {journal} {\bibinfo  {journal} {Phys. Rev. Lett.}\ }\textbf {\bibinfo {volume} {40}},\ \bibinfo {pages} {279} (\bibinfo {year} {1978})}\BibitemShut {NoStop}%
\bibitem [{\citenamefont {Davidson}\ and\ \citenamefont {Wali}(1982)}]{Davidson:1981zd}%
  \BibitemOpen
  \bibfield  {author} {\bibinfo {author} {\bibfnamefont {A.}~\bibnamefont {Davidson}}\ and\ \bibinfo {author} {\bibfnamefont {K.~C.}\ \bibnamefont {Wali}},\ }\href {\doibase 10.1103/PhysRevLett.48.11} {\bibfield  {journal} {\bibinfo  {journal} {Phys. Rev. Lett.}\ }\textbf {\bibinfo {volume} {48}},\ \bibinfo {pages} {11} (\bibinfo {year} {1982})}\BibitemShut {NoStop}%
\bibitem [{\citenamefont {Reiss}(1982)}]{Reiss:1982sq}%
  \BibitemOpen
  \bibfield  {author} {\bibinfo {author} {\bibfnamefont {D.~B.}\ \bibnamefont {Reiss}},\ }\href {\doibase 10.1016/0370-2693(82)90647-5} {\bibfield  {journal} {\bibinfo  {journal} {Phys. Lett. B}\ }\textbf {\bibinfo {volume} {115}},\ \bibinfo {pages} {217} (\bibinfo {year} {1982})}\BibitemShut {NoStop}%
\bibitem [{\citenamefont {Wilczek}(1982)}]{Wilczek:1982rv}%
  \BibitemOpen
  \bibfield  {author} {\bibinfo {author} {\bibfnamefont {F.}~\bibnamefont {Wilczek}},\ }\href {\doibase 10.1103/PhysRevLett.49.1549} {\bibfield  {journal} {\bibinfo  {journal} {Phys. Rev. Lett.}\ }\textbf {\bibinfo {volume} {49}},\ \bibinfo {pages} {1549} (\bibinfo {year} {1982})}\BibitemShut {NoStop}%
\bibitem [{\citenamefont {Davidson}\ \emph {et~al.}(1984{\natexlab{a}})\citenamefont {Davidson}, \citenamefont {Nair},\ and\ \citenamefont {Wali}}]{Davidson:1983fy}%
  \BibitemOpen
  \bibfield  {author} {\bibinfo {author} {\bibfnamefont {A.}~\bibnamefont {Davidson}}, \bibinfo {author} {\bibfnamefont {V.~P.}\ \bibnamefont {Nair}}, \ and\ \bibinfo {author} {\bibfnamefont {K.~C.}\ \bibnamefont {Wali}},\ }\href {\doibase 10.1103/PhysRevD.29.1504} {\bibfield  {journal} {\bibinfo  {journal} {Phys. Rev. D}\ }\textbf {\bibinfo {volume} {29}},\ \bibinfo {pages} {1504} (\bibinfo {year} {1984}{\natexlab{a}})}\BibitemShut {NoStop}%
\bibitem [{\citenamefont {Davidson}\ \emph {et~al.}(1984{\natexlab{b}})\citenamefont {Davidson}, \citenamefont {Nair},\ and\ \citenamefont {Wali}}]{Davidson:1983fe}%
  \BibitemOpen
  \bibfield  {author} {\bibinfo {author} {\bibfnamefont {A.}~\bibnamefont {Davidson}}, \bibinfo {author} {\bibfnamefont {V.~P.}\ \bibnamefont {Nair}}, \ and\ \bibinfo {author} {\bibfnamefont {K.~C.}\ \bibnamefont {Wali}},\ }\href {\doibase 10.1103/PhysRevD.29.1513} {\bibfield  {journal} {\bibinfo  {journal} {Phys. Rev. D}\ }\textbf {\bibinfo {volume} {29}},\ \bibinfo {pages} {1513} (\bibinfo {year} {1984}{\natexlab{b}})}\BibitemShut {NoStop}%
\bibitem [{\citenamefont {Arvanitaki}\ \emph {et~al.}(2010)\citenamefont {Arvanitaki}, \citenamefont {Dimopoulos}, \citenamefont {Dubovsky}, \citenamefont {Kaloper},\ and\ \citenamefont {March-Russell}}]{Arvanitaki:2009fg}%
  \BibitemOpen
  \bibfield  {author} {\bibinfo {author} {\bibfnamefont {A.}~\bibnamefont {Arvanitaki}}, \bibinfo {author} {\bibfnamefont {S.}~\bibnamefont {Dimopoulos}}, \bibinfo {author} {\bibfnamefont {S.}~\bibnamefont {Dubovsky}}, \bibinfo {author} {\bibfnamefont {N.}~\bibnamefont {Kaloper}}, \ and\ \bibinfo {author} {\bibfnamefont {J.}~\bibnamefont {March-Russell}},\ }\href {\doibase 10.1103/PhysRevD.81.123530} {\bibfield  {journal} {\bibinfo  {journal} {Phys. Rev. D}\ }\textbf {\bibinfo {volume} {81}},\ \bibinfo {pages} {123530} (\bibinfo {year} {2010})},\ \Eprint {http://arxiv.org/abs/0905.4720} {arXiv:0905.4720 [hep-th]} \BibitemShut {NoStop}%
\bibitem [{\citenamefont {Georgi}\ \emph {et~al.}(1986)\citenamefont {Georgi}, \citenamefont {Kaplan},\ and\ \citenamefont {Randall}}]{Georgi:1986df}%
  \BibitemOpen
  \bibfield  {author} {\bibinfo {author} {\bibfnamefont {H.}~\bibnamefont {Georgi}}, \bibinfo {author} {\bibfnamefont {D.~B.}\ \bibnamefont {Kaplan}}, \ and\ \bibinfo {author} {\bibfnamefont {L.}~\bibnamefont {Randall}},\ }\href {\doibase 10.1016/0370-2693(86)90688-X} {\bibfield  {journal} {\bibinfo  {journal} {Phys. Lett. B}\ }\textbf {\bibinfo {volume} {169}},\ \bibinfo {pages} {73} (\bibinfo {year} {1986})}\BibitemShut {NoStop}%
\bibitem [{\citenamefont {Jaeckel}\ and\ \citenamefont {Ringwald}(2010)}]{Jaeckel:2010ni}%
  \BibitemOpen
  \bibfield  {author} {\bibinfo {author} {\bibfnamefont {J.}~\bibnamefont {Jaeckel}}\ and\ \bibinfo {author} {\bibfnamefont {A.}~\bibnamefont {Ringwald}},\ }\href {\doibase 10.1146/annurev.nucl.012809.104433} {\bibfield  {journal} {\bibinfo  {journal} {Ann. Rev. Nucl. Part. Sci.}\ }\textbf {\bibinfo {volume} {60}},\ \bibinfo {pages} {405} (\bibinfo {year} {2010})},\ \Eprint {http://arxiv.org/abs/1002.0329} {arXiv:1002.0329 [hep-ph]} \BibitemShut {NoStop}%
\bibitem [{\citenamefont {Ringwald}(2014)}]{Ringwald:2014vqa}%
  \BibitemOpen
  \bibfield  {author} {\bibinfo {author} {\bibfnamefont {A.}~\bibnamefont {Ringwald}},\ }in\ \href@noop {} {\emph {\bibinfo {booktitle} {{49th Rencontres de Moriond on Electroweak Interactions and Unified Theories}}}}\ (\bibinfo {year} {2014})\ pp.\ \bibinfo {pages} {223--230},\ \Eprint {http://arxiv.org/abs/1407.0546} {arXiv:1407.0546 [hep-ph]} \BibitemShut {NoStop}%
\bibitem [{\citenamefont {Bauer}\ \emph {et~al.}(2017)\citenamefont {Bauer}, \citenamefont {Neubert},\ and\ \citenamefont {Thamm}}]{Bauer:2017ris}%
  \BibitemOpen
  \bibfield  {author} {\bibinfo {author} {\bibfnamefont {M.}~\bibnamefont {Bauer}}, \bibinfo {author} {\bibfnamefont {M.}~\bibnamefont {Neubert}}, \ and\ \bibinfo {author} {\bibfnamefont {A.}~\bibnamefont {Thamm}},\ }\href {\doibase 10.1007/JHEP12(2017)044} {\bibfield  {journal} {\bibinfo  {journal} {JHEP}\ }\textbf {\bibinfo {volume} {12}},\ \bibinfo {pages} {044} (\bibinfo {year} {2017})},\ \Eprint {http://arxiv.org/abs/1708.00443} {arXiv:1708.00443 [hep-ph]} \BibitemShut {NoStop}%
\bibitem [{\citenamefont {Brivio}\ \emph {et~al.}(2017)\citenamefont {Brivio}, \citenamefont {Gavela}, \citenamefont {Merlo}, \citenamefont {Mimasu}, \citenamefont {No}, \citenamefont {del Rey},\ and\ \citenamefont {Sanz}}]{Brivio:2017ije}%
  \BibitemOpen
  \bibfield  {author} {\bibinfo {author} {\bibfnamefont {I.}~\bibnamefont {Brivio}}, \bibinfo {author} {\bibfnamefont {M.~B.}\ \bibnamefont {Gavela}}, \bibinfo {author} {\bibfnamefont {L.}~\bibnamefont {Merlo}}, \bibinfo {author} {\bibfnamefont {K.}~\bibnamefont {Mimasu}}, \bibinfo {author} {\bibfnamefont {J.~M.}\ \bibnamefont {No}}, \bibinfo {author} {\bibfnamefont {R.}~\bibnamefont {del Rey}}, \ and\ \bibinfo {author} {\bibfnamefont {V.}~\bibnamefont {Sanz}},\ }\href {\doibase 10.1140/epjc/s10052-017-5111-3} {\bibfield  {journal} {\bibinfo  {journal} {Eur. Phys. J. C}\ }\textbf {\bibinfo {volume} {77}},\ \bibinfo {pages} {572} (\bibinfo {year} {2017})},\ \Eprint {http://arxiv.org/abs/1701.05379} {arXiv:1701.05379 [hep-ph]} \BibitemShut {NoStop}%
\bibitem [{\citenamefont {Choi}\ \emph {et~al.}(2021)\citenamefont {Choi}, \citenamefont {Im},\ and\ \citenamefont {Sub~Shin}}]{Choi:2020rgn}%
  \BibitemOpen
  \bibfield  {author} {\bibinfo {author} {\bibfnamefont {K.}~\bibnamefont {Choi}}, \bibinfo {author} {\bibfnamefont {S.~H.}\ \bibnamefont {Im}}, \ and\ \bibinfo {author} {\bibfnamefont {C.}~\bibnamefont {Sub~Shin}},\ }\href {\doibase 10.1146/annurev-nucl-120720-031147} {\bibfield  {journal} {\bibinfo  {journal} {Ann. Rev. Nucl. Part. Sci.}\ }\textbf {\bibinfo {volume} {71}},\ \bibinfo {pages} {225} (\bibinfo {year} {2021})},\ \Eprint {http://arxiv.org/abs/2012.05029} {arXiv:2012.05029 [hep-ph]} \BibitemShut {NoStop}%
\bibitem [{\citenamefont {Giannotti}(2023)}]{Giannotti:2022euq}%
  \BibitemOpen
  \bibfield  {author} {\bibinfo {author} {\bibfnamefont {M.}~\bibnamefont {Giannotti}},\ }\href {\doibase 10.1088/1742-6596/2502/1/012003} {\bibfield  {journal} {\bibinfo  {journal} {J. Phys. Conf. Ser.}\ }\textbf {\bibinfo {volume} {2502}},\ \bibinfo {pages} {012003} (\bibinfo {year} {2023})},\ \Eprint {http://arxiv.org/abs/2205.06831} {arXiv:2205.06831 [hep-ph]} \BibitemShut {NoStop}%
\bibitem [{\citenamefont {Dalla Valle~Garcia}\ \emph {et~al.}(2024)\citenamefont {Dalla Valle~Garcia}, \citenamefont {Kahlhoefer}, \citenamefont {Ovchynnikov},\ and\ \citenamefont {Zaporozhchenko}}]{DallaValleGarcia:2023xhh}%
  \BibitemOpen
  \bibfield  {author} {\bibinfo {author} {\bibfnamefont {G.}~\bibnamefont {Dalla Valle~Garcia}}, \bibinfo {author} {\bibfnamefont {F.}~\bibnamefont {Kahlhoefer}}, \bibinfo {author} {\bibfnamefont {M.}~\bibnamefont {Ovchynnikov}}, \ and\ \bibinfo {author} {\bibfnamefont {A.}~\bibnamefont {Zaporozhchenko}},\ }\href {\doibase 10.1103/PhysRevD.109.055042} {\bibfield  {journal} {\bibinfo  {journal} {Phys. Rev. D}\ }\textbf {\bibinfo {volume} {109}},\ \bibinfo {pages} {055042} (\bibinfo {year} {2024})},\ \Eprint {http://arxiv.org/abs/2310.03524} {arXiv:2310.03524 [hep-ph]} \BibitemShut {NoStop}%
\bibitem [{\citenamefont {Domcke}\ \emph {et~al.}(2020)\citenamefont {Domcke}, \citenamefont {Ema}, \citenamefont {Mukaida},\ and\ \citenamefont {Yamada}}]{Domcke:2020kcp}%
  \BibitemOpen
  \bibfield  {author} {\bibinfo {author} {\bibfnamefont {V.}~\bibnamefont {Domcke}}, \bibinfo {author} {\bibfnamefont {Y.}~\bibnamefont {Ema}}, \bibinfo {author} {\bibfnamefont {K.}~\bibnamefont {Mukaida}}, \ and\ \bibinfo {author} {\bibfnamefont {M.}~\bibnamefont {Yamada}},\ }\href {\doibase 10.1007/JHEP08(2020)096} {\bibfield  {journal} {\bibinfo  {journal} {JHEP}\ }\textbf {\bibinfo {volume} {08}},\ \bibinfo {pages} {096} (\bibinfo {year} {2020})},\ \Eprint {http://arxiv.org/abs/2006.03148} {arXiv:2006.03148 [hep-ph]} \BibitemShut {NoStop}%
\bibitem [{\citenamefont {Chun}\ and\ \citenamefont {Jung}(2024)}]{Chun:2023eqc}%
  \BibitemOpen
  \bibfield  {author} {\bibinfo {author} {\bibfnamefont {E.~J.}\ \bibnamefont {Chun}}\ and\ \bibinfo {author} {\bibfnamefont {T.~H.}\ \bibnamefont {Jung}},\ }\href {\doibase 10.1103/PhysRevD.109.095004} {\bibfield  {journal} {\bibinfo  {journal} {Phys. Rev. D}\ }\textbf {\bibinfo {volume} {109}},\ \bibinfo {pages} {095004} (\bibinfo {year} {2024})},\ \Eprint {http://arxiv.org/abs/2311.09005} {arXiv:2311.09005 [hep-ph]} \BibitemShut {NoStop}%
\bibitem [{\citenamefont {Fong}\ \emph {et~al.}(2023)\citenamefont {Fong}, \citenamefont {Ghoshal}, \citenamefont {Naskar}, \citenamefont {Rahat},\ and\ \citenamefont {Saad}}]{Fong:2023egk}%
  \BibitemOpen
  \bibfield  {author} {\bibinfo {author} {\bibfnamefont {C.~S.}\ \bibnamefont {Fong}}, \bibinfo {author} {\bibfnamefont {A.}~\bibnamefont {Ghoshal}}, \bibinfo {author} {\bibfnamefont {A.}~\bibnamefont {Naskar}}, \bibinfo {author} {\bibfnamefont {M.~H.}\ \bibnamefont {Rahat}}, \ and\ \bibinfo {author} {\bibfnamefont {S.}~\bibnamefont {Saad}},\ }\href {\doibase 10.1007/JHEP11(2023)182} {\bibfield  {journal} {\bibinfo  {journal} {JHEP}\ }\textbf {\bibinfo {volume} {11}},\ \bibinfo {pages} {182} (\bibinfo {year} {2023})},\ \Eprint {http://arxiv.org/abs/2307.07550} {arXiv:2307.07550 [hep-ph]} \BibitemShut {NoStop}%
\bibitem [{\citenamefont {Datta}\ \emph {et~al.}(2024)\citenamefont {Datta}, \citenamefont {Manna},\ and\ \citenamefont {Sil}}]{Datta:2024xhg}%
  \BibitemOpen
  \bibfield  {author} {\bibinfo {author} {\bibfnamefont {A.}~\bibnamefont {Datta}}, \bibinfo {author} {\bibfnamefont {S.~K.}\ \bibnamefont {Manna}}, \ and\ \bibinfo {author} {\bibfnamefont {A.}~\bibnamefont {Sil}},\ }\href {\doibase 10.1103/PhysRevD.110.095035} {\bibfield  {journal} {\bibinfo  {journal} {Phys. Rev. D}\ }\textbf {\bibinfo {volume} {110}},\ \bibinfo {pages} {095035} (\bibinfo {year} {2024})},\ \Eprint {http://arxiv.org/abs/2405.07003} {arXiv:2405.07003 [hep-ph]} \BibitemShut {NoStop}%
\bibitem [{\citenamefont {Preskill}\ \emph {et~al.}(1983)\citenamefont {Preskill}, \citenamefont {Wise},\ and\ \citenamefont {Wilczek}}]{Preskill:1982cy}%
  \BibitemOpen
  \bibfield  {author} {\bibinfo {author} {\bibfnamefont {J.}~\bibnamefont {Preskill}}, \bibinfo {author} {\bibfnamefont {M.~B.}\ \bibnamefont {Wise}}, \ and\ \bibinfo {author} {\bibfnamefont {F.}~\bibnamefont {Wilczek}},\ }\href {\doibase 10.1016/0370-2693(83)90637-8} {\bibfield  {journal} {\bibinfo  {journal} {Phys. Lett. B}\ }\textbf {\bibinfo {volume} {120}},\ \bibinfo {pages} {127} (\bibinfo {year} {1983})}\BibitemShut {NoStop}%
\bibitem [{\citenamefont {Abbott}\ and\ \citenamefont {Sikivie}(1983)}]{Abbott:1982af}%
  \BibitemOpen
  \bibfield  {author} {\bibinfo {author} {\bibfnamefont {L.~F.}\ \bibnamefont {Abbott}}\ and\ \bibinfo {author} {\bibfnamefont {P.}~\bibnamefont {Sikivie}},\ }\href {\doibase 10.1016/0370-2693(83)90638-X} {\bibfield  {journal} {\bibinfo  {journal} {Phys. Lett. B}\ }\textbf {\bibinfo {volume} {120}},\ \bibinfo {pages} {133} (\bibinfo {year} {1983})}\BibitemShut {NoStop}%
\bibitem [{\citenamefont {Dine}\ and\ \citenamefont {Fischler}(1983)}]{Dine:1982ah}%
  \BibitemOpen
  \bibfield  {author} {\bibinfo {author} {\bibfnamefont {M.}~\bibnamefont {Dine}}\ and\ \bibinfo {author} {\bibfnamefont {W.}~\bibnamefont {Fischler}},\ }\href {\doibase 10.1016/0370-2693(83)90639-1} {\bibfield  {journal} {\bibinfo  {journal} {Phys. Lett. B}\ }\textbf {\bibinfo {volume} {120}},\ \bibinfo {pages} {137} (\bibinfo {year} {1983})}\BibitemShut {NoStop}%
\bibitem [{\citenamefont {Li}\ \emph {et~al.}(2014)\citenamefont {Li}, \citenamefont {Rindler-Daller},\ and\ \citenamefont {Shapiro}}]{Li:2013nal}%
  \BibitemOpen
  \bibfield  {author} {\bibinfo {author} {\bibfnamefont {B.}~\bibnamefont {Li}}, \bibinfo {author} {\bibfnamefont {T.}~\bibnamefont {Rindler-Daller}}, \ and\ \bibinfo {author} {\bibfnamefont {P.~R.}\ \bibnamefont {Shapiro}},\ }\href {\doibase 10.1103/PhysRevD.89.083536} {\bibfield  {journal} {\bibinfo  {journal} {Phys. Rev. D}\ }\textbf {\bibinfo {volume} {89}},\ \bibinfo {pages} {083536} (\bibinfo {year} {2014})},\ \Eprint {http://arxiv.org/abs/1310.6061} {arXiv:1310.6061 [astro-ph.CO]} \BibitemShut {NoStop}%
\bibitem [{\citenamefont {Co}\ \emph {et~al.}(2022{\natexlab{a}})\citenamefont {Co}, \citenamefont {Dunsky}, \citenamefont {Fernandez}, \citenamefont {Ghalsasi}, \citenamefont {Hall}, \citenamefont {Harigaya},\ and\ \citenamefont {Shelton}}]{Co:2021lkc}%
  \BibitemOpen
  \bibfield  {author} {\bibinfo {author} {\bibfnamefont {R.~T.}\ \bibnamefont {Co}}, \bibinfo {author} {\bibfnamefont {D.}~\bibnamefont {Dunsky}}, \bibinfo {author} {\bibfnamefont {N.}~\bibnamefont {Fernandez}}, \bibinfo {author} {\bibfnamefont {A.}~\bibnamefont {Ghalsasi}}, \bibinfo {author} {\bibfnamefont {L.~J.}\ \bibnamefont {Hall}}, \bibinfo {author} {\bibfnamefont {K.}~\bibnamefont {Harigaya}}, \ and\ \bibinfo {author} {\bibfnamefont {J.}~\bibnamefont {Shelton}},\ }\href {\doibase 10.1007/JHEP09(2022)116} {\bibfield  {journal} {\bibinfo  {journal} {JHEP}\ }\textbf {\bibinfo {volume} {09}},\ \bibinfo {pages} {116} (\bibinfo {year} {2022}{\natexlab{a}})},\ \Eprint {http://arxiv.org/abs/2108.09299} {arXiv:2108.09299 [hep-ph]} \BibitemShut {NoStop}%
\bibitem [{\citenamefont {Gouttenoire}\ \emph {et~al.}(2021{\natexlab{b}})\citenamefont {Gouttenoire}, \citenamefont {Servant},\ and\ \citenamefont {Simakachorn}}]{Gouttenoire:2021wzu}%
  \BibitemOpen
  \bibfield  {author} {\bibinfo {author} {\bibfnamefont {Y.}~\bibnamefont {Gouttenoire}}, \bibinfo {author} {\bibfnamefont {G.}~\bibnamefont {Servant}}, \ and\ \bibinfo {author} {\bibfnamefont {P.}~\bibnamefont {Simakachorn}},\ }\href@noop {} {\  (\bibinfo {year} {2021}{\natexlab{b}})},\ \Eprint {http://arxiv.org/abs/2108.10328} {arXiv:2108.10328 [hep-ph]} \BibitemShut {NoStop}%
\bibitem [{\citenamefont {Harigaya}\ \emph {et~al.}(2023{\natexlab{a}})\citenamefont {Harigaya}, \citenamefont {Inomata},\ and\ \citenamefont {Terada}}]{Harigaya:2023mhl}%
  \BibitemOpen
  \bibfield  {author} {\bibinfo {author} {\bibfnamefont {K.}~\bibnamefont {Harigaya}}, \bibinfo {author} {\bibfnamefont {K.}~\bibnamefont {Inomata}}, \ and\ \bibinfo {author} {\bibfnamefont {T.}~\bibnamefont {Terada}},\ }\href {\doibase 10.1103/PhysRevD.108.L081303} {\bibfield  {journal} {\bibinfo  {journal} {Phys. Rev. D}\ }\textbf {\bibinfo {volume} {108}},\ \bibinfo {pages} {L081303} (\bibinfo {year} {2023}{\natexlab{a}})},\ \Eprint {http://arxiv.org/abs/2305.14242} {arXiv:2305.14242 [hep-ph]} \BibitemShut {NoStop}%
\bibitem [{\citenamefont {Harigaya}\ \emph {et~al.}(2023{\natexlab{b}})\citenamefont {Harigaya}, \citenamefont {Inomata},\ and\ \citenamefont {Terada}}]{Harigaya:2023pmw}%
  \BibitemOpen
  \bibfield  {author} {\bibinfo {author} {\bibfnamefont {K.}~\bibnamefont {Harigaya}}, \bibinfo {author} {\bibfnamefont {K.}~\bibnamefont {Inomata}}, \ and\ \bibinfo {author} {\bibfnamefont {T.}~\bibnamefont {Terada}},\ }\href {\doibase 10.1103/PhysRevD.108.123538} {\bibfield  {journal} {\bibinfo  {journal} {Phys. Rev. D}\ }\textbf {\bibinfo {volume} {108}},\ \bibinfo {pages} {123538} (\bibinfo {year} {2023}{\natexlab{b}})},\ \Eprint {http://arxiv.org/abs/2309.00228} {arXiv:2309.00228 [astro-ph.CO]} \BibitemShut {NoStop}%
\bibitem [{\citenamefont {Chung}\ and\ \citenamefont {Tadepalli}(2024)}]{Chung:2024ctx}%
  \BibitemOpen
  \bibfield  {author} {\bibinfo {author} {\bibfnamefont {D.~J.~H.}\ \bibnamefont {Chung}}\ and\ \bibinfo {author} {\bibfnamefont {S.~C.}\ \bibnamefont {Tadepalli}},\ }\href@noop {} {\  (\bibinfo {year} {2024})},\ \Eprint {http://arxiv.org/abs/2406.12976} {arXiv:2406.12976 [astro-ph.CO]} \BibitemShut {NoStop}%
\bibitem [{\citenamefont {Duval}\ \emph {et~al.}(2024)\citenamefont {Duval}, \citenamefont {Kuroyanagi}, \citenamefont {Mariotti}, \citenamefont {Romero-Rodr\'\i{}guez},\ and\ \citenamefont {Sakellariadou}}]{Duval:2024jsg}%
  \BibitemOpen
  \bibfield  {author} {\bibinfo {author} {\bibfnamefont {H.}~\bibnamefont {Duval}}, \bibinfo {author} {\bibfnamefont {S.}~\bibnamefont {Kuroyanagi}}, \bibinfo {author} {\bibfnamefont {A.}~\bibnamefont {Mariotti}}, \bibinfo {author} {\bibfnamefont {A.}~\bibnamefont {Romero-Rodr\'\i{}guez}}, \ and\ \bibinfo {author} {\bibfnamefont {M.}~\bibnamefont {Sakellariadou}},\ }\href {\doibase 10.1103/PhysRevD.110.103503} {\bibfield  {journal} {\bibinfo  {journal} {Phys. Rev. D}\ }\textbf {\bibinfo {volume} {110}},\ \bibinfo {pages} {103503} (\bibinfo {year} {2024})},\ \Eprint {http://arxiv.org/abs/2405.10201} {arXiv:2405.10201 [gr-qc]} \BibitemShut {NoStop}%
\bibitem [{\citenamefont {Affleck}\ and\ \citenamefont {Dine}(1985)}]{Affleck:1984fy}%
  \BibitemOpen
  \bibfield  {author} {\bibinfo {author} {\bibfnamefont {I.}~\bibnamefont {Affleck}}\ and\ \bibinfo {author} {\bibfnamefont {M.}~\bibnamefont {Dine}},\ }\href {\doibase 10.1016/0550-3213(85)90021-5} {\bibfield  {journal} {\bibinfo  {journal} {Nucl. Phys. B}\ }\textbf {\bibinfo {volume} {249}},\ \bibinfo {pages} {361} (\bibinfo {year} {1985})}\BibitemShut {NoStop}%
\bibitem [{\citenamefont {Co}\ and\ \citenamefont {Harigaya}(2020)}]{Co:2019wyp}%
  \BibitemOpen
  \bibfield  {author} {\bibinfo {author} {\bibfnamefont {R.~T.}\ \bibnamefont {Co}}\ and\ \bibinfo {author} {\bibfnamefont {K.}~\bibnamefont {Harigaya}},\ }\href {\doibase 10.1103/PhysRevLett.124.111602} {\bibfield  {journal} {\bibinfo  {journal} {Phys. Rev. Lett.}\ }\textbf {\bibinfo {volume} {124}},\ \bibinfo {pages} {111602} (\bibinfo {year} {2020})},\ \Eprint {http://arxiv.org/abs/1910.02080} {arXiv:1910.02080 [hep-ph]} \BibitemShut {NoStop}%
\bibitem [{\citenamefont {Co}\ \emph {et~al.}(2020{\natexlab{a}})\citenamefont {Co}, \citenamefont {Hall},\ and\ \citenamefont {Harigaya}}]{Co:2019jts}%
  \BibitemOpen
  \bibfield  {author} {\bibinfo {author} {\bibfnamefont {R.~T.}\ \bibnamefont {Co}}, \bibinfo {author} {\bibfnamefont {L.~J.}\ \bibnamefont {Hall}}, \ and\ \bibinfo {author} {\bibfnamefont {K.}~\bibnamefont {Harigaya}},\ }\href {\doibase 10.1103/PhysRevLett.124.251802} {\bibfield  {journal} {\bibinfo  {journal} {Phys. Rev. Lett.}\ }\textbf {\bibinfo {volume} {124}},\ \bibinfo {pages} {251802} (\bibinfo {year} {2020}{\natexlab{a}})},\ \Eprint {http://arxiv.org/abs/1910.14152} {arXiv:1910.14152 [hep-ph]} \BibitemShut {NoStop}%
\bibitem [{\citenamefont {Co}\ \emph {et~al.}(2020{\natexlab{b}})\citenamefont {Co}, \citenamefont {Hall}, \citenamefont {Harigaya}, \citenamefont {Olive},\ and\ \citenamefont {Verner}}]{Co:2020dya}%
  \BibitemOpen
  \bibfield  {author} {\bibinfo {author} {\bibfnamefont {R.~T.}\ \bibnamefont {Co}}, \bibinfo {author} {\bibfnamefont {L.~J.}\ \bibnamefont {Hall}}, \bibinfo {author} {\bibfnamefont {K.}~\bibnamefont {Harigaya}}, \bibinfo {author} {\bibfnamefont {K.~A.}\ \bibnamefont {Olive}}, \ and\ \bibinfo {author} {\bibfnamefont {S.}~\bibnamefont {Verner}},\ }\href {\doibase 10.1088/1475-7516/2020/08/036} {\bibfield  {journal} {\bibinfo  {journal} {JCAP}\ }\textbf {\bibinfo {volume} {08}},\ \bibinfo {pages} {036} (\bibinfo {year} {2020}{\natexlab{b}})},\ \Eprint {http://arxiv.org/abs/2004.00629} {arXiv:2004.00629 [hep-ph]} \BibitemShut {NoStop}%
\bibitem [{\citenamefont {Co}\ \emph {et~al.}(2021{\natexlab{a}})\citenamefont {Co}, \citenamefont {Hall},\ and\ \citenamefont {Harigaya}}]{Co:2020xlh}%
  \BibitemOpen
  \bibfield  {author} {\bibinfo {author} {\bibfnamefont {R.~T.}\ \bibnamefont {Co}}, \bibinfo {author} {\bibfnamefont {L.~J.}\ \bibnamefont {Hall}}, \ and\ \bibinfo {author} {\bibfnamefont {K.}~\bibnamefont {Harigaya}},\ }\href {\doibase 10.1007/JHEP01(2021)172} {\bibfield  {journal} {\bibinfo  {journal} {JHEP}\ }\textbf {\bibinfo {volume} {01}},\ \bibinfo {pages} {172} (\bibinfo {year} {2021}{\natexlab{a}})},\ \Eprint {http://arxiv.org/abs/2006.04809} {arXiv:2006.04809 [hep-ph]} \BibitemShut {NoStop}%
\bibitem [{\citenamefont {Co}\ \emph {et~al.}(2021{\natexlab{b}})\citenamefont {Co}, \citenamefont {Fernandez}, \citenamefont {Ghalsasi}, \citenamefont {Hall},\ and\ \citenamefont {Harigaya}}]{Co:2020jtv}%
  \BibitemOpen
  \bibfield  {author} {\bibinfo {author} {\bibfnamefont {R.~T.}\ \bibnamefont {Co}}, \bibinfo {author} {\bibfnamefont {N.}~\bibnamefont {Fernandez}}, \bibinfo {author} {\bibfnamefont {A.}~\bibnamefont {Ghalsasi}}, \bibinfo {author} {\bibfnamefont {L.~J.}\ \bibnamefont {Hall}}, \ and\ \bibinfo {author} {\bibfnamefont {K.}~\bibnamefont {Harigaya}},\ }\href {\doibase 10.1007/JHEP03(2021)017} {\bibfield  {journal} {\bibinfo  {journal} {JHEP}\ }\textbf {\bibinfo {volume} {03}},\ \bibinfo {pages} {017} (\bibinfo {year} {2021}{\natexlab{b}})},\ \Eprint {http://arxiv.org/abs/2006.05687} {arXiv:2006.05687 [hep-ph]} \BibitemShut {NoStop}%
\bibitem [{\citenamefont {Jeong}\ and\ \citenamefont {Takahashi}(2013{\natexlab{a}})}]{Jeong:2013axf}%
  \BibitemOpen
  \bibfield  {author} {\bibinfo {author} {\bibfnamefont {K.~S.}\ \bibnamefont {Jeong}}\ and\ \bibinfo {author} {\bibfnamefont {F.}~\bibnamefont {Takahashi}},\ }\href {\doibase 10.1007/JHEP04(2013)121} {\bibfield  {journal} {\bibinfo  {journal} {JHEP}\ }\textbf {\bibinfo {volume} {04}},\ \bibinfo {pages} {121} (\bibinfo {year} {2013}{\natexlab{a}})},\ \Eprint {http://arxiv.org/abs/1302.1486} {arXiv:1302.1486 [hep-ph]} \BibitemShut {NoStop}%
\bibitem [{\citenamefont {Jeong}\ and\ \citenamefont {Takahashi}(2013{\natexlab{b}})}]{Jeong:2013xta}%
  \BibitemOpen
  \bibfield  {author} {\bibinfo {author} {\bibfnamefont {K.~S.}\ \bibnamefont {Jeong}}\ and\ \bibinfo {author} {\bibfnamefont {F.}~\bibnamefont {Takahashi}},\ }\href {\doibase 10.1016/j.physletb.2013.10.061} {\bibfield  {journal} {\bibinfo  {journal} {Phys. Lett. B}\ }\textbf {\bibinfo {volume} {727}},\ \bibinfo {pages} {448} (\bibinfo {year} {2013}{\natexlab{b}})},\ \Eprint {http://arxiv.org/abs/1304.8131} {arXiv:1304.8131 [hep-ph]} \BibitemShut {NoStop}%
\bibitem [{\citenamefont {Higaki}\ \emph {et~al.}(2014)\citenamefont {Higaki}, \citenamefont {Jeong},\ and\ \citenamefont {Takahashi}}]{Higaki:2014ooa}%
  \BibitemOpen
  \bibfield  {author} {\bibinfo {author} {\bibfnamefont {T.}~\bibnamefont {Higaki}}, \bibinfo {author} {\bibfnamefont {K.~S.}\ \bibnamefont {Jeong}}, \ and\ \bibinfo {author} {\bibfnamefont {F.}~\bibnamefont {Takahashi}},\ }\href {\doibase 10.1016/j.physletb.2014.05.014} {\bibfield  {journal} {\bibinfo  {journal} {Phys. Lett. B}\ }\textbf {\bibinfo {volume} {734}},\ \bibinfo {pages} {21} (\bibinfo {year} {2014})},\ \Eprint {http://arxiv.org/abs/1403.4186} {arXiv:1403.4186 [hep-ph]} \BibitemShut {NoStop}%
\bibitem [{\citenamefont {Co}\ \emph {et~al.}(2021{\natexlab{c}})\citenamefont {Co}, \citenamefont {Harigaya},\ and\ \citenamefont {Pierce}}]{Co:2021rhi}%
  \BibitemOpen
  \bibfield  {author} {\bibinfo {author} {\bibfnamefont {R.~T.}\ \bibnamefont {Co}}, \bibinfo {author} {\bibfnamefont {K.}~\bibnamefont {Harigaya}}, \ and\ \bibinfo {author} {\bibfnamefont {A.}~\bibnamefont {Pierce}},\ }\href {\doibase 10.1007/JHEP12(2021)099} {\bibfield  {journal} {\bibinfo  {journal} {JHEP}\ }\textbf {\bibinfo {volume} {12}},\ \bibinfo {pages} {099} (\bibinfo {year} {2021}{\natexlab{c}})},\ \Eprint {http://arxiv.org/abs/2104.02077} {arXiv:2104.02077 [hep-ph]} \BibitemShut {NoStop}%
\bibitem [{\citenamefont {Kawamura}\ and\ \citenamefont {Raby}(2022)}]{Kawamura:2021xpu}%
  \BibitemOpen
  \bibfield  {author} {\bibinfo {author} {\bibfnamefont {J.}~\bibnamefont {Kawamura}}\ and\ \bibinfo {author} {\bibfnamefont {S.}~\bibnamefont {Raby}},\ }\href {\doibase 10.1007/JHEP04(2022)116} {\bibfield  {journal} {\bibinfo  {journal} {JHEP}\ }\textbf {\bibinfo {volume} {04}},\ \bibinfo {pages} {116} (\bibinfo {year} {2022})},\ \Eprint {http://arxiv.org/abs/2109.08605} {arXiv:2109.08605 [hep-ph]} \BibitemShut {NoStop}%
\bibitem [{\citenamefont {Co}\ \emph {et~al.}(2021{\natexlab{d}})\citenamefont {Co}, \citenamefont {Harigaya}, \citenamefont {Johnson},\ and\ \citenamefont {Pierce}}]{Co:2021qgl}%
  \BibitemOpen
  \bibfield  {author} {\bibinfo {author} {\bibfnamefont {R.~T.}\ \bibnamefont {Co}}, \bibinfo {author} {\bibfnamefont {K.}~\bibnamefont {Harigaya}}, \bibinfo {author} {\bibfnamefont {Z.}~\bibnamefont {Johnson}}, \ and\ \bibinfo {author} {\bibfnamefont {A.}~\bibnamefont {Pierce}},\ }\href {\doibase 10.1007/JHEP11(2021)210} {\bibfield  {journal} {\bibinfo  {journal} {JHEP}\ }\textbf {\bibinfo {volume} {11}},\ \bibinfo {pages} {210} (\bibinfo {year} {2021}{\natexlab{d}})},\ \Eprint {http://arxiv.org/abs/2110.05487} {arXiv:2110.05487 [hep-ph]} \BibitemShut {NoStop}%
\bibitem [{\citenamefont {Co}\ \emph {et~al.}(2022{\natexlab{b}})\citenamefont {Co}, \citenamefont {Gherghetta},\ and\ \citenamefont {Harigaya}}]{Co:2022aav}%
  \BibitemOpen
  \bibfield  {author} {\bibinfo {author} {\bibfnamefont {R.~T.}\ \bibnamefont {Co}}, \bibinfo {author} {\bibfnamefont {T.}~\bibnamefont {Gherghetta}}, \ and\ \bibinfo {author} {\bibfnamefont {K.}~\bibnamefont {Harigaya}},\ }\href {\doibase 10.1007/JHEP10(2022)121} {\bibfield  {journal} {\bibinfo  {journal} {JHEP}\ }\textbf {\bibinfo {volume} {10}},\ \bibinfo {pages} {121} (\bibinfo {year} {2022}{\natexlab{b}})},\ \Eprint {http://arxiv.org/abs/2206.00678} {arXiv:2206.00678 [hep-ph]} \BibitemShut {NoStop}%
\bibitem [{\citenamefont {Barnes}\ \emph {et~al.}(2023)\citenamefont {Barnes}, \citenamefont {Co}, \citenamefont {Harigaya},\ and\ \citenamefont {Pierce}}]{Barnes:2022ren}%
  \BibitemOpen
  \bibfield  {author} {\bibinfo {author} {\bibfnamefont {P.}~\bibnamefont {Barnes}}, \bibinfo {author} {\bibfnamefont {R.~T.}\ \bibnamefont {Co}}, \bibinfo {author} {\bibfnamefont {K.}~\bibnamefont {Harigaya}}, \ and\ \bibinfo {author} {\bibfnamefont {A.}~\bibnamefont {Pierce}},\ }\href {\doibase 10.1007/JHEP05(2023)114} {\bibfield  {journal} {\bibinfo  {journal} {JHEP}\ }\textbf {\bibinfo {volume} {05}},\ \bibinfo {pages} {114} (\bibinfo {year} {2023})},\ \Eprint {http://arxiv.org/abs/2208.07878} {arXiv:2208.07878 [hep-ph]} \BibitemShut {NoStop}%
\bibitem [{\citenamefont {Co}\ \emph {et~al.}(2023)\citenamefont {Co}, \citenamefont {Domcke},\ and\ \citenamefont {Harigaya}}]{Co:2022kul}%
  \BibitemOpen
  \bibfield  {author} {\bibinfo {author} {\bibfnamefont {R.~T.}\ \bibnamefont {Co}}, \bibinfo {author} {\bibfnamefont {V.}~\bibnamefont {Domcke}}, \ and\ \bibinfo {author} {\bibfnamefont {K.}~\bibnamefont {Harigaya}},\ }\href {\doibase 10.1007/JHEP07(2023)179} {\bibfield  {journal} {\bibinfo  {journal} {JHEP}\ }\textbf {\bibinfo {volume} {07}},\ \bibinfo {pages} {179} (\bibinfo {year} {2023})},\ \Eprint {http://arxiv.org/abs/2211.12517} {arXiv:2211.12517 [hep-ph]} \BibitemShut {NoStop}%
\bibitem [{\citenamefont {Badziak}\ and\ \citenamefont {Harigaya}(2023)}]{Badziak:2023fsc}%
  \BibitemOpen
  \bibfield  {author} {\bibinfo {author} {\bibfnamefont {M.}~\bibnamefont {Badziak}}\ and\ \bibinfo {author} {\bibfnamefont {K.}~\bibnamefont {Harigaya}},\ }\href {\doibase 10.1007/JHEP06(2023)014} {\bibfield  {journal} {\bibinfo  {journal} {JHEP}\ }\textbf {\bibinfo {volume} {06}},\ \bibinfo {pages} {014} (\bibinfo {year} {2023})},\ \Eprint {http://arxiv.org/abs/2301.09647} {arXiv:2301.09647 [hep-ph]} \BibitemShut {NoStop}%
\bibitem [{\citenamefont {Berbig}(2024{\natexlab{a}})}]{Berbig:2023uzs}%
  \BibitemOpen
  \bibfield  {author} {\bibinfo {author} {\bibfnamefont {M.}~\bibnamefont {Berbig}},\ }\href {\doibase 10.1007/JHEP01(2024)061} {\bibfield  {journal} {\bibinfo  {journal} {JHEP}\ }\textbf {\bibinfo {volume} {01}},\ \bibinfo {pages} {061} (\bibinfo {year} {2024}{\natexlab{a}})},\ \Eprint {http://arxiv.org/abs/2307.14121} {arXiv:2307.14121 [hep-ph]} \BibitemShut {NoStop}%
\bibitem [{\citenamefont {Chao}\ and\ \citenamefont {Peng}(2023)}]{Chao:2023ojl}%
  \BibitemOpen
  \bibfield  {author} {\bibinfo {author} {\bibfnamefont {W.}~\bibnamefont {Chao}}\ and\ \bibinfo {author} {\bibfnamefont {Y.-Q.}\ \bibnamefont {Peng}},\ }\href@noop {} {\  (\bibinfo {year} {2023})},\ \Eprint {http://arxiv.org/abs/2311.06469} {arXiv:2311.06469 [hep-ph]} \BibitemShut {NoStop}%
\bibitem [{\citenamefont {Barnes}\ \emph {et~al.}(2024)\citenamefont {Barnes}, \citenamefont {Co}, \citenamefont {Harigaya},\ and\ \citenamefont {Pierce}}]{Barnes:2024jap}%
  \BibitemOpen
  \bibfield  {author} {\bibinfo {author} {\bibfnamefont {P.}~\bibnamefont {Barnes}}, \bibinfo {author} {\bibfnamefont {R.~T.}\ \bibnamefont {Co}}, \bibinfo {author} {\bibfnamefont {K.}~\bibnamefont {Harigaya}}, \ and\ \bibinfo {author} {\bibfnamefont {A.}~\bibnamefont {Pierce}},\ }\href@noop {} {\  (\bibinfo {year} {2024})},\ \Eprint {http://arxiv.org/abs/2402.10263} {arXiv:2402.10263 [hep-ph]} \BibitemShut {NoStop}%
\bibitem [{\citenamefont {Figueroa}\ and\ \citenamefont {Byrnes}(2017)}]{Figueroa:2016dsc}%
  \BibitemOpen
  \bibfield  {author} {\bibinfo {author} {\bibfnamefont {D.~G.}\ \bibnamefont {Figueroa}}\ and\ \bibinfo {author} {\bibfnamefont {C.~T.}\ \bibnamefont {Byrnes}},\ }\href {\doibase 10.1016/j.physletb.2017.01.059} {\bibfield  {journal} {\bibinfo  {journal} {Phys. Lett. B}\ }\textbf {\bibinfo {volume} {767}},\ \bibinfo {pages} {272} (\bibinfo {year} {2017})},\ \Eprint {http://arxiv.org/abs/1604.03905} {arXiv:1604.03905 [hep-ph]} \BibitemShut {NoStop}%
\bibitem [{\citenamefont {Nakama}\ and\ \citenamefont {Yokoyama}(2019)}]{Nakama:2018gll}%
  \BibitemOpen
  \bibfield  {author} {\bibinfo {author} {\bibfnamefont {T.}~\bibnamefont {Nakama}}\ and\ \bibinfo {author} {\bibfnamefont {J.}~\bibnamefont {Yokoyama}},\ }\href {\doibase 10.1093/ptep/ptz014} {\bibfield  {journal} {\bibinfo  {journal} {PTEP}\ }\textbf {\bibinfo {volume} {2019}},\ \bibinfo {pages} {033E02} (\bibinfo {year} {2019})},\ \Eprint {http://arxiv.org/abs/1803.07111} {arXiv:1803.07111 [gr-qc]} \BibitemShut {NoStop}%
\bibitem [{\citenamefont {Chun}\ \emph {et~al.}(2024)\citenamefont {Chun}, \citenamefont {Jyoti~Das}, \citenamefont {He}, \citenamefont {Jung},\ and\ \citenamefont {Sun}}]{Chun:2024gvp}%
  \BibitemOpen
  \bibfield  {author} {\bibinfo {author} {\bibfnamefont {E.~J.}\ \bibnamefont {Chun}}, \bibinfo {author} {\bibfnamefont {S.}~\bibnamefont {Jyoti~Das}}, \bibinfo {author} {\bibfnamefont {M.}~\bibnamefont {He}}, \bibinfo {author} {\bibfnamefont {T.~H.}\ \bibnamefont {Jung}}, \ and\ \bibinfo {author} {\bibfnamefont {J.}~\bibnamefont {Sun}},\ }\href@noop {} {\  (\bibinfo {year} {2024})},\ \Eprint {http://arxiv.org/abs/2406.04180} {arXiv:2406.04180 [hep-ph]} \BibitemShut {NoStop}%
\bibitem [{\citenamefont {Ferreira}\ \emph {et~al.}(2018)\citenamefont {Ferreira}, \citenamefont {Notari},\ and\ \citenamefont {Simeon}}]{Ferreira:2018nav}%
  \BibitemOpen
  \bibfield  {author} {\bibinfo {author} {\bibfnamefont {R.~Z.}\ \bibnamefont {Ferreira}}, \bibinfo {author} {\bibfnamefont {A.}~\bibnamefont {Notari}}, \ and\ \bibinfo {author} {\bibfnamefont {G.}~\bibnamefont {Simeon}},\ }\href {\doibase 10.1088/1475-7516/2018/11/021} {\bibfield  {journal} {\bibinfo  {journal} {JCAP}\ }\textbf {\bibinfo {volume} {11}},\ \bibinfo {pages} {021} (\bibinfo {year} {2018})},\ \Eprint {http://arxiv.org/abs/1806.05511} {arXiv:1806.05511 [astro-ph.CO]} \BibitemShut {NoStop}%
\bibitem [{\citenamefont {Takahashi}\ and\ \citenamefont {Yin}(2019)}]{Takahashi:2019pqf}%
  \BibitemOpen
  \bibfield  {author} {\bibinfo {author} {\bibfnamefont {F.}~\bibnamefont {Takahashi}}\ and\ \bibinfo {author} {\bibfnamefont {W.}~\bibnamefont {Yin}},\ }\href {\doibase 10.1007/JHEP10(2019)120} {\bibfield  {journal} {\bibinfo  {journal} {JHEP}\ }\textbf {\bibinfo {volume} {10}},\ \bibinfo {pages} {120} (\bibinfo {year} {2019})},\ \Eprint {http://arxiv.org/abs/1908.06071} {arXiv:1908.06071 [hep-ph]} \BibitemShut {NoStop}%
\bibitem [{\citenamefont {Huang}\ \emph {et~al.}(2020)\citenamefont {Huang}, \citenamefont {Madden}, \citenamefont {Racco},\ and\ \citenamefont {Reig}}]{Huang:2020etx}%
  \BibitemOpen
  \bibfield  {author} {\bibinfo {author} {\bibfnamefont {J.}~\bibnamefont {Huang}}, \bibinfo {author} {\bibfnamefont {A.}~\bibnamefont {Madden}}, \bibinfo {author} {\bibfnamefont {D.}~\bibnamefont {Racco}}, \ and\ \bibinfo {author} {\bibfnamefont {M.}~\bibnamefont {Reig}},\ }\href {\doibase 10.1007/JHEP10(2020)143} {\bibfield  {journal} {\bibinfo  {journal} {JHEP}\ }\textbf {\bibinfo {volume} {10}},\ \bibinfo {pages} {143} (\bibinfo {year} {2020})},\ \Eprint {http://arxiv.org/abs/2006.07379} {arXiv:2006.07379 [hep-ph]} \BibitemShut {NoStop}%
\bibitem [{\citenamefont {Cohen}\ and\ \citenamefont {Kaplan}(1987)}]{Cohen:1987vi}%
  \BibitemOpen
  \bibfield  {author} {\bibinfo {author} {\bibfnamefont {A.~G.}\ \bibnamefont {Cohen}}\ and\ \bibinfo {author} {\bibfnamefont {D.~B.}\ \bibnamefont {Kaplan}},\ }\href {\doibase 10.1016/0370-2693(87)91369-4} {\bibfield  {journal} {\bibinfo  {journal} {Phys. Lett. B}\ }\textbf {\bibinfo {volume} {199}},\ \bibinfo {pages} {251} (\bibinfo {year} {1987})}\BibitemShut {NoStop}%
\bibitem [{\citenamefont {Cohen}\ and\ \citenamefont {Kaplan}(1988)}]{Cohen:1988kt}%
  \BibitemOpen
  \bibfield  {author} {\bibinfo {author} {\bibfnamefont {A.~G.}\ \bibnamefont {Cohen}}\ and\ \bibinfo {author} {\bibfnamefont {D.~B.}\ \bibnamefont {Kaplan}},\ }\href {\doibase 10.1016/0550-3213(88)90134-4} {\bibfield  {journal} {\bibinfo  {journal} {Nucl. Phys. B}\ }\textbf {\bibinfo {volume} {308}},\ \bibinfo {pages} {913} (\bibinfo {year} {1988})}\BibitemShut {NoStop}%
\bibitem [{\citenamefont {Salvio}(2021)}]{Salvio:2021lka}%
  \BibitemOpen
  \bibfield  {author} {\bibinfo {author} {\bibfnamefont {A.}~\bibnamefont {Salvio}},\ }\href {\doibase 10.1088/1475-7516/2021/10/011} {\bibfield  {journal} {\bibinfo  {journal} {JCAP}\ }\textbf {\bibinfo {volume} {10}},\ \bibinfo {pages} {011} (\bibinfo {year} {2021})},\ \Eprint {http://arxiv.org/abs/2107.03389} {arXiv:2107.03389 [hep-ph]} \BibitemShut {NoStop}%
\bibitem [{\citenamefont {Ghoshal}\ \emph {et~al.}(2023)\citenamefont {Ghoshal}, \citenamefont {Khlopov}, \citenamefont {Lalak},\ and\ \citenamefont {Porey}}]{Ghoshal:2023jvf}%
  \BibitemOpen
  \bibfield  {author} {\bibinfo {author} {\bibfnamefont {A.}~\bibnamefont {Ghoshal}}, \bibinfo {author} {\bibfnamefont {M.~Y.}\ \bibnamefont {Khlopov}}, \bibinfo {author} {\bibfnamefont {Z.}~\bibnamefont {Lalak}}, \ and\ \bibinfo {author} {\bibfnamefont {S.}~\bibnamefont {Porey}},\ }\href@noop {} {\  (\bibinfo {year} {2023})},\ \Eprint {http://arxiv.org/abs/2306.08675} {arXiv:2306.08675 [hep-ph]} \BibitemShut {NoStop}%
\bibitem [{\citenamefont {Takahashi}\ and\ \citenamefont {Yamada}(2015)}]{Takahashi:2015waa}%
  \BibitemOpen
  \bibfield  {author} {\bibinfo {author} {\bibfnamefont {F.}~\bibnamefont {Takahashi}}\ and\ \bibinfo {author} {\bibfnamefont {M.}~\bibnamefont {Yamada}},\ }\href {\doibase 10.1088/1475-7516/2015/10/010} {\bibfield  {journal} {\bibinfo  {journal} {JCAP}\ }\textbf {\bibinfo {volume} {10}},\ \bibinfo {pages} {010} (\bibinfo {year} {2015})},\ \Eprint {http://arxiv.org/abs/1507.06387} {arXiv:1507.06387 [hep-ph]} \BibitemShut {NoStop}%
\bibitem [{\citenamefont {Berbig}(2024{\natexlab{b}})}]{Berbig:2024ufe}%
  \BibitemOpen
  \bibfield  {author} {\bibinfo {author} {\bibfnamefont {M.}~\bibnamefont {Berbig}},\ }\href {\doibase 10.1103/PhysRevD.110.095008} {\bibfield  {journal} {\bibinfo  {journal} {Phys. Rev. D}\ }\textbf {\bibinfo {volume} {110}},\ \bibinfo {pages} {095008} (\bibinfo {year} {2024}{\natexlab{b}})},\ \Eprint {http://arxiv.org/abs/2404.06441} {arXiv:2404.06441 [hep-ph]} \BibitemShut {NoStop}%
\bibitem [{\citenamefont {Bettoni}\ and\ \citenamefont {Rubio}(2018)}]{Bettoni:2018utf}%
  \BibitemOpen
  \bibfield  {author} {\bibinfo {author} {\bibfnamefont {D.}~\bibnamefont {Bettoni}}\ and\ \bibinfo {author} {\bibfnamefont {J.}~\bibnamefont {Rubio}},\ }\href {\doibase 10.1016/j.physletb.2018.07.046} {\bibfield  {journal} {\bibinfo  {journal} {Phys. Lett. B}\ }\textbf {\bibinfo {volume} {784}},\ \bibinfo {pages} {122} (\bibinfo {year} {2018})},\ \Eprint {http://arxiv.org/abs/1805.02669} {arXiv:1805.02669 [astro-ph.CO]} \BibitemShut {NoStop}%
\bibitem [{\citenamefont {Bettoni}\ \emph {et~al.}(2024)\citenamefont {Bettoni}, \citenamefont {Laverda}, \citenamefont {Eiguren},\ and\ \citenamefont {Rubio}}]{Bettoni:2024ixe}%
  \BibitemOpen
  \bibfield  {author} {\bibinfo {author} {\bibfnamefont {D.}~\bibnamefont {Bettoni}}, \bibinfo {author} {\bibfnamefont {G.}~\bibnamefont {Laverda}}, \bibinfo {author} {\bibfnamefont {A.~L.}\ \bibnamefont {Eiguren}}, \ and\ \bibinfo {author} {\bibfnamefont {J.}~\bibnamefont {Rubio}},\ }\href@noop {} {\  (\bibinfo {year} {2024})},\ \Eprint {http://arxiv.org/abs/2409.15450} {arXiv:2409.15450 [gr-qc]} \BibitemShut {NoStop}%
\bibitem [{\citenamefont {Felder}\ \emph {et~al.}(1999{\natexlab{b}})\citenamefont {Felder}, \citenamefont {Kofman},\ and\ \citenamefont {Linde}}]{Felder:1998vq}%
  \BibitemOpen
  \bibfield  {author} {\bibinfo {author} {\bibfnamefont {G.~N.}\ \bibnamefont {Felder}}, \bibinfo {author} {\bibfnamefont {L.}~\bibnamefont {Kofman}}, \ and\ \bibinfo {author} {\bibfnamefont {A.~D.}\ \bibnamefont {Linde}},\ }\href {\doibase 10.1103/PhysRevD.59.123523} {\bibfield  {journal} {\bibinfo  {journal} {Phys. Rev. D}\ }\textbf {\bibinfo {volume} {59}},\ \bibinfo {pages} {123523} (\bibinfo {year} {1999}{\natexlab{b}})},\ \Eprint {http://arxiv.org/abs/hep-ph/9812289} {arXiv:hep-ph/9812289} \BibitemShut {NoStop}%
\bibitem [{\citenamefont {Dimopoulos}\ \emph {et~al.}(2018)\citenamefont {Dimopoulos}, \citenamefont {Donaldson~Wood},\ and\ \citenamefont {Owen}}]{Dimopoulos:2017tud}%
  \BibitemOpen
  \bibfield  {author} {\bibinfo {author} {\bibfnamefont {K.}~\bibnamefont {Dimopoulos}}, \bibinfo {author} {\bibfnamefont {L.}~\bibnamefont {Donaldson~Wood}}, \ and\ \bibinfo {author} {\bibfnamefont {C.}~\bibnamefont {Owen}},\ }\href {\doibase 10.1103/PhysRevD.97.063525} {\bibfield  {journal} {\bibinfo  {journal} {Phys. Rev. D}\ }\textbf {\bibinfo {volume} {97}},\ \bibinfo {pages} {063525} (\bibinfo {year} {2018})},\ \Eprint {http://arxiv.org/abs/1712.01760} {arXiv:1712.01760 [astro-ph.CO]} \BibitemShut {NoStop}%
\bibitem [{\citenamefont {Feng}\ and\ \citenamefont {Li}(2003)}]{Feng:2002nb}%
  \BibitemOpen
  \bibfield  {author} {\bibinfo {author} {\bibfnamefont {B.}~\bibnamefont {Feng}}\ and\ \bibinfo {author} {\bibfnamefont {M.-z.}\ \bibnamefont {Li}},\ }\href {\doibase 10.1016/S0370-2693(03)00589-6} {\bibfield  {journal} {\bibinfo  {journal} {Phys. Lett. B}\ }\textbf {\bibinfo {volume} {564}},\ \bibinfo {pages} {169} (\bibinfo {year} {2003})},\ \Eprint {http://arxiv.org/abs/hep-ph/0212213} {arXiv:hep-ph/0212213} \BibitemShut {NoStop}%
\bibitem [{\citenamefont {Bueno~Sanchez}\ and\ \citenamefont {Dimopoulos}(2007)}]{BuenoSanchez:2007jxm}%
  \BibitemOpen
  \bibfield  {author} {\bibinfo {author} {\bibfnamefont {J.~C.}\ \bibnamefont {Bueno~Sanchez}}\ and\ \bibinfo {author} {\bibfnamefont {K.}~\bibnamefont {Dimopoulos}},\ }\href {\doibase 10.1088/1475-7516/2007/11/007} {\bibfield  {journal} {\bibinfo  {journal} {JCAP}\ }\textbf {\bibinfo {volume} {11}},\ \bibinfo {pages} {007} (\bibinfo {year} {2007})},\ \Eprint {http://arxiv.org/abs/0707.3967} {arXiv:0707.3967 [hep-ph]} \BibitemShut {NoStop}%
\bibitem [{\citenamefont {Dalianis}\ and\ \citenamefont {Kodaxis}(2022)}]{Dalianis:2021dbs}%
  \BibitemOpen
  \bibfield  {author} {\bibinfo {author} {\bibfnamefont {I.}~\bibnamefont {Dalianis}}\ and\ \bibinfo {author} {\bibfnamefont {G.~P.}\ \bibnamefont {Kodaxis}},\ }\href {\doibase 10.3390/galaxies10010031} {\bibfield  {journal} {\bibinfo  {journal} {Galaxies}\ }\textbf {\bibinfo {volume} {10}},\ \bibinfo {pages} {31} (\bibinfo {year} {2022})},\ \Eprint {http://arxiv.org/abs/2112.15576} {arXiv:2112.15576 [astro-ph.CO]} \BibitemShut {NoStop}%
\bibitem [{\citenamefont {Felder}\ \emph {et~al.}(2001)\citenamefont {Felder}, \citenamefont {Kofman},\ and\ \citenamefont {Linde}}]{Felder:2001kt}%
  \BibitemOpen
  \bibfield  {author} {\bibinfo {author} {\bibfnamefont {G.~N.}\ \bibnamefont {Felder}}, \bibinfo {author} {\bibfnamefont {L.}~\bibnamefont {Kofman}}, \ and\ \bibinfo {author} {\bibfnamefont {A.~D.}\ \bibnamefont {Linde}},\ }\href {\doibase 10.1103/PhysRevD.64.123517} {\bibfield  {journal} {\bibinfo  {journal} {Phys. Rev. D}\ }\textbf {\bibinfo {volume} {64}},\ \bibinfo {pages} {123517} (\bibinfo {year} {2001})},\ \Eprint {http://arxiv.org/abs/hep-th/0106179} {arXiv:hep-th/0106179} \BibitemShut {NoStop}%
\bibitem [{\citenamefont {Harigaya}(2019)}]{Harigaya:2019emn}%
  \BibitemOpen
  \bibfield  {author} {\bibinfo {author} {\bibfnamefont {K.}~\bibnamefont {Harigaya}},\ }\href {\doibase 10.1007/JHEP08(2019)085} {\bibfield  {journal} {\bibinfo  {journal} {JHEP}\ }\textbf {\bibinfo {volume} {08}},\ \bibinfo {pages} {085} (\bibinfo {year} {2019})},\ \Eprint {http://arxiv.org/abs/1906.05286} {arXiv:1906.05286 [hep-ph]} \BibitemShut {NoStop}%
\bibitem [{\citenamefont {Barrie}\ \emph {et~al.}(2022{\natexlab{a}})\citenamefont {Barrie}, \citenamefont {Han},\ and\ \citenamefont {Murayama}}]{Barrie:2021mwi}%
  \BibitemOpen
  \bibfield  {author} {\bibinfo {author} {\bibfnamefont {N.~D.}\ \bibnamefont {Barrie}}, \bibinfo {author} {\bibfnamefont {C.}~\bibnamefont {Han}}, \ and\ \bibinfo {author} {\bibfnamefont {H.}~\bibnamefont {Murayama}},\ }\href {\doibase 10.1103/PhysRevLett.128.141801} {\bibfield  {journal} {\bibinfo  {journal} {Phys. Rev. Lett.}\ }\textbf {\bibinfo {volume} {128}},\ \bibinfo {pages} {141801} (\bibinfo {year} {2022}{\natexlab{a}})},\ \Eprint {http://arxiv.org/abs/2106.03381} {arXiv:2106.03381 [hep-ph]} \BibitemShut {NoStop}%
\bibitem [{\citenamefont {Barrie}\ \emph {et~al.}(2022{\natexlab{b}})\citenamefont {Barrie}, \citenamefont {Han},\ and\ \citenamefont {Murayama}}]{Barrie:2022cub}%
  \BibitemOpen
  \bibfield  {author} {\bibinfo {author} {\bibfnamefont {N.~D.}\ \bibnamefont {Barrie}}, \bibinfo {author} {\bibfnamefont {C.}~\bibnamefont {Han}}, \ and\ \bibinfo {author} {\bibfnamefont {H.}~\bibnamefont {Murayama}},\ }\href {\doibase 10.1007/JHEP05(2022)160} {\bibfield  {journal} {\bibinfo  {journal} {JHEP}\ }\textbf {\bibinfo {volume} {05}},\ \bibinfo {pages} {160} (\bibinfo {year} {2022}{\natexlab{b}})},\ \Eprint {http://arxiv.org/abs/2204.08202} {arXiv:2204.08202 [hep-ph]} \BibitemShut {NoStop}%
\bibitem [{\citenamefont {Barrie}\ and\ \citenamefont {Han}(2024)}]{Barrie:2024yhj}%
  \BibitemOpen
  \bibfield  {author} {\bibinfo {author} {\bibfnamefont {N.~D.}\ \bibnamefont {Barrie}}\ and\ \bibinfo {author} {\bibfnamefont {C.}~\bibnamefont {Han}},\ }\href@noop {} {\  (\bibinfo {year} {2024})},\ \Eprint {http://arxiv.org/abs/2402.15245} {arXiv:2402.15245 [hep-ph]} \BibitemShut {NoStop}%
\bibitem [{\citenamefont {Kajantie}\ \emph {et~al.}(2003)\citenamefont {Kajantie}, \citenamefont {Laine}, \citenamefont {Rummukainen},\ and\ \citenamefont {Schroder}}]{Kajantie:2002wa}%
  \BibitemOpen
  \bibfield  {author} {\bibinfo {author} {\bibfnamefont {K.}~\bibnamefont {Kajantie}}, \bibinfo {author} {\bibfnamefont {M.}~\bibnamefont {Laine}}, \bibinfo {author} {\bibfnamefont {K.}~\bibnamefont {Rummukainen}}, \ and\ \bibinfo {author} {\bibfnamefont {Y.}~\bibnamefont {Schroder}},\ }\href {\doibase 10.1103/PhysRevD.67.105008} {\bibfield  {journal} {\bibinfo  {journal} {Phys. Rev. D}\ }\textbf {\bibinfo {volume} {67}},\ \bibinfo {pages} {105008} (\bibinfo {year} {2003})},\ \Eprint {http://arxiv.org/abs/hep-ph/0211321} {arXiv:hep-ph/0211321} \BibitemShut {NoStop}%
\bibitem [{\citenamefont {Davoudiasl}\ \emph {et~al.}(2004)\citenamefont {Davoudiasl}, \citenamefont {Kitano}, \citenamefont {Kribs}, \citenamefont {Murayama},\ and\ \citenamefont {Steinhardt}}]{Davoudiasl:2004gf}%
  \BibitemOpen
  \bibfield  {author} {\bibinfo {author} {\bibfnamefont {H.}~\bibnamefont {Davoudiasl}}, \bibinfo {author} {\bibfnamefont {R.}~\bibnamefont {Kitano}}, \bibinfo {author} {\bibfnamefont {G.~D.}\ \bibnamefont {Kribs}}, \bibinfo {author} {\bibfnamefont {H.}~\bibnamefont {Murayama}}, \ and\ \bibinfo {author} {\bibfnamefont {P.~J.}\ \bibnamefont {Steinhardt}},\ }\href {\doibase 10.1103/PhysRevLett.93.201301} {\bibfield  {journal} {\bibinfo  {journal} {Phys. Rev. Lett.}\ }\textbf {\bibinfo {volume} {93}},\ \bibinfo {pages} {201301} (\bibinfo {year} {2004})},\ \Eprint {http://arxiv.org/abs/hep-ph/0403019} {arXiv:hep-ph/0403019} \BibitemShut {NoStop}%
\bibitem [{\citenamefont {Saikawa}\ and\ \citenamefont {Shirai}(2018)}]{Saikawa:2018rcs}%
  \BibitemOpen
  \bibfield  {author} {\bibinfo {author} {\bibfnamefont {K.}~\bibnamefont {Saikawa}}\ and\ \bibinfo {author} {\bibfnamefont {S.}~\bibnamefont {Shirai}},\ }\href {\doibase 10.1088/1475-7516/2018/05/035} {\bibfield  {journal} {\bibinfo  {journal} {JCAP}\ }\textbf {\bibinfo {volume} {05}},\ \bibinfo {pages} {035} (\bibinfo {year} {2018})},\ \Eprint {http://arxiv.org/abs/1803.01038} {arXiv:1803.01038 [hep-ph]} \BibitemShut {NoStop}%
\bibitem [{\citenamefont {Huston}\ \emph {et~al.}(2023)\citenamefont {Huston}, \citenamefont {Rabbertz},\ and\ \citenamefont {Zanderighi}}]{Huston:2023ofk}%
  \BibitemOpen
  \bibfield  {author} {\bibinfo {author} {\bibfnamefont {J.}~\bibnamefont {Huston}}, \bibinfo {author} {\bibfnamefont {K.}~\bibnamefont {Rabbertz}}, \ and\ \bibinfo {author} {\bibfnamefont {G.}~\bibnamefont {Zanderighi}},\ }\href@noop {} {\  (\bibinfo {year} {2023})},\ \Eprint {http://arxiv.org/abs/2312.14015} {arXiv:2312.14015 [hep-ph]} \BibitemShut {NoStop}%
\bibitem [{\citenamefont {Deur}\ \emph {et~al.}(2022)\citenamefont {Deur}, \citenamefont {Burkert}, \citenamefont {Chen},\ and\ \citenamefont {Korsch}}]{Deur:2022msf}%
  \BibitemOpen
  \bibfield  {author} {\bibinfo {author} {\bibfnamefont {A.}~\bibnamefont {Deur}}, \bibinfo {author} {\bibfnamefont {V.}~\bibnamefont {Burkert}}, \bibinfo {author} {\bibfnamefont {J.~P.}\ \bibnamefont {Chen}}, \ and\ \bibinfo {author} {\bibfnamefont {W.}~\bibnamefont {Korsch}},\ }\href {\doibase 10.3390/particles5020015} {\bibfield  {journal} {\bibinfo  {journal} {Particles}\ }\textbf {\bibinfo {volume} {5}},\ \bibinfo {pages} {171} (\bibinfo {year} {2022})},\ \Eprint {http://arxiv.org/abs/2205.01169} {arXiv:2205.01169 [hep-ph]} \BibitemShut {NoStop}%
\bibitem [{\citenamefont {Giovannini}(1998)}]{Giovannini:1998bp}%
  \BibitemOpen
  \bibfield  {author} {\bibinfo {author} {\bibfnamefont {M.}~\bibnamefont {Giovannini}},\ }\href {\doibase 10.1103/PhysRevD.58.083504} {\bibfield  {journal} {\bibinfo  {journal} {Phys. Rev. D}\ }\textbf {\bibinfo {volume} {58}},\ \bibinfo {pages} {083504} (\bibinfo {year} {1998})},\ \Eprint {http://arxiv.org/abs/hep-ph/9806329} {arXiv:hep-ph/9806329} \BibitemShut {NoStop}%
\bibitem [{\citenamefont {Haque}\ \emph {et~al.}(2021)\citenamefont {Haque}, \citenamefont {Maity}, \citenamefont {Paul},\ and\ \citenamefont {Sriramkumar}}]{Haque:2021dha}%
  \BibitemOpen
  \bibfield  {author} {\bibinfo {author} {\bibfnamefont {M.~R.}\ \bibnamefont {Haque}}, \bibinfo {author} {\bibfnamefont {D.}~\bibnamefont {Maity}}, \bibinfo {author} {\bibfnamefont {T.}~\bibnamefont {Paul}}, \ and\ \bibinfo {author} {\bibfnamefont {L.}~\bibnamefont {Sriramkumar}},\ }\href {\doibase 10.1103/PhysRevD.104.063513} {\bibfield  {journal} {\bibinfo  {journal} {Phys. Rev. D}\ }\textbf {\bibinfo {volume} {104}},\ \bibinfo {pages} {063513} (\bibinfo {year} {2021})},\ \Eprint {http://arxiv.org/abs/2105.09242} {arXiv:2105.09242 [astro-ph.CO]} \BibitemShut {NoStop}%
\bibitem [{\citenamefont {Figueroa}\ and\ \citenamefont {Tanin}(2019)}]{Figueroa:2019paj}%
  \BibitemOpen
  \bibfield  {author} {\bibinfo {author} {\bibfnamefont {D.~G.}\ \bibnamefont {Figueroa}}\ and\ \bibinfo {author} {\bibfnamefont {E.~H.}\ \bibnamefont {Tanin}},\ }\href {\doibase 10.1088/1475-7516/2019/08/011} {\bibfield  {journal} {\bibinfo  {journal} {JCAP}\ }\textbf {\bibinfo {volume} {08}},\ \bibinfo {pages} {011} (\bibinfo {year} {2019})},\ \Eprint {http://arxiv.org/abs/1905.11960} {arXiv:1905.11960 [astro-ph.CO]} \BibitemShut {NoStop}%
\bibitem [{\citenamefont {Chen}\ \emph {et~al.}(2024)\citenamefont {Chen}, \citenamefont {Dimopoulos}, \citenamefont {Er\"oncel},\ and\ \citenamefont {Ghoshal}}]{Chen:2024roo}%
  \BibitemOpen
  \bibfield  {author} {\bibinfo {author} {\bibfnamefont {C.}~\bibnamefont {Chen}}, \bibinfo {author} {\bibfnamefont {K.}~\bibnamefont {Dimopoulos}}, \bibinfo {author} {\bibfnamefont {C.}~\bibnamefont {Er\"oncel}}, \ and\ \bibinfo {author} {\bibfnamefont {A.}~\bibnamefont {Ghoshal}},\ }\href {\doibase 10.1103/PhysRevD.110.063554} {\bibfield  {journal} {\bibinfo  {journal} {Phys. Rev. D}\ }\textbf {\bibinfo {volume} {110}},\ \bibinfo {pages} {063554} (\bibinfo {year} {2024})},\ \Eprint {http://arxiv.org/abs/2405.01679} {arXiv:2405.01679 [hep-ph]} \BibitemShut {NoStop}%
\bibitem [{\citenamefont {Maggiore}(2000)}]{Maggiore:1999vm}%
  \BibitemOpen
  \bibfield  {author} {\bibinfo {author} {\bibfnamefont {M.}~\bibnamefont {Maggiore}},\ }\href {\doibase 10.1016/S0370-1573(99)00102-7} {\bibfield  {journal} {\bibinfo  {journal} {Phys. Rept.}\ }\textbf {\bibinfo {volume} {331}},\ \bibinfo {pages} {283} (\bibinfo {year} {2000})},\ \Eprint {http://arxiv.org/abs/gr-qc/9909001} {arXiv:gr-qc/9909001} \BibitemShut {NoStop}%
\bibitem [{\citenamefont {Boyle}\ and\ \citenamefont {Buonanno}(2008)}]{Boyle:2007zx}%
  \BibitemOpen
  \bibfield  {author} {\bibinfo {author} {\bibfnamefont {L.~A.}\ \bibnamefont {Boyle}}\ and\ \bibinfo {author} {\bibfnamefont {A.}~\bibnamefont {Buonanno}},\ }\href {\doibase 10.1103/PhysRevD.78.043531} {\bibfield  {journal} {\bibinfo  {journal} {Phys. Rev. D}\ }\textbf {\bibinfo {volume} {78}},\ \bibinfo {pages} {043531} (\bibinfo {year} {2008})},\ \Eprint {http://arxiv.org/abs/0708.2279} {arXiv:0708.2279 [astro-ph]} \BibitemShut {NoStop}%
\bibitem [{\citenamefont {Caprini}\ and\ \citenamefont {Figueroa}(2018)}]{Caprini:2018mtu}%
  \BibitemOpen
  \bibfield  {author} {\bibinfo {author} {\bibfnamefont {C.}~\bibnamefont {Caprini}}\ and\ \bibinfo {author} {\bibfnamefont {D.~G.}\ \bibnamefont {Figueroa}},\ }\href {\doibase 10.1088/1361-6382/aac608} {\bibfield  {journal} {\bibinfo  {journal} {Class. Quant. Grav.}\ }\textbf {\bibinfo {volume} {35}},\ \bibinfo {pages} {163001} (\bibinfo {year} {2018})},\ \Eprint {http://arxiv.org/abs/1801.04268} {arXiv:1801.04268 [astro-ph.CO]} \BibitemShut {NoStop}%
\bibitem [{\citenamefont {Yeh}\ \emph {et~al.}(2022)\citenamefont {Yeh}, \citenamefont {Shelton}, \citenamefont {Olive},\ and\ \citenamefont {Fields}}]{Yeh:2022heq}%
  \BibitemOpen
  \bibfield  {author} {\bibinfo {author} {\bibfnamefont {T.-H.}\ \bibnamefont {Yeh}}, \bibinfo {author} {\bibfnamefont {J.}~\bibnamefont {Shelton}}, \bibinfo {author} {\bibfnamefont {K.~A.}\ \bibnamefont {Olive}}, \ and\ \bibinfo {author} {\bibfnamefont {B.~D.}\ \bibnamefont {Fields}},\ }\href {\doibase 10.1088/1475-7516/2022/10/046} {\bibfield  {journal} {\bibinfo  {journal} {JCAP}\ }\textbf {\bibinfo {volume} {10}},\ \bibinfo {pages} {046} (\bibinfo {year} {2022})},\ \Eprint {http://arxiv.org/abs/2207.13133} {arXiv:2207.13133 [astro-ph.CO]} \BibitemShut {NoStop}%
\bibitem [{\citenamefont {Abazajian}\ \emph {et~al.}(2019)\citenamefont {Abazajian} \emph {et~al.}}]{Abazajian:2019eic}%
  \BibitemOpen
  \bibfield  {author} {\bibinfo {author} {\bibfnamefont {K.}~\bibnamefont {Abazajian}} \emph {et~al.},\ }\href@noop {} {\  (\bibinfo {year} {2019})},\ \Eprint {http://arxiv.org/abs/1907.04473} {arXiv:1907.04473 [astro-ph.IM]} \BibitemShut {NoStop}%
\bibitem [{\citenamefont {Hanany}\ \emph {et~al.}(2019)\citenamefont {Hanany} \emph {et~al.}}]{NASAPICO:2019thw}%
  \BibitemOpen
  \bibfield  {author} {\bibinfo {author} {\bibfnamefont {S.}~\bibnamefont {Hanany}} \emph {et~al.} (\bibinfo {collaboration} {NASA PICO}),\ }\href@noop {} {\  (\bibinfo {year} {2019})},\ \Eprint {http://arxiv.org/abs/1902.10541} {arXiv:1902.10541 [astro-ph.IM]} \BibitemShut {NoStop}%
\bibitem [{\citenamefont {Aiola}\ \emph {et~al.}(2022)\citenamefont {Aiola} \emph {et~al.}}]{CMB-HD:2022bsz}%
  \BibitemOpen
  \bibfield  {author} {\bibinfo {author} {\bibfnamefont {S.}~\bibnamefont {Aiola}} \emph {et~al.} (\bibinfo {collaboration} {CMB-HD}),\ }\href@noop {} {\  (\bibinfo {year} {2022})},\ \Eprint {http://arxiv.org/abs/2203.05728} {arXiv:2203.05728 [astro-ph.CO]} \BibitemShut {NoStop}%
\bibitem [{\citenamefont {Bouchet}\ \emph {et~al.}(2011)\citenamefont {Bouchet} \emph {et~al.}}]{COrE:2011bfs}%
  \BibitemOpen
  \bibfield  {author} {\bibinfo {author} {\bibfnamefont {F.~R.}\ \bibnamefont {Bouchet}} \emph {et~al.} (\bibinfo {collaboration} {COrE}),\ }\href@noop {} {\  (\bibinfo {year} {2011})},\ \Eprint {http://arxiv.org/abs/1102.2181} {arXiv:1102.2181 [astro-ph.CO]} \BibitemShut {NoStop}%
\bibitem [{\citenamefont {Laureijs}\ \emph {et~al.}(2011)\citenamefont {Laureijs} \emph {et~al.}}]{EUCLID:2011zbd}%
  \BibitemOpen
  \bibfield  {author} {\bibinfo {author} {\bibfnamefont {R.}~\bibnamefont {Laureijs}} \emph {et~al.} (\bibinfo {collaboration} {EUCLID}),\ }\href@noop {} {\  (\bibinfo {year} {2011})},\ \Eprint {http://arxiv.org/abs/1110.3193} {arXiv:1110.3193 [astro-ph.CO]} \BibitemShut {NoStop}%
\bibitem [{\citenamefont {Gelmini}\ and\ \citenamefont {Roncadelli}(1981)}]{Gelmini:1980re}%
  \BibitemOpen
  \bibfield  {author} {\bibinfo {author} {\bibfnamefont {G.~B.}\ \bibnamefont {Gelmini}}\ and\ \bibinfo {author} {\bibfnamefont {M.}~\bibnamefont {Roncadelli}},\ }\href {\doibase 10.1016/0370-2693(81)90559-1} {\bibfield  {journal} {\bibinfo  {journal} {Phys. Lett. B}\ }\textbf {\bibinfo {volume} {99}},\ \bibinfo {pages} {411} (\bibinfo {year} {1981})}\BibitemShut {NoStop}%
\bibitem [{\citenamefont {Buchmuller}\ \emph {et~al.}(2005)\citenamefont {Buchmuller}, \citenamefont {Di~Bari},\ and\ \citenamefont {Plumacher}}]{Buchmuller:2004nz}%
  \BibitemOpen
  \bibfield  {author} {\bibinfo {author} {\bibfnamefont {W.}~\bibnamefont {Buchmuller}}, \bibinfo {author} {\bibfnamefont {P.}~\bibnamefont {Di~Bari}}, \ and\ \bibinfo {author} {\bibfnamefont {M.}~\bibnamefont {Plumacher}},\ }\href {\doibase 10.1016/j.aop.2004.02.003} {\bibfield  {journal} {\bibinfo  {journal} {Annals Phys.}\ }\textbf {\bibinfo {volume} {315}},\ \bibinfo {pages} {305} (\bibinfo {year} {2005})},\ \Eprint {http://arxiv.org/abs/hep-ph/0401240} {arXiv:hep-ph/0401240} \BibitemShut {NoStop}%
\bibitem [{\citenamefont {Chen}\ \emph {et~al.}(2020)\citenamefont {Chen}, \citenamefont {Dutta~Banik},\ and\ \citenamefont {Liu}}]{Chen:2019etb}%
  \BibitemOpen
  \bibfield  {author} {\bibinfo {author} {\bibfnamefont {S.-L.}\ \bibnamefont {Chen}}, \bibinfo {author} {\bibfnamefont {A.}~\bibnamefont {Dutta~Banik}}, \ and\ \bibinfo {author} {\bibfnamefont {Z.-K.}\ \bibnamefont {Liu}},\ }\href {\doibase 10.1088/1475-7516/2020/03/009} {\bibfield  {journal} {\bibinfo  {journal} {JCAP}\ }\textbf {\bibinfo {volume} {03}},\ \bibinfo {pages} {009} (\bibinfo {year} {2020})},\ \Eprint {http://arxiv.org/abs/1912.07185} {arXiv:1912.07185 [hep-ph]} \BibitemShut {NoStop}%
\bibitem [{\citenamefont {Audren}\ \emph {et~al.}(2014)\citenamefont {Audren}, \citenamefont {Lesgourgues}, \citenamefont {Mangano}, \citenamefont {Serpico},\ and\ \citenamefont {Tram}}]{Audren:2014bca}%
  \BibitemOpen
  \bibfield  {author} {\bibinfo {author} {\bibfnamefont {B.}~\bibnamefont {Audren}}, \bibinfo {author} {\bibfnamefont {J.}~\bibnamefont {Lesgourgues}}, \bibinfo {author} {\bibfnamefont {G.}~\bibnamefont {Mangano}}, \bibinfo {author} {\bibfnamefont {P.~D.}\ \bibnamefont {Serpico}}, \ and\ \bibinfo {author} {\bibfnamefont {T.}~\bibnamefont {Tram}},\ }\href {\doibase 10.1088/1475-7516/2014/12/028} {\bibfield  {journal} {\bibinfo  {journal} {JCAP}\ }\textbf {\bibinfo {volume} {12}},\ \bibinfo {pages} {028} (\bibinfo {year} {2014})},\ \Eprint {http://arxiv.org/abs/1407.2418} {arXiv:1407.2418 [astro-ph.CO]} \BibitemShut {NoStop}%
\bibitem [{\citenamefont {Enqvist}\ \emph {et~al.}(2020)\citenamefont {Enqvist}, \citenamefont {Nadathur}, \citenamefont {Sekiguchi},\ and\ \citenamefont {Takahashi}}]{Enqvist:2019tsa}%
  \BibitemOpen
  \bibfield  {author} {\bibinfo {author} {\bibfnamefont {K.}~\bibnamefont {Enqvist}}, \bibinfo {author} {\bibfnamefont {S.}~\bibnamefont {Nadathur}}, \bibinfo {author} {\bibfnamefont {T.}~\bibnamefont {Sekiguchi}}, \ and\ \bibinfo {author} {\bibfnamefont {T.}~\bibnamefont {Takahashi}},\ }\href {\doibase 10.1088/1475-7516/2020/04/015} {\bibfield  {journal} {\bibinfo  {journal} {JCAP}\ }\textbf {\bibinfo {volume} {04}},\ \bibinfo {pages} {015} (\bibinfo {year} {2020})},\ \Eprint {http://arxiv.org/abs/1906.09112} {arXiv:1906.09112 [astro-ph.CO]} \BibitemShut {NoStop}%
\bibitem [{\citenamefont {Nygaard}\ \emph {et~al.}(2021)\citenamefont {Nygaard}, \citenamefont {Tram},\ and\ \citenamefont {Hannestad}}]{Nygaard:2020sow}%
  \BibitemOpen
  \bibfield  {author} {\bibinfo {author} {\bibfnamefont {A.}~\bibnamefont {Nygaard}}, \bibinfo {author} {\bibfnamefont {T.}~\bibnamefont {Tram}}, \ and\ \bibinfo {author} {\bibfnamefont {S.}~\bibnamefont {Hannestad}},\ }\href {\doibase 10.1088/1475-7516/2021/05/017} {\bibfield  {journal} {\bibinfo  {journal} {JCAP}\ }\textbf {\bibinfo {volume} {05}},\ \bibinfo {pages} {017} (\bibinfo {year} {2021})},\ \Eprint {http://arxiv.org/abs/2011.01632} {arXiv:2011.01632 [astro-ph.CO]} \BibitemShut {NoStop}%
\bibitem [{\citenamefont {Alvi}\ \emph {et~al.}(2022)\citenamefont {Alvi}, \citenamefont {Brinckmann}, \citenamefont {Gerbino}, \citenamefont {Lattanzi},\ and\ \citenamefont {Pagano}}]{Alvi:2022aam}%
  \BibitemOpen
  \bibfield  {author} {\bibinfo {author} {\bibfnamefont {S.}~\bibnamefont {Alvi}}, \bibinfo {author} {\bibfnamefont {T.}~\bibnamefont {Brinckmann}}, \bibinfo {author} {\bibfnamefont {M.}~\bibnamefont {Gerbino}}, \bibinfo {author} {\bibfnamefont {M.}~\bibnamefont {Lattanzi}}, \ and\ \bibinfo {author} {\bibfnamefont {L.}~\bibnamefont {Pagano}},\ }\href {\doibase 10.1088/1475-7516/2022/11/015} {\bibfield  {journal} {\bibinfo  {journal} {JCAP}\ }\textbf {\bibinfo {volume} {11}},\ \bibinfo {pages} {015} (\bibinfo {year} {2022})},\ \Eprint {http://arxiv.org/abs/2205.05636} {arXiv:2205.05636 [astro-ph.CO]} \BibitemShut {NoStop}%
\bibitem [{\citenamefont {Simon}\ \emph {et~al.}(2022)\citenamefont {Simon}, \citenamefont {Franco~Abell\'an}, \citenamefont {Du}, \citenamefont {Poulin},\ and\ \citenamefont {Tsai}}]{Simon:2022ftd}%
  \BibitemOpen
  \bibfield  {author} {\bibinfo {author} {\bibfnamefont {T.}~\bibnamefont {Simon}}, \bibinfo {author} {\bibfnamefont {G.}~\bibnamefont {Franco~Abell\'an}}, \bibinfo {author} {\bibfnamefont {P.}~\bibnamefont {Du}}, \bibinfo {author} {\bibfnamefont {V.}~\bibnamefont {Poulin}}, \ and\ \bibinfo {author} {\bibfnamefont {Y.}~\bibnamefont {Tsai}},\ }\href {\doibase 10.1103/PhysRevD.106.023516} {\bibfield  {journal} {\bibinfo  {journal} {Phys. Rev. D}\ }\textbf {\bibinfo {volume} {106}},\ \bibinfo {pages} {023516} (\bibinfo {year} {2022})},\ \Eprint {http://arxiv.org/abs/2203.07440} {arXiv:2203.07440 [astro-ph.CO]} \BibitemShut {NoStop}%
\bibitem [{\citenamefont {Crowder}\ and\ \citenamefont {Cornish}(2005)}]{Crowder:2005nr}%
  \BibitemOpen
  \bibfield  {author} {\bibinfo {author} {\bibfnamefont {J.}~\bibnamefont {Crowder}}\ and\ \bibinfo {author} {\bibfnamefont {N.~J.}\ \bibnamefont {Cornish}},\ }\href {\doibase 10.1103/PhysRevD.72.083005} {\bibfield  {journal} {\bibinfo  {journal} {Phys. Rev. D}\ }\textbf {\bibinfo {volume} {72}},\ \bibinfo {pages} {083005} (\bibinfo {year} {2005})},\ \Eprint {http://arxiv.org/abs/gr-qc/0506015} {arXiv:gr-qc/0506015} \BibitemShut {NoStop}%
\bibitem [{\citenamefont {Corbin}\ and\ \citenamefont {Cornish}(2006)}]{Corbin:2005ny}%
  \BibitemOpen
  \bibfield  {author} {\bibinfo {author} {\bibfnamefont {V.}~\bibnamefont {Corbin}}\ and\ \bibinfo {author} {\bibfnamefont {N.~J.}\ \bibnamefont {Cornish}},\ }\href {\doibase 10.1088/0264-9381/23/7/014} {\bibfield  {journal} {\bibinfo  {journal} {Class. Quant. Grav.}\ }\textbf {\bibinfo {volume} {23}},\ \bibinfo {pages} {2435} (\bibinfo {year} {2006})},\ \Eprint {http://arxiv.org/abs/gr-qc/0512039} {arXiv:gr-qc/0512039} \BibitemShut {NoStop}%
\bibitem [{\citenamefont {Sato}\ \emph {et~al.}(2017)\citenamefont {Sato} \emph {et~al.}}]{Sato:2017dkf}%
  \BibitemOpen
  \bibfield  {author} {\bibinfo {author} {\bibfnamefont {S.}~\bibnamefont {Sato}} \emph {et~al.},\ }\href {\doibase 10.1088/1742-6596/840/1/012010} {\bibfield  {journal} {\bibinfo  {journal} {J. Phys. Conf. Ser.}\ }\textbf {\bibinfo {volume} {840}},\ \bibinfo {pages} {012010} (\bibinfo {year} {2017})}\BibitemShut {NoStop}%
\bibitem [{\citenamefont {Ishikawa}\ \emph {et~al.}(2021)\citenamefont {Ishikawa} \emph {et~al.}}]{Ishikawa:2020hlo}%
  \BibitemOpen
  \bibfield  {author} {\bibinfo {author} {\bibfnamefont {T.}~\bibnamefont {Ishikawa}} \emph {et~al.},\ }\href {\doibase 10.3390/galaxies9010014} {\bibfield  {journal} {\bibinfo  {journal} {Galaxies}\ }\textbf {\bibinfo {volume} {9}},\ \bibinfo {pages} {14} (\bibinfo {year} {2021})},\ \Eprint {http://arxiv.org/abs/2012.11859} {arXiv:2012.11859 [gr-qc]} \BibitemShut {NoStop}%
\bibitem [{\citenamefont {Ringwald}\ \emph {et~al.}(2021)\citenamefont {Ringwald}, \citenamefont {Sch\"utte-Engel},\ and\ \citenamefont {Tamarit}}]{Ringwald:2020ist}%
  \BibitemOpen
  \bibfield  {author} {\bibinfo {author} {\bibfnamefont {A.}~\bibnamefont {Ringwald}}, \bibinfo {author} {\bibfnamefont {J.}~\bibnamefont {Sch\"utte-Engel}}, \ and\ \bibinfo {author} {\bibfnamefont {C.}~\bibnamefont {Tamarit}},\ }\href {\doibase 10.1088/1475-7516/2021/03/054} {\bibfield  {journal} {\bibinfo  {journal} {JCAP}\ }\textbf {\bibinfo {volume} {03}},\ \bibinfo {pages} {054} (\bibinfo {year} {2021})},\ \Eprint {http://arxiv.org/abs/2011.04731} {arXiv:2011.04731 [hep-ph]} \BibitemShut {NoStop}%
\bibitem [{\citenamefont {Ringwald}\ and\ \citenamefont {Tamarit}(2022)}]{Ringwald:2022xif}%
  \BibitemOpen
  \bibfield  {author} {\bibinfo {author} {\bibfnamefont {A.}~\bibnamefont {Ringwald}}\ and\ \bibinfo {author} {\bibfnamefont {C.}~\bibnamefont {Tamarit}},\ }\href {\doibase 10.1103/PhysRevD.106.063027} {\bibfield  {journal} {\bibinfo  {journal} {Phys. Rev. D}\ }\textbf {\bibinfo {volume} {106}},\ \bibinfo {pages} {063027} (\bibinfo {year} {2022})},\ \Eprint {http://arxiv.org/abs/2203.00621} {arXiv:2203.00621 [hep-ph]} \BibitemShut {NoStop}%
\bibitem [{\citenamefont {Aggarwal}\ \emph {et~al.}(2021)\citenamefont {Aggarwal} \emph {et~al.}}]{Aggarwal:2020olq}%
  \BibitemOpen
  \bibfield  {author} {\bibinfo {author} {\bibfnamefont {N.}~\bibnamefont {Aggarwal}} \emph {et~al.},\ }\href {\doibase 10.1007/s41114-021-00032-5} {\bibfield  {journal} {\bibinfo  {journal} {Living Rev. Rel.}\ }\textbf {\bibinfo {volume} {24}},\ \bibinfo {pages} {4} (\bibinfo {year} {2021})},\ \Eprint {http://arxiv.org/abs/2011.12414} {arXiv:2011.12414 [gr-qc]} \BibitemShut {NoStop}%
\bibitem [{\citenamefont {Aker}\ \emph {et~al.}(2022)\citenamefont {Aker} \emph {et~al.}}]{KATRIN:2021uub}%
  \BibitemOpen
  \bibfield  {author} {\bibinfo {author} {\bibfnamefont {M.}~\bibnamefont {Aker}} \emph {et~al.} (\bibinfo {collaboration} {KATRIN}),\ }\href {\doibase 10.1038/s41567-021-01463-1} {\bibfield  {journal} {\bibinfo  {journal} {Nature Phys.}\ }\textbf {\bibinfo {volume} {18}},\ \bibinfo {pages} {160} (\bibinfo {year} {2022})},\ \Eprint {http://arxiv.org/abs/2105.08533} {arXiv:2105.08533 [hep-ex]} \BibitemShut {NoStop}%
\bibitem [{\citenamefont {Fonseca}\ \emph {et~al.}(2020)\citenamefont {Fonseca}, \citenamefont {Morgante}, \citenamefont {Sato},\ and\ \citenamefont {Servant}}]{Fonseca:2019ypl}%
  \BibitemOpen
  \bibfield  {author} {\bibinfo {author} {\bibfnamefont {N.}~\bibnamefont {Fonseca}}, \bibinfo {author} {\bibfnamefont {E.}~\bibnamefont {Morgante}}, \bibinfo {author} {\bibfnamefont {R.}~\bibnamefont {Sato}}, \ and\ \bibinfo {author} {\bibfnamefont {G.}~\bibnamefont {Servant}},\ }\href {\doibase 10.1007/JHEP04(2020)010} {\bibfield  {journal} {\bibinfo  {journal} {JHEP}\ }\textbf {\bibinfo {volume} {04}},\ \bibinfo {pages} {010} (\bibinfo {year} {2020})},\ \Eprint {http://arxiv.org/abs/1911.08472} {arXiv:1911.08472 [hep-ph]} \BibitemShut {NoStop}%
\bibitem [{\citenamefont {Er\"oncel}\ \emph {et~al.}(2022)\citenamefont {Er\"oncel}, \citenamefont {Sato}, \citenamefont {Servant},\ and\ \citenamefont {S\o{}rensen}}]{Eroncel:2022vjg}%
  \BibitemOpen
  \bibfield  {author} {\bibinfo {author} {\bibfnamefont {C.}~\bibnamefont {Er\"oncel}}, \bibinfo {author} {\bibfnamefont {R.}~\bibnamefont {Sato}}, \bibinfo {author} {\bibfnamefont {G.}~\bibnamefont {Servant}}, \ and\ \bibinfo {author} {\bibfnamefont {P.}~\bibnamefont {S\o{}rensen}},\ }\href {\doibase 10.1088/1475-7516/2022/10/053} {\bibfield  {journal} {\bibinfo  {journal} {JCAP}\ }\textbf {\bibinfo {volume} {10}},\ \bibinfo {pages} {053} (\bibinfo {year} {2022})},\ \Eprint {http://arxiv.org/abs/2206.14259} {arXiv:2206.14259 [hep-ph]} \BibitemShut {NoStop}%
\bibitem [{\citenamefont {Er\"oncel}\ and\ \citenamefont {Servant}(2023)}]{Eroncel:2022efc}%
  \BibitemOpen
  \bibfield  {author} {\bibinfo {author} {\bibfnamefont {C.}~\bibnamefont {Er\"oncel}}\ and\ \bibinfo {author} {\bibfnamefont {G.}~\bibnamefont {Servant}},\ }\href {\doibase 10.1088/1475-7516/2023/01/009} {\bibfield  {journal} {\bibinfo  {journal} {JCAP}\ }\textbf {\bibinfo {volume} {01}},\ \bibinfo {pages} {009} (\bibinfo {year} {2023})},\ \Eprint {http://arxiv.org/abs/2207.10111} {arXiv:2207.10111 [hep-ph]} \BibitemShut {NoStop}%
\bibitem [{\citenamefont {Mukhanov}(2005)}]{Mukhanov:2005sc}%
  \BibitemOpen
  \bibfield  {author} {\bibinfo {author} {\bibfnamefont {V.}~\bibnamefont {Mukhanov}},\ }\href {\doibase 10.1017/CBO9780511790553} {\emph {\bibinfo {title} {{Physical Foundations of Cosmology}}}}\ (\bibinfo  {publisher} {Cambridge University Press},\ \bibinfo {address} {Oxford},\ \bibinfo {year} {2005})\BibitemShut {NoStop}%
\bibitem [{\citenamefont {Bunch}\ and\ \citenamefont {Davies}(1978)}]{Bunch:1978yq}%
  \BibitemOpen
  \bibfield  {author} {\bibinfo {author} {\bibfnamefont {T.~S.}\ \bibnamefont {Bunch}}\ and\ \bibinfo {author} {\bibfnamefont {P.~C.~W.}\ \bibnamefont {Davies}},\ }\href {\doibase 10.1098/rspa.1978.0060} {\bibfield  {journal} {\bibinfo  {journal} {Proc. Roy. Soc. Lond. A}\ }\textbf {\bibinfo {volume} {360}},\ \bibinfo {pages} {117} (\bibinfo {year} {1978})}\BibitemShut {NoStop}%
\bibitem [{\citenamefont {Mijic}(1994)}]{Mijic:1994vv}%
  \BibitemOpen
  \bibfield  {author} {\bibinfo {author} {\bibfnamefont {M.}~\bibnamefont {Mijic}},\ }\href {\doibase 10.1103/PhysRevD.49.6434} {\bibfield  {journal} {\bibinfo  {journal} {Phys. Rev. D}\ }\textbf {\bibinfo {volume} {49}},\ \bibinfo {pages} {6434} (\bibinfo {year} {1994})},\ \Eprint {http://arxiv.org/abs/gr-qc/9401030} {arXiv:gr-qc/9401030} \BibitemShut {NoStop}%
\bibitem [{\citenamefont {Lyth}\ \emph {et~al.}(2003)\citenamefont {Lyth}, \citenamefont {Ungarelli},\ and\ \citenamefont {Wands}}]{Lyth:2002my}%
  \BibitemOpen
  \bibfield  {author} {\bibinfo {author} {\bibfnamefont {D.~H.}\ \bibnamefont {Lyth}}, \bibinfo {author} {\bibfnamefont {C.}~\bibnamefont {Ungarelli}}, \ and\ \bibinfo {author} {\bibfnamefont {D.}~\bibnamefont {Wands}},\ }\href {\doibase 10.1103/PhysRevD.67.023503} {\bibfield  {journal} {\bibinfo  {journal} {Phys. Rev. D}\ }\textbf {\bibinfo {volume} {67}},\ \bibinfo {pages} {023503} (\bibinfo {year} {2003})},\ \Eprint {http://arxiv.org/abs/astro-ph/0208055} {arXiv:astro-ph/0208055} \BibitemShut {NoStop}%
\bibitem [{\citenamefont {Lyth}\ and\ \citenamefont {Wands}(2003)}]{Lyth:2003ip}%
  \BibitemOpen
  \bibfield  {author} {\bibinfo {author} {\bibfnamefont {D.~H.}\ \bibnamefont {Lyth}}\ and\ \bibinfo {author} {\bibfnamefont {D.}~\bibnamefont {Wands}},\ }\href {\doibase 10.1103/PhysRevD.68.103516} {\bibfield  {journal} {\bibinfo  {journal} {Phys. Rev. D}\ }\textbf {\bibinfo {volume} {68}},\ \bibinfo {pages} {103516} (\bibinfo {year} {2003})},\ \Eprint {http://arxiv.org/abs/astro-ph/0306500} {arXiv:astro-ph/0306500} \BibitemShut {NoStop}%
\bibitem [{\citenamefont {Bezrukov}\ and\ \citenamefont {Shaposhnikov}(2014)}]{Bezrukov:2014bra}%
  \BibitemOpen
  \bibfield  {author} {\bibinfo {author} {\bibfnamefont {F.}~\bibnamefont {Bezrukov}}\ and\ \bibinfo {author} {\bibfnamefont {M.}~\bibnamefont {Shaposhnikov}},\ }\href {\doibase 10.1016/j.physletb.2014.05.074} {\bibfield  {journal} {\bibinfo  {journal} {Phys. Lett. B}\ }\textbf {\bibinfo {volume} {734}},\ \bibinfo {pages} {249} (\bibinfo {year} {2014})},\ \Eprint {http://arxiv.org/abs/1403.6078} {arXiv:1403.6078 [hep-ph]} \BibitemShut {NoStop}%
\bibitem [{\citenamefont {Hamada}\ \emph {et~al.}(2015)\citenamefont {Hamada}, \citenamefont {Kawai}, \citenamefont {Oda},\ and\ \citenamefont {Park}}]{Hamada:2014wna}%
  \BibitemOpen
  \bibfield  {author} {\bibinfo {author} {\bibfnamefont {Y.}~\bibnamefont {Hamada}}, \bibinfo {author} {\bibfnamefont {H.}~\bibnamefont {Kawai}}, \bibinfo {author} {\bibfnamefont {K.-y.}\ \bibnamefont {Oda}}, \ and\ \bibinfo {author} {\bibfnamefont {S.~C.}\ \bibnamefont {Park}},\ }\href {\doibase 10.1103/PhysRevD.91.053008} {\bibfield  {journal} {\bibinfo  {journal} {Phys. Rev. D}\ }\textbf {\bibinfo {volume} {91}},\ \bibinfo {pages} {053008} (\bibinfo {year} {2015})},\ \Eprint {http://arxiv.org/abs/1408.4864} {arXiv:1408.4864 [hep-ph]} \BibitemShut {NoStop}%
\bibitem [{\citenamefont {Ezquiaga}\ \emph {et~al.}(2018)\citenamefont {Ezquiaga}, \citenamefont {Garcia-Bellido},\ and\ \citenamefont {Ruiz~Morales}}]{Ezquiaga:2017fvi}%
  \BibitemOpen
  \bibfield  {author} {\bibinfo {author} {\bibfnamefont {J.~M.}\ \bibnamefont {Ezquiaga}}, \bibinfo {author} {\bibfnamefont {J.}~\bibnamefont {Garcia-Bellido}}, \ and\ \bibinfo {author} {\bibfnamefont {E.}~\bibnamefont {Ruiz~Morales}},\ }\href {\doibase 10.1016/j.physletb.2017.11.039} {\bibfield  {journal} {\bibinfo  {journal} {Phys. Lett. B}\ }\textbf {\bibinfo {volume} {776}},\ \bibinfo {pages} {345} (\bibinfo {year} {2018})},\ \Eprint {http://arxiv.org/abs/1705.04861} {arXiv:1705.04861 [astro-ph.CO]} \BibitemShut {NoStop}%
\bibitem [{\citenamefont {Drees}\ and\ \citenamefont {Xu}(2021)}]{Drees:2019xpp}%
  \BibitemOpen
  \bibfield  {author} {\bibinfo {author} {\bibfnamefont {M.}~\bibnamefont {Drees}}\ and\ \bibinfo {author} {\bibfnamefont {Y.}~\bibnamefont {Xu}},\ }\href {\doibase 10.1140/epjc/s10052-021-08976-2} {\bibfield  {journal} {\bibinfo  {journal} {Eur. Phys. J. C}\ }\textbf {\bibinfo {volume} {81}},\ \bibinfo {pages} {182} (\bibinfo {year} {2021})},\ \Eprint {http://arxiv.org/abs/1905.13581} {arXiv:1905.13581 [hep-ph]} \BibitemShut {NoStop}%
\bibitem [{\citenamefont {Cheong}\ \emph {et~al.}(2021)\citenamefont {Cheong}, \citenamefont {Lee},\ and\ \citenamefont {Park}}]{Cheong:2021vdb}%
  \BibitemOpen
  \bibfield  {author} {\bibinfo {author} {\bibfnamefont {D.~Y.}\ \bibnamefont {Cheong}}, \bibinfo {author} {\bibfnamefont {S.~M.}\ \bibnamefont {Lee}}, \ and\ \bibinfo {author} {\bibfnamefont {S.~C.}\ \bibnamefont {Park}},\ }\href {\doibase 10.1007/s40042-021-00086-2} {\bibfield  {journal} {\bibinfo  {journal} {J. Korean Phys. Soc.}\ }\textbf {\bibinfo {volume} {78}},\ \bibinfo {pages} {897} (\bibinfo {year} {2021})},\ \Eprint {http://arxiv.org/abs/2103.00177} {arXiv:2103.00177 [hep-ph]} \BibitemShut {NoStop}%
\bibitem [{\citenamefont {Ghoshal}\ \emph {et~al.}(2024)\citenamefont {Ghoshal}, \citenamefont {Okada}, \citenamefont {Paul},\ and\ \citenamefont {Raut}}]{Ghoshal:2024hfk}%
  \BibitemOpen
  \bibfield  {author} {\bibinfo {author} {\bibfnamefont {A.}~\bibnamefont {Ghoshal}}, \bibinfo {author} {\bibfnamefont {N.}~\bibnamefont {Okada}}, \bibinfo {author} {\bibfnamefont {A.}~\bibnamefont {Paul}}, \ and\ \bibinfo {author} {\bibfnamefont {D.}~\bibnamefont {Raut}},\ }\href@noop {} {\  (\bibinfo {year} {2024})},\ \Eprint {http://arxiv.org/abs/2405.10537} {arXiv:2405.10537 [astro-ph.CO]} \BibitemShut {NoStop}%
\end{thebibliography}%

\end{document}